\documentclass[11pt,a4paper]{article}
\usepackage[utf8]{inputenc}
\usepackage{jheppub}
\usepackage{color}

\usepackage{amsmath}
\usepackage{amssymb}
\usepackage{hyperref}

\newcommand{\bmat}{\left ( \begin{array}{cc}}
\newcommand{\emat}{ \end{array}\right )}

\renewcommand\epsilon\varepsilon
\renewcommand\phi\varphi

\newcommand\be{\begin{eqnarray}}
\newcommand\ee{\end{eqnarray}}
\newcommand{\nn}{\nonumber}
\newcommand{\eref}[1]{(\ref{#1})}

\DeclareMathOperator{\tr}{tr}

\DeclareMathOperator*{\argmin}{arg\,min}

\graphicspath{{figures/}}

\newcommand{\UN}{\mathrm{U}(N)}
\newcommand{\GLN}{\mathrm{GL}(N,\mathbb{C})}

\begin{document}

\title{\boldmath Complex Langevin Simulation of a Random Matrix Model at Nonzero Chemical Potential}
%%%
\author[a]{J.~Bloch} 
\author[b]{, J.~Glesaaen} 
\author[c]{, J. J. M.~Verbaarschot}
\author[d,e,f]{and S.~Zafeiropoulos.}
\affiliation[a]{Department of Physics, University of Regensburg, Regensburg, Germany}
\affiliation[b]{Department of Physics, Swansea University, Swansea, United Kingdom}
\affiliation[c]{Department of Physics and Astronomy, Stony Brook University, Stony Brook, New York 11794, USA}
\affiliation[d]{Institute for Theoretical Physics, Heidelberg University,
Philosophenweg 12, 69120 Heidelberg, Germany}
\affiliation[e]{Department of Physics, The College of William \& Mary, Williamsburg, VA 23187, USA}
\affiliation[f]{Thomas Jefferson National Accelerator Facility, Newport News, VA 23606, USA}
\emailAdd{jacques.bloch@ur.de}
\emailAdd{jonas.glesaaen@swansea.ac.uk}
\emailAdd{jacobus.verbaarschot@stonybrook.edu}
\emailAdd{zafeiropoulos@thphys.uni-heidelberg.de}
%%%%

\date{\today}

\abstract{
In this paper we test the complex Langevin algorithm for numerical simulations of a random matrix model of QCD with a first order phase transition to a phase of finite baryon density. We observe that a naive implementation of the algorithm leads to phase quenched results, which were also derived analytically  in this article.  We test several fixes for the convergence issues of the algorithm, in particular the method of gauge cooling, the shifted representation, the deformation technique and reweighted complex Langevin, but only the latter method reproduces the correct analytical results in the region where the quark mass is inside the domain of the eigenvalues. In order to shed more light on the issues of the methods we  also apply them to a similar random matrix model with a milder sign problem and no phase transition, and in that case gauge cooling cooling solves the convergence problems as was shown before in the literature.
}

\maketitle

\allowdisplaybreaks[3]

% !TEX root = ../cl.tex
\section{Introduction}

A first principles study of the QCD phase diagram in the plane of temperature
($T$) and baryon chemical potential ($\mu$) is one of the most challenging
problems of modern high energy physics. Its understanding will lead to profound
answers ranging from cosmology and the early universe to the
physics of neutron stars.  Analytical approaches tend to fail because the theory is strongly interacting, and only for extreme values of the
temperature and/or the baryon chemical potential can the theory be studied
perturbatively due to asymptotic freedom.  Lattice numerical simulations have
contributed tremendously to the understanding of the vacuum properties of the
theory and have also firmly established that with physical quark masses the
deconfinement transition at zero baryon chemical potential is a crossover.
Despite all these celebrated results the situation at finite baryon density is
very different \cite{Philipsen:2010gj} and our knowledge is mainly based on
models for QCD or lattice simulations at small values of $\mu$ (more precisely
$\mu/T<1$ and $\mu < m_\pi/2$ \cite{Splittorff:2006fu}).

It is well known that the culprit behind this lack of results is the infamous
sign problem which is prohibiting numerical simulations when $\mu/T >1$ or
$\mu>m_\pi/2$. The determinant of the Dirac operator becomes complex for
SU$(N_c)$  Yang-Mills theories with $N_c \ge 3$ and quarks in the fundamental
representation. Consequently, standard Markov Chain Monte Carlo (MCMC) methods,
which require a real and positive probability weight, cannot be applied.
There have been many attempts at tackling this problem by the QCD community.
Some of them try to circumvent the sign problem, while others study related
theories which have no sign problem. To circumvent the sign problem one can
perform a Taylor expansion around $\mu=0$ \cite{Allton:2002zi} or use
reweighting methods \cite{Barbour:1991vs}, however, these methods cannot go
beyond $\mu/T \geq1$ at physical quark masses due to serious problems such as
the limited radius of convergence of the Taylor series or the exponentially
small reweighting factor. Alternatively, one can study QCD with imaginary baryon
chemical potential \cite{DElia:2002tig, deForcrand:2002hgr}, or perform
simulations of two-color QCD or of QCD with adjoint quarks
\cite{Kogut:2000ek,Allton:2002zi,Kogut:2002zg}, which have no sign problem at
all. However, as these theories have a different phase diagram from the one of
QCD, one can at best extract qualitative information regarding the QCD phase
diagram. 

A method that has attracted a great deal of attention recently, and which is not
based on MCMC methods, is the method of stochastic quantization, also called
Langevin method. For the case of complex actions, the complex Langevin (CL)
method  was pioneered independently by Parisi~\cite{Parisi:1984cs} and Klauder
~\cite{Klauder:1983nn} more than 30 years ago. Despite the fact
that stochastic quantization yields the same results as path integral quantization for systems with a real action, this is, unfortunately,
not always the case when the action is complex. One of the most serious problems
is that the method sometimes converges towards the wrong limit. Convergence
criteria have been established \cite{Aarts:2009uq}, however, these are not
fulfilled in realistic QCD simulations, at least for the range of parameters
that are of interest for mapping the unknown part of the QCD phase diagram
\cite{Fodor:2015doa,Bloch:2017jzi}. Nevertheless, the CL algorithm has given
correct results in many non-trivial systems for which we know the solution, and
it seems to be quite successful for QCD simulations in the deconfined phase
\cite{Fodor:2015doa,Bloch:2017jzi}, as well as in simulations for heavy quarks
\cite{Seiler:2012wz,Sexty:2013ica,Aarts:2014bwa,Aarts:2017vrv}, where the
results can be validated by other methods.   

In this article we are attempting to understand the properties of the algorithm
very close to the chiral limit in the cold and dense regime. To achieve that, we
are studying a random matrix theory (RMT) model which shares many key features
of QCD such as spontaneous breaking of chiral symmetry, a finite density phase
transition, as well as a complex fermion determinant, which causes a strong sign
problem. The model that we have been studying was introduced by Stephanov
\cite{Stephanov:1996ki} based on a random matrix model for the finite
temperature chiral phase transition \cite{Jackson:1995nf}. There is a
significant literature studying the convergence properties of the CL algorithm
in RMT but all the existing studies are based on a finite density model
introduced by Osborn \cite{Osborn:2004rf} (or an improved version thereof
\cite{Bloch:2012bh}), which possesses many similarities with the one by
Stephanov but also has big differences, most notably the lack of a phase
transition to a nonzero baryon density phase. \textit{Sensu stricto} the Osborn
model is only a model of QCD in the confined phase at small chemical potential.

A great deal of analytical knowledge for non-perturbative aspects of QCD came
from RMT studies. These include among others finite density results
\cite{Halasz:1997he}, lattice spacing effects on the lowest eigenvalues of the
Dirac operator for Wilson fermions \cite{Damgaard:2010cz,
  Akemann:2010em,Kieburg:2011uf, Kieburg:2013xta, Kieburg:2015vqa,
  Cichy:2016tyj}, and the effect of topology on the Dirac spectrum
\cite{Verbaarschot:1994qf}. In this article we are addressing the convergence
properties of the CL algorithm for a model of continuum QCD at nonzero chemical
potential. The model has a known analytic solution, and by simulating it
numerically we can get an explicit handle on the various issues of the
algorithm.

This article starts out with the definition of the random matrix models that will be
studied by the CL algorithm which is introduced in section 3. In section 4, we discuss
the fermion determinant and the spectrum of the Dirac operator. The CL reweighting method is analyzed in section 5, while the shifted representation, in which the
chemical potential is shifted to the bosonic part of the action, is discussed in section 6.
Cooling methods are investigated for two different random matrix models in section 7. As a last attempt to fix the convergence problems of the CL algorithm, we
study the deformation method in section 8. Concluding remarks are made in section
9, and analytical results for the phase quenched partition function are worked out
in Appendix. A preliminary account of some of the results in this paper
appeared as conference proceedings \cite{Bloch:2016jwt,Bloch:2017LAT}.
% !TEX root = ../cl.tex

\section{Random Matrix Model}
\label{Section:RMT}

In this section we discuss a random matrix theory inspired model \cite{Shuryak:1992pi,Jackson:1995nf} for QCD at finite baryon density originally proposed by Stephanov \cite{Stephanov:1996ki}. The model's partition function reads
\begin{equation}
  \mathcal{Z}^{N_f}_N=e^{N\mu^2}\int dW dW^{\dagger}{\det}^{N_f}(D+m)e^{-N\,\tr WW^\dagger}.
  \label{Zst}
\end{equation}
The Dirac operator $D$ has the form
\begin{equation}
  D=
  \begin{pmatrix}
    0 & iW+\mu\\
    iW^\dagger+\mu & 0
  \end{pmatrix}
  ,
  \label{Dsteph}
\end{equation}
where a term containing the baryon chemical potential $\mu\gamma_0$ has been
coupled to the chRMT Dirac operator  proposed in
\cite{Shuryak:1992pi,Verbaarschot:1994qf}. The $N\times (N+\nu)$ matrix elements
of $W$ are complex numbers, $N$ is the size of the block matrix $W$ and the
index $\nu$ of the Dirac Matrix is the analogue of the topological charge. This
model was first introduced for imaginary chemical potential
\cite{Jackson:1995nf} to study the QCD chiral phase transition at nonzero
temperature (which in the model appears as an imaginary chemical potential).

In this article we choose $\nu=0$, since the topological charge does not have a
significant effect on the quantities of interest. For $\mu >0$, the eigenvalues
of $D$ become complex, and are roughly distributed homogeneously inside a strip
of width $\sim \mu^2$ (for finite $N$ it is an ellipse). Similarly to QCD,
numerical simulations of this random matrix theory have an exponentially hard
sign problem, especially when the quark mass is inside the cloud of eigenvalues
of the Dirac operator.

Our attention will be focused mainly on two observables, the mass dependent
chiral condensate (note that the physical chiral condensate is $-\Sigma$)
defined by
\begin{equation}
  \Sigma=\frac 1{2N}\frac{\partial \log{\mathcal{Z}^{N_f}_N}}{\partial m},
  \label{PBP}
\end{equation}
and the baryon number density given by
\begin{equation}
  n_B=\frac 1{2N}\frac{\partial \log{\mathcal{Z}^{N_f}_N}}{\partial \mu}.
  \label{nb}
\end{equation}
There is no unique way of introducing a chemical potential in a random matrix
model. Some alternatives turn to be advantageous from a symmetry point of view
\cite{Osborn:2004rf}.  In particular, the Osborn model \cite{Osborn:2004rf} has
a $\UN\times\UN$ symmetry which makes it possible to obtain analytical results
for the joint probability distribution function of the eigenvalues, which is the
starting point of many powerful random matrix methods. The Stephanov model has only a $\UN$ invariance, and it is not possible to obtain an analytical solution
for the joint eigenvalue density. In the case of the Osborn model, by extending
the method of orthogonal polynomials to bi-orthogonal polynomials, all $n$-point
spectral correlators can be obtained.

The Osborn model is a two-matrix model that has a form similar to the Stephanov
model. In the chiral basis it is given by
\begin{equation}
  \mathcal{Z}^{N_f}_N=e^{N\mu^2}\int dWdW' dW^{\dagger}dW'^{\dagger}{\det}^{N_f}(D+m)e^{-N\,\tr (WW^\dagger+W'W'^\dagger)},
  \label{Zosb}
\end{equation}
where the Dirac operator $D$ has the form
\begin{equation}
  D=
  \begin{pmatrix}
    0        & iW+\mu W'\\
    iW^\dagger+\mu W'^{\dagger} & 0
  \end{pmatrix}
  .
  \label{Dosb}
\end{equation}
Remarkably, the partition function at finite baryon density can be related to
the one at zero baryon density by introducing a trivial multiplicative factor
and a mass rescaling as follows \cite{Osborn:2004rf,Bloch:2012ye},
\begin{equation}
  \label{mufactorization}
  \mathcal{Z}^{N_f}_N(m,\mu) = (1-\mu^2)^{N_f N}\mathcal{Z}^{N_f}_N\left(\frac{m}{\sqrt{1-\mu^2}},0\right).
\end{equation}
Consequently, it is natural to expect that the Osborn model does not possess the
rich phenomenological structure of the Stephanov model, which exhibits a phase
transition separating a phase with zero baryon density from a phase with
nonzero baryon density.  Strictly speaking the Osborn model should only be
considered as a model for QCD  at small chemical potential, precisely due to the
absence of a phase transition to a phase with nonzero baryon density. In
addition, one can conclude that the sign problem of the Osborn model is of a weaker nature and therefore may be remedied by some clever techniques
\cite{Bloch:2011jx,Bloch:2012ye,Bloch:2012bh,Mollgaard:2013qra,Mollgaard:2014mga,Nagata:2016alq}.
Both random matrix models possess
the same global symmetries with the same spontaneous symmetry breaking pattern
as in QCD and  yield the $epsilon$ limit of the QCD chiral Lagrangian.  
It is noteworthy that in case of QCD with three colors in the fundamental
representation this chiral Lagrangian does
not have a dependence on the baryon chemical potential. The reason is that
the Goldstone bosons, i.e., the pions, do not carry baryon charge.
The chiral Lagrangian of phase quenched QCD, where $\mu$ becomes the isospin
chemical potential, has a nontrivial $\mu$-dependence which, at the
mean field level, or in the $\epsilon$ domain, is given by the $\mu$ dependence
of the large $N$ limit of the partition functions \eref{Zst} or \eref{Zosb}.

We could contemplate other random matrix models where the dependence of the
chemical potential is integrated out in the evaluation of the partition
function. For example, the Dirac operator
\begin{equation}
  D=
  \begin{pmatrix}
    0        & iW+\mu e^{i\phi}\\
    iW^\dagger+\mu e^{i\phi} & 0
  \end{pmatrix} ,
\end{equation}
where $\phi$ is uniformly random in $[-\pi,\pi]$, and the partition function is
defined by
\begin{equation}
  \mathcal{Z}^{N_f}_N=e^{N\mu^2}\int d\phi dW dW^{\dagger}{\det}^{N_f}(D+m)e^{-N\,\tr WW^\dagger}.
  \label{new}
\end{equation}
It is clear that the partition function does not depend on $\mu$ while the
eigenvalues of $D$ are complex. Since the chemical potential can be eliminated
by changing the integration contour of the $\phi$ integral, the CL algorithm
should be able to solve this problem correctly. We will, however, not study this
model in this paper.

The unquenched partition function of the Stephanov model was cast analytically
in a form that allows for either an easy numerical evaluation at finite $N$, or
that allows for a complete analytical solution via a saddle point approximation in the
thermodynamic limit where $N\to\infty$ \cite{Stephanov:1996ki,Halasz:1997he}.
For the $N_f=1$ case, the partition function, in units where the chiral
condensate $\Sigma=1$, takes the following $\sigma$-model form via bosonization
methods 
\begin{equation}
  \mathcal{Z}^{N_f=1}_N(m,\mu) = e^{N\mu^2}\int d\sigma d\sigma^* e^{-N\sigma^2} (\sigma \sigma^* +
   m(\sigma+\sigma^*) + m^2 -\mu^2)^N \ ,
  \label{zoneflavor}
\end{equation}
where $\sigma$ is the bosonized version of $\bar{\psi}_L \psi_R$. A change of
variables to polar coordinates renders the angular integral calculable
analytically and yields a modified Bessel function, such that the partition
function can be written as a one-fold integral,
\begin{equation}
  \mathcal{Z}^{N_f=1}_N(m,\mu) = \pi e^{-Nm^2+N\mu^2}\int_0^\infty du
  (u-\mu^2)^N I_0(2mN\sqrt u) e^{-Nu} \ .
  \label{zi0}
\end{equation}

A saddle point analysis of the partition function can be performed in the
thermodynamic limit. This was analyzed in detail in \cite{Halasz:1997he} but we
will repeat some of the main steps here for the convenience of the reader. The
saddle point equation reads
\begin{equation}
  \frac 1{u-\mu^2} = 1- \frac m{\sqrt u}.
\end{equation}
The Stephanov model exhibits a first order phase transition which takes place
when $|Z_{u=u_b}|= |Z_{u=u_r}|$, with $u_b$ and $u_r$ being two different
solutions of the saddle-point equation giving the same free-energy. One can
rewrite this condition 
\begin{equation}
  |(u_b-\mu^2) e^{2m\sqrt u_b -u_b}|= |(\mu^2-u_r) e^{2m\sqrt u_r -u_r}|. 
\end{equation}
This is a transcendental equation that, in the chiral limit, has the solutions
$u_r = 0$ and $u_b = 1+\mu^2$. Therefore in this limit one has the critical
curve
\begin{equation}
  {\rm Re}\left [ 1+\mu^2 +\log \mu^2 \right ] = 0 ,
\end{equation}
which for the case of a real baryon chemical potential leads to the critical
value $\mu_c = 0.527\dots$ in the chiral limit. This critical curve is also
valid for the case of a complex chemical potential. In particular, for an imaginary chemical
potential, there is a second order phase transition to a restored phase at $\mu
=1 $.

\begin{figure}[t!]
  \centering
  \includegraphics[width=8cm]{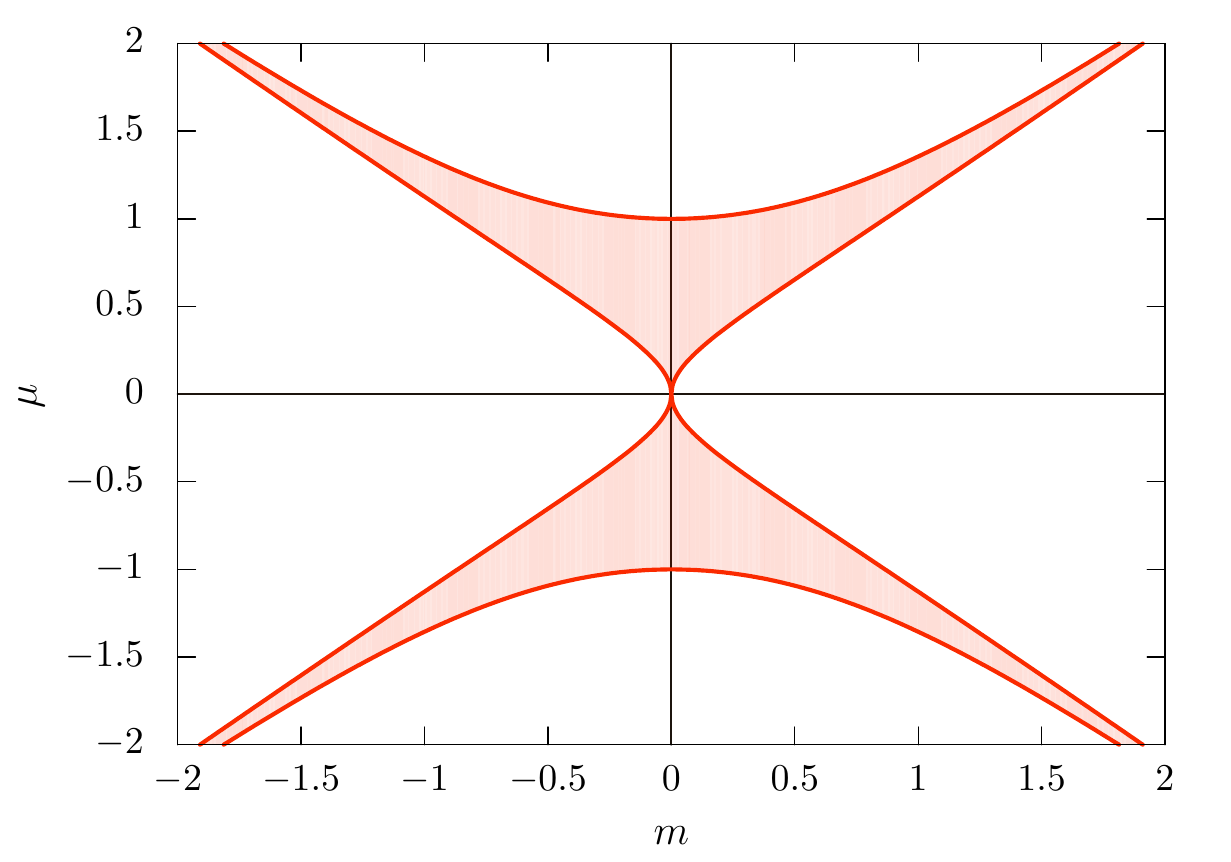}
  \caption{Phase diagram of the two flavor phase quenched random matrix theory
    in the mass-chemical potential plane.
    The shaded area shows the region of the phase diagram with a nonzero pion
    condensate which lies in between the pion condensation transition and a chiral phase transition.
  } 
  \label{fig1}
\end{figure}
 
We will also compare our results to the two-flavor phase quenched RMT partition
function \cite{Toublan:1999hx} or the partition function at nonzero isospin
chemical potential \cite{Alford:1998sd,Son:2000xc}. The two-flavor partition
function is an eight dimensional integral and is much more complicated than the
one-flavor partition function, which is only a two-dimensional integral.
However, in the large $N$ limit, it can be evaluated by a saddle-point
approximation, see Appendix \ref{app1}. It has a pion-condensation phase for
$\mu > m_\pi/2$ corresponding to the parameter domain when the quark mass is
inside the support of the eigenvalues. In Fig.~\ref{fig1} we show the phase
diagram in the plane of the chemical potential and quark masses (which are taken
to be equal for the two flavors). For nonzero mass and increasing chemical
potential, we find a phase transition to a pion condensation phase at
$\mu=m_\pi/2$, and for larger $\mu$, a second phase transition to a chirally
restored phase. In the region between the curves, the quark mass is in the
domain of the eigenvalues and CL is expected to fail. In the outside region, the
mean field result for full QCD and phase quenched QCD coincide, and CL is
expected to work. For QCD we expect a similar forbidden region.

In order to study the properties of the Langevin algorithm we will perform
numerical simulations of the Stephanov model employing the CL algorithm and test
its convergence properties by comparing the obtained numerical data for the
chiral condensate and the baryon density with analytical results computed using
the partition function (\ref{zi0}). In several cases we will also simulate the
Osborn model in order to display potential issues that might arise for the CL
method when switching from a model without a phase transition to one where
the sign problem triggers a phase transition.

\section{Complex Langevin}
\label{CL}

Stochastic quantization and the Langevin equation form a natural bridge between
quantum field theory (QFT) and statistical mechanics. In the case of a real
action, expectation values of the path integral  can be obtained by averaging
over an ensemble of configurations that have been generated by the Langevin
evolution. Here, in order to set the stage and define our notation, we will
consider the one degree of freedom, trivial ``QFT" whose partition function has
the following path integral form $\mathcal{Z}=\int e^{-S(x)} dx$. The
discretized real Langevin equation for updating the dynamical variable $x$ is
\begin{equation}
  x(t+\Delta t)=x(t)-\partial_x S(x(t))\Delta t +\Delta \xi, 
\end{equation}
where the noise term $\Delta \xi$ is a stochastic variable with zero mean and
variance given by $2\sqrt {\Delta t}$. Generalizing the concept of stochastic
quantization to the case of complex actions, requires us to promote every real
degree of freedom to its complex counterpart. This complexification will
naturally occur when evolving the degrees of freedom according to the Langevin
equation, as the derivative of the action, usually coined as the drift term
($\partial_x S(x(t))\Delta t$), is complex and will push the dynamical variables
into the complex plane. In this case,  $x$ will give its place to $z=x+iy$,
whose evolution as a function of the Langevin time $t$ will be given by the
following update equation
\begin{equation}
  z(t+\Delta t)=z(t)-\partial_z S(z(t))\Delta t +\Delta \xi.
\end{equation}
The Langevin equation thus generates a probabilistic ensemble $\{z(t)\}$ where
observables are calculated by averaging along the Langevin trajectory. One can
quite easily generalize the Langevin equation from systems with one degree of
freedom to more complicated systems, such as field theories with an infinite
number of
degrees of freedom. In our case we need to modify this formalism for the case of
an RMT model, which can be done in a straightforward way as is shown below. 

The complex random matrix $W$ in the random matrix model \eref{Dsteph}, in its
Cartesian representation, can be decomposed as $W=A+iB$ where
$W^{\dagger}=A^{\top}-iB^{\top}$ with $A$ and $B$ both real.  In this case the
measure of integration $dW dW^{\dagger}$ becomes $dA dB$. The action
corresponding to the partition function \eref{Zst} reads 
\begin{equation}
  S=N\tr(W^{\dagger}W)-N_f \tr(\log(m^2-\mu^2 +W^{\dagger}W-i\mu (W+W^{\dagger}))).
  \label{action}
\end{equation}
At finite chemical potential the matrices $A$ and $B$ will take on complex
values due to the complex Langevin flow. We therefore introduce the complexified
matrices $X = A + i B$ and $Y = A^{\top} - i B^{\top}$ which will replace $W$
and $W^{\dagger}$ in the following expressions. At $\mu = 0$ we have
$X^{\dagger} = Y$, however this will not be the case as $A$ and $B$ become
complex.  The matrices $A$ and $B$ will have the following Langevin evolution
\begin{align}
  A^{(n+1)}_{mn} &= A^{(n)}_{mn}-2N\Delta t A_{mn}+N_f \Delta t %
    [(X G)_{mn}+(G Y)^{\top}_{mn} - i\mu (G_{mn}+G_{mn}^{\top})]+\Delta \xi , 
  \label{cartesianAupd} \\
  B^{(n+1)}_{mn} &= B^{(n)}_{mn}-2N\Delta t B_{mn}+N_f \Delta t %
    [(X G)_{mn}-(G Y)^{\top}_{mn}-i\mu (G_{mn}^{\top}-G_{mn})]+\Delta \xi .
  \label{cartesianBupd}
\end{align}
where we have simplified the notation by introducing the matrix $G$,
\begin{equation}
  G=(m^2-\mu^2 + Y X -i\mu (X+Y))^{-1}.
\end{equation}

\begin{figure}[t!]
  \centering
  \includegraphics[width=0.48\linewidth]{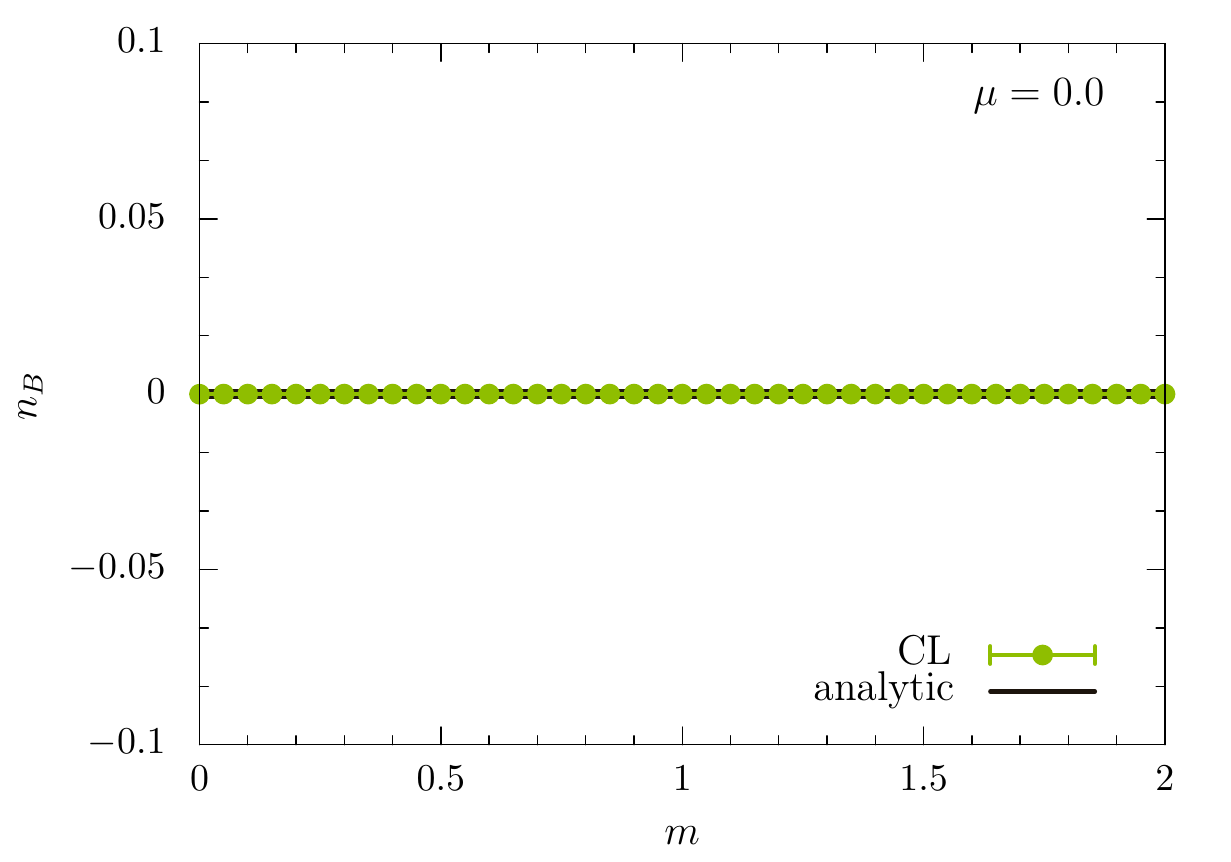}
  \hspace{1mm}
  \includegraphics[width=0.48\linewidth]{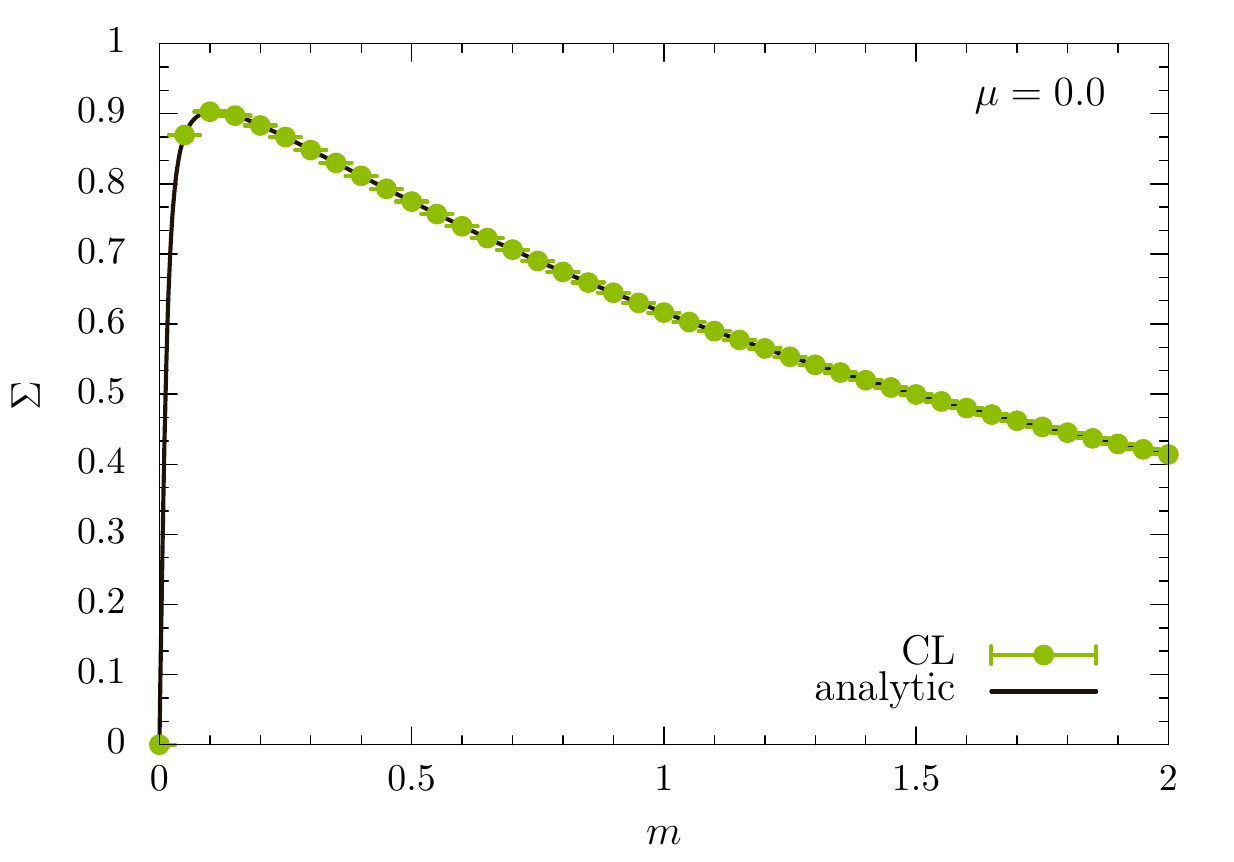}
  \caption{The chiral condensate, $\Sigma$ (RHS), and
    the baryon number density, $n_B$  (LHS), for the
    random matrix model \eref{Dsteph} plotted as a function of $m$ for $\mu=0$.}
  \label{numsanity0}
\end{figure}

\begin{figure}[t!]
  \centering
  \includegraphics[width=0.48\linewidth]{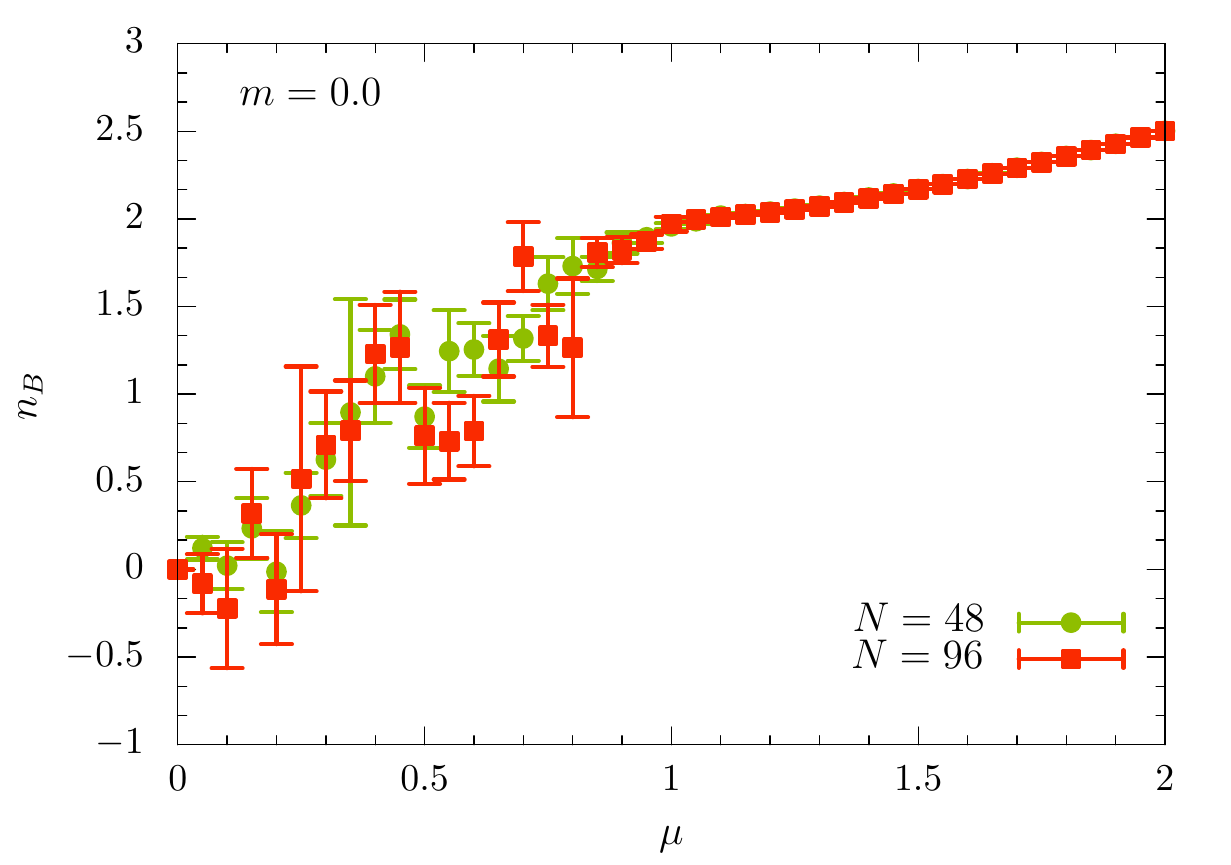}
  \hspace{1mm}
  \includegraphics[width=0.48\linewidth]{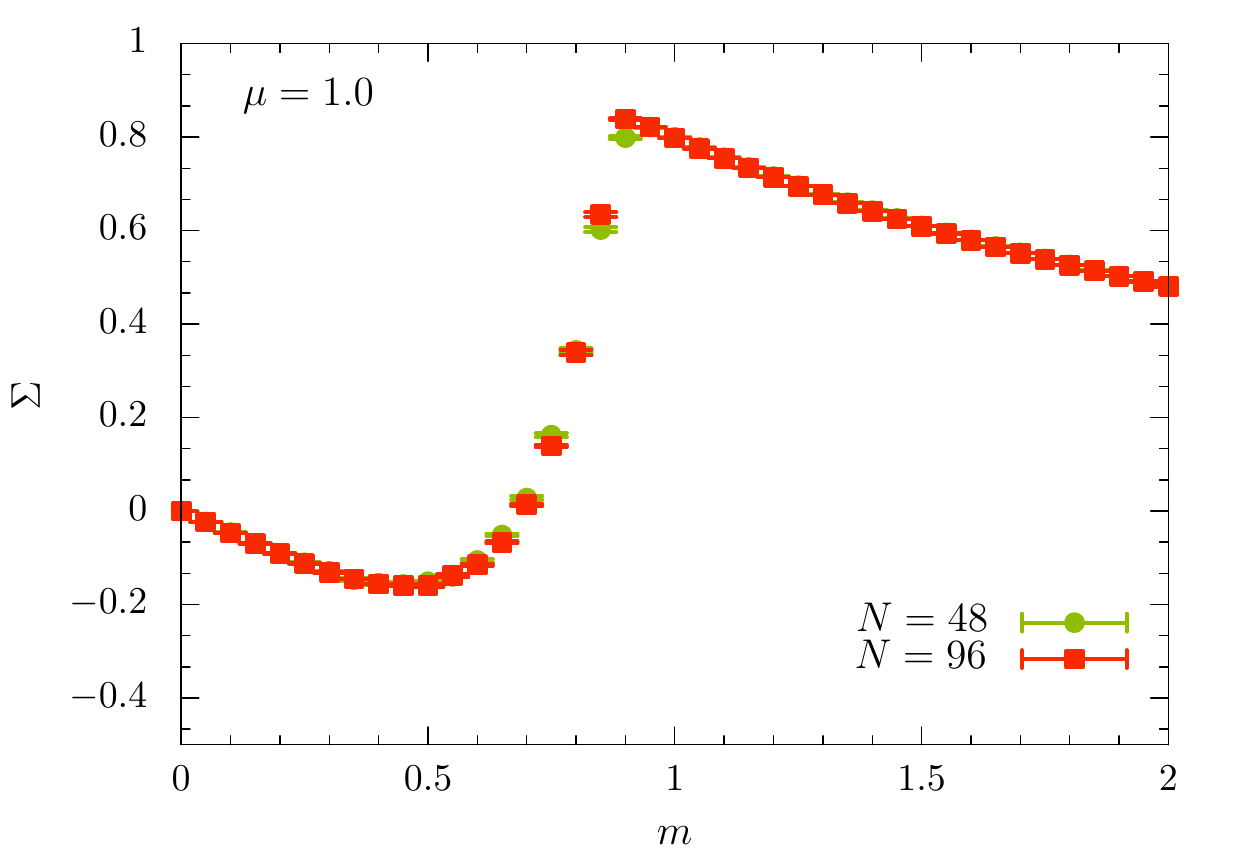}
  \caption{Matrix size sensitivity for the random matrix model \eref{Dsteph}:
    the chiral condensate, $\Sigma$, versus $m$ for $\mu=1$ (RHS), and the
    baryon number density, $n_B$, versus $\mu$ for $m=0$ (LHS).}
  \label{numsanity}
\end{figure}

\begin{figure}[t!]
  \centering
  \includegraphics[width=0.48\linewidth]{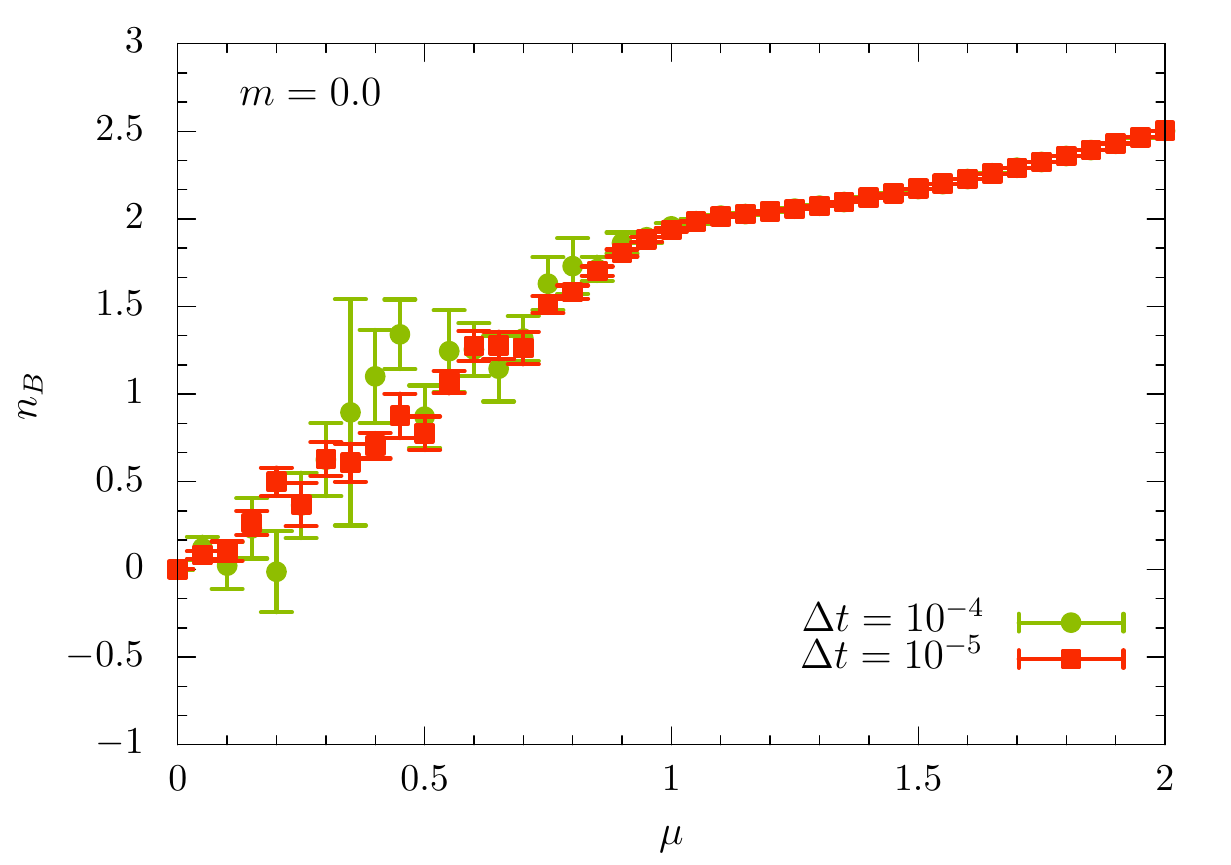}
  \hspace{1mm}
  \includegraphics[width=0.48\linewidth]{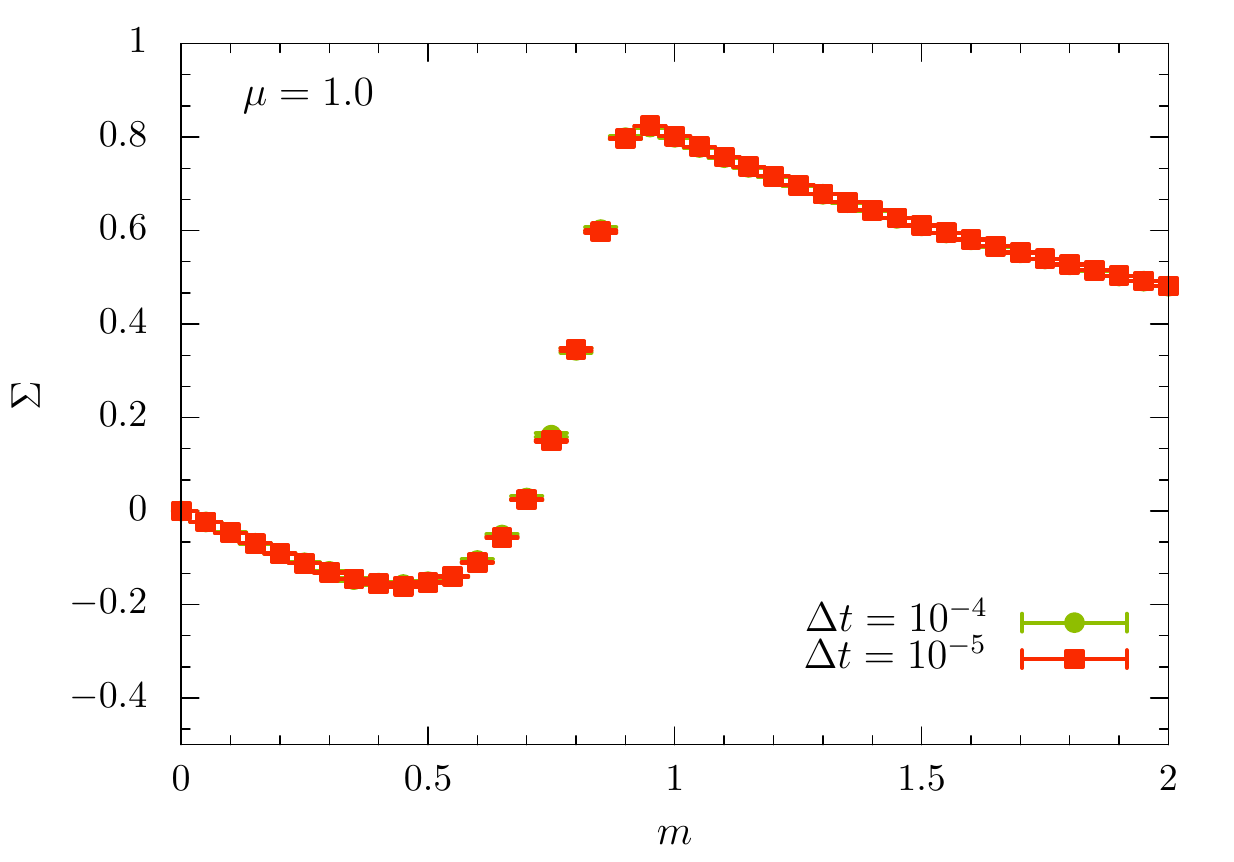}
  \caption{Step size sensitivity for the random matrix model \eref{Dsteph}:  the chiral condensate, $\Sigma$, versus $m$ for $\mu=1$ (RHS), and the
    baryon number density, $n_B$, versus $\mu$ for $m=0$(LHS).}
  \label{numsanity2}
\end{figure}

%%%%mu-scan compare with pq
%%%%%%%%%%
\begin{figure}[t!] 
  \centering
  \includegraphics[width=.48\linewidth]{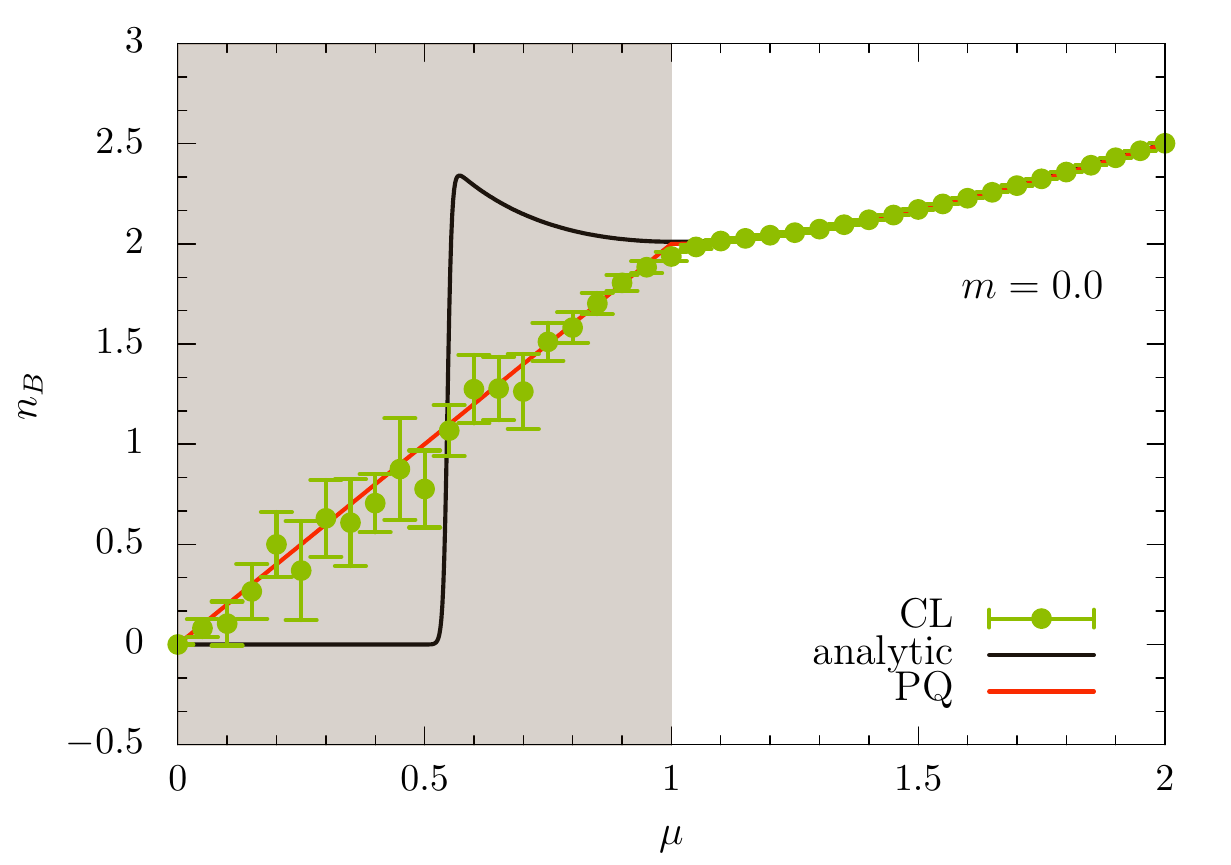} 
  \includegraphics[width=.48\linewidth]{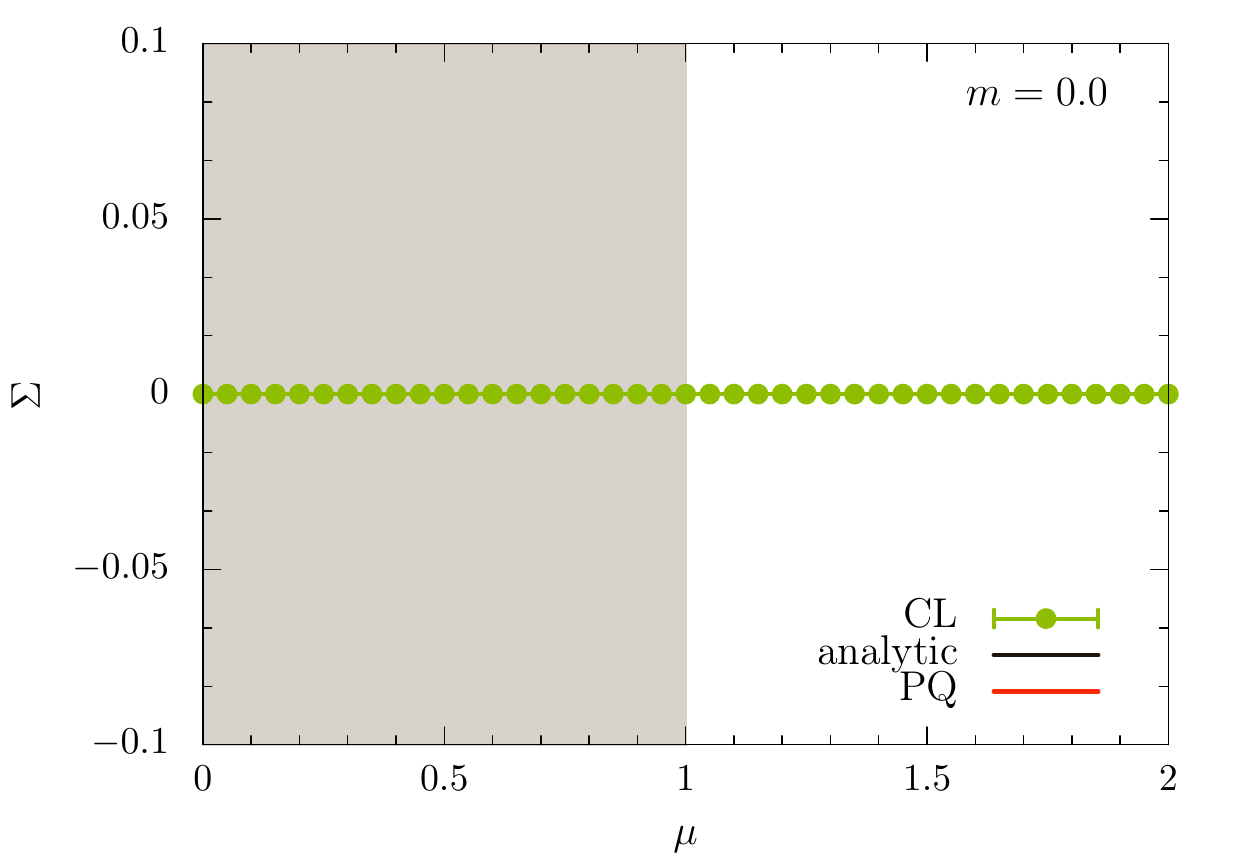}
  \\[1ex]
  \includegraphics[width=.48\linewidth]{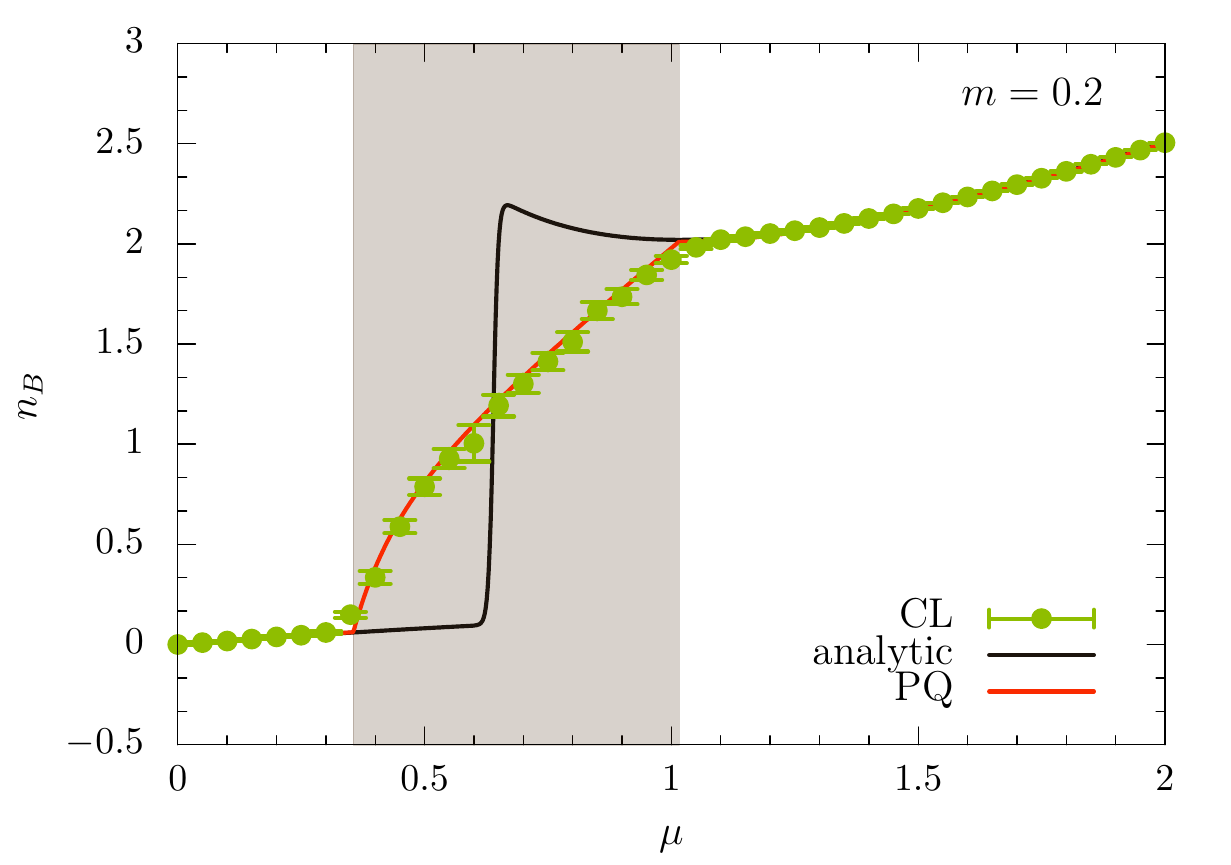} 
  \includegraphics[width=.48\linewidth]{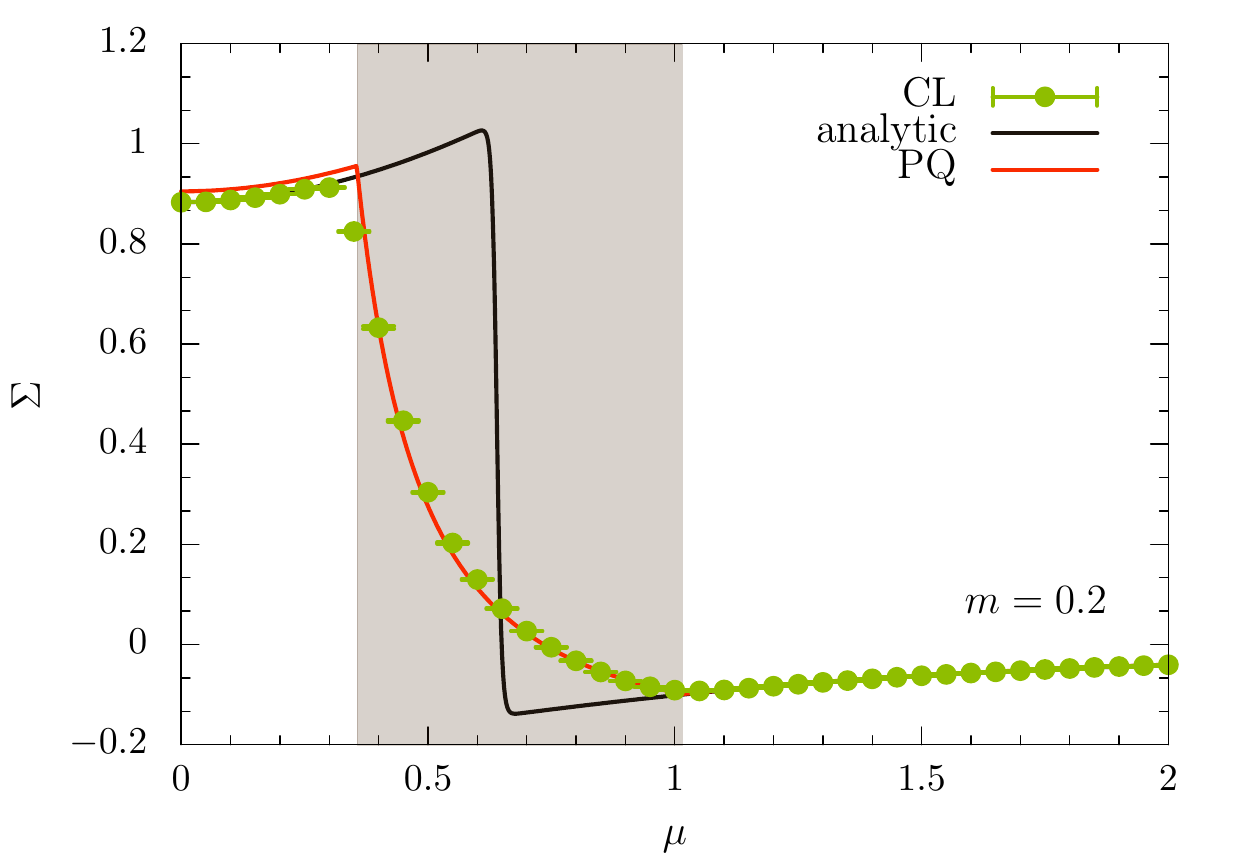} 
  \\[1ex]
  \includegraphics[width=.48\linewidth]{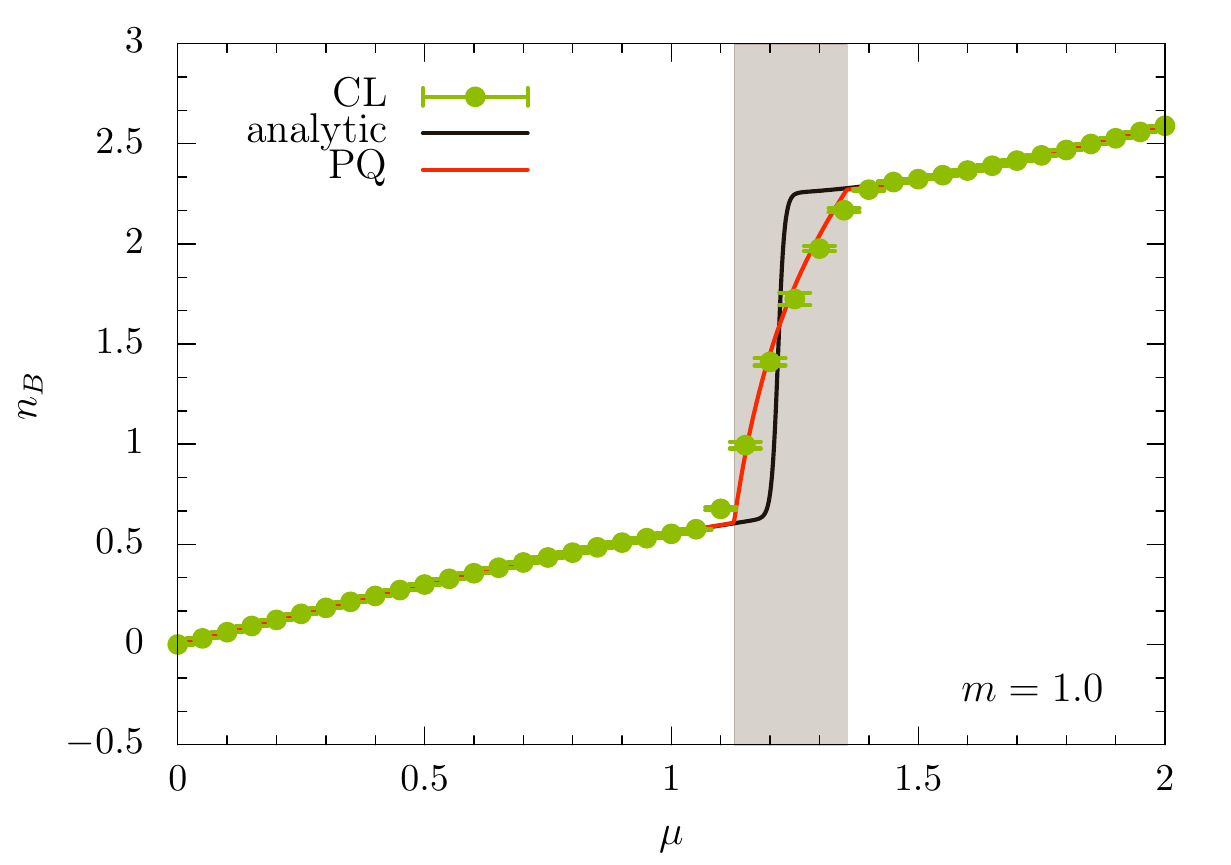} 
  \includegraphics[width=.48\linewidth]{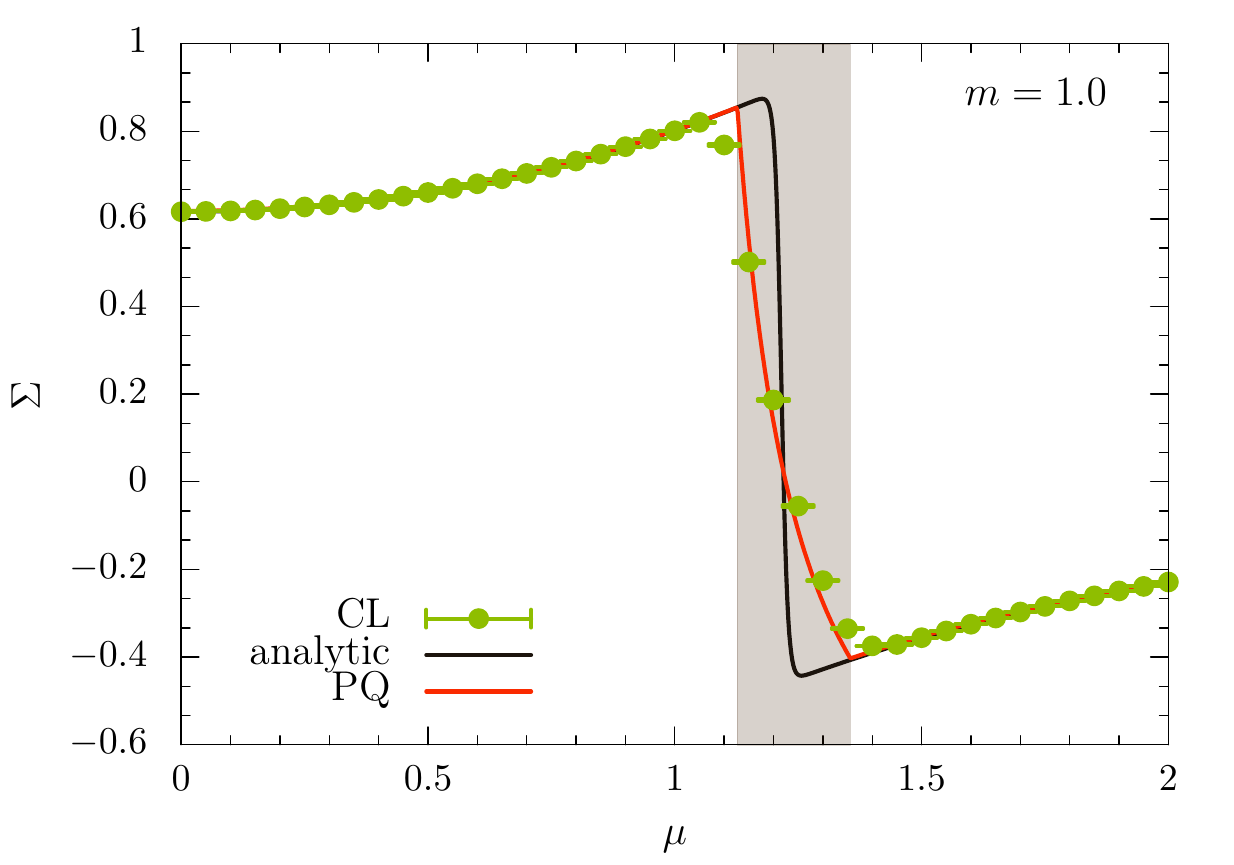} 
  \caption{$\mu$-scan for $m=0$ (upper panels), $m=0.2$ (middle panels) and $m=1$
    (lower panels) for the random matrix model \eref{Dsteph} with matrix size
    $N=48$. The shaded region corresponds to that outlined in Fig.~\ref{fig1}.
    Again we show the baryon number density on the left and the chiral
    condensate on the right.}
  \label{muscan}
\end{figure}

The first step in our simulations is to establish  that in the absence of a
baryon chemical potential, when the action is real, the real Langevin
simulations give the correct analytical answer. This is shown  in
Fig.~\ref{numsanity0} where the baryon number (left) and the chiral condensate
(right) are plotted as a function of the mass $m$. We also check the $N$
dependence of our results to confirm that the results are independent of the
matrix size. This is also important when comparing our results to those from
the phase quenched analytic results, as these are computed in the large $N$
limit.  Fig.~\ref{numsanity} shows the baryon number (left) and the chiral
condensate (right) at two different matrix sizes, and we conclude that $N=48$ is
sufficient. Finally, we show that our results do not depend on the step size of
the discretized Langevin equation. As is demonstrated in Fig.~\ref{numsanity2}
for the baryon number (left) and the chiral condensate (right), a choice of
$\Delta t=10^{-4}$ is sufficient to eliminate discretization errors.

\begin{figure}[t!] 
  \centering
  \includegraphics[width=.48\linewidth]{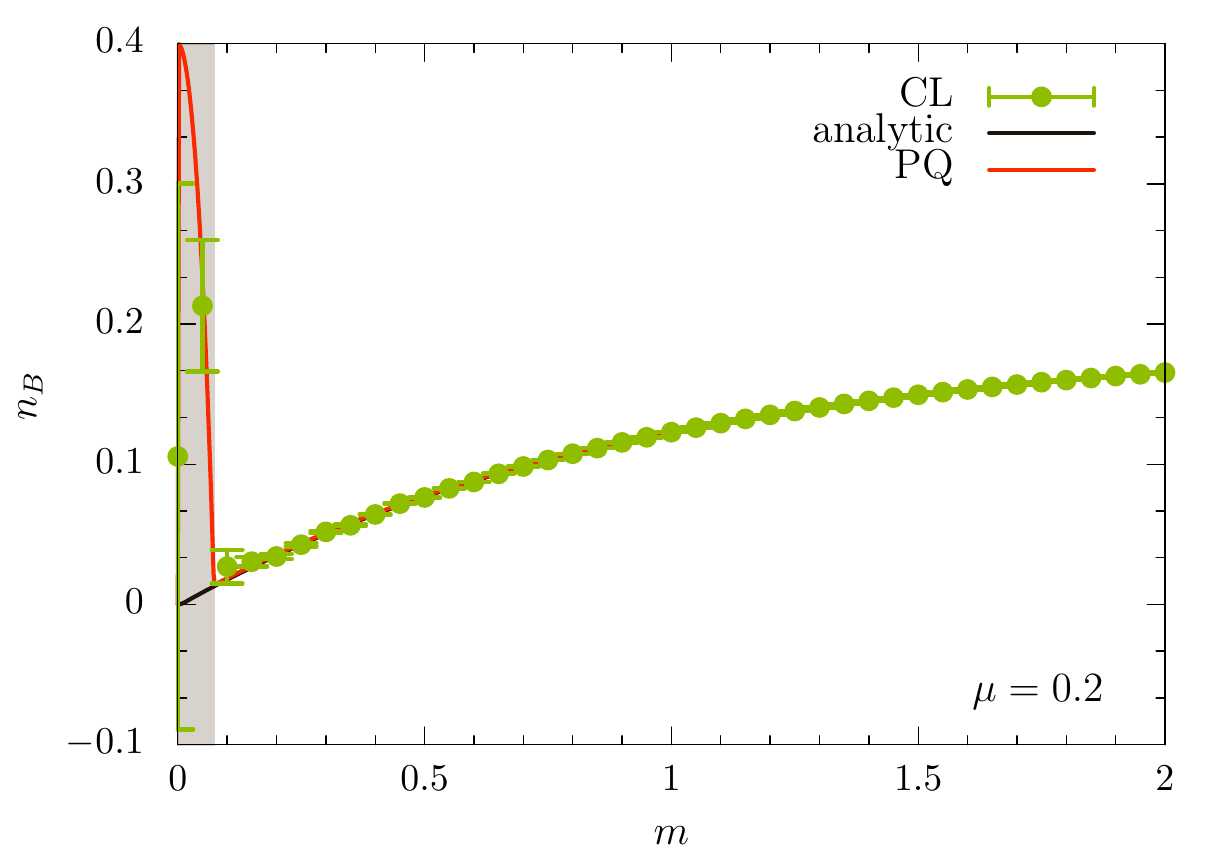} 
  \includegraphics[width=.48\linewidth]{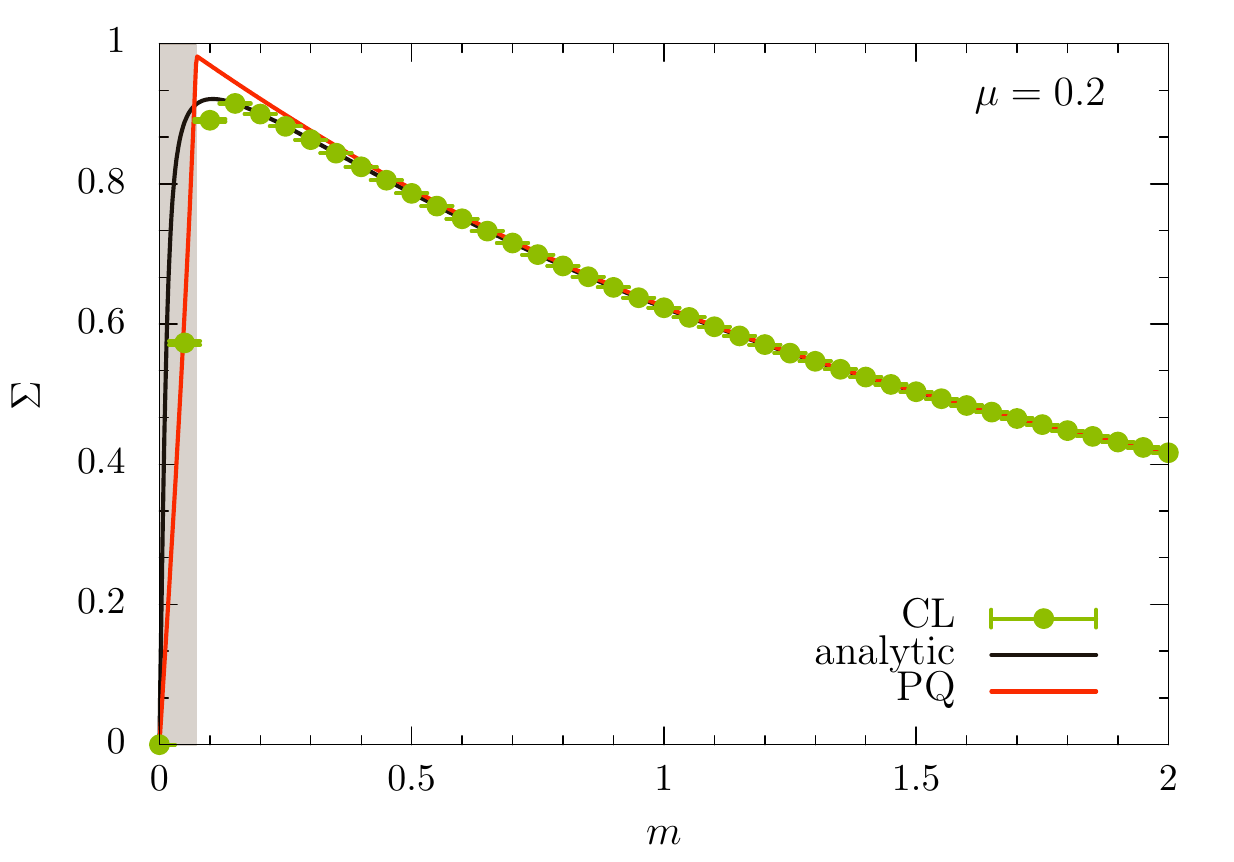} 
  \\[1ex]
  \includegraphics[width=.48\linewidth]{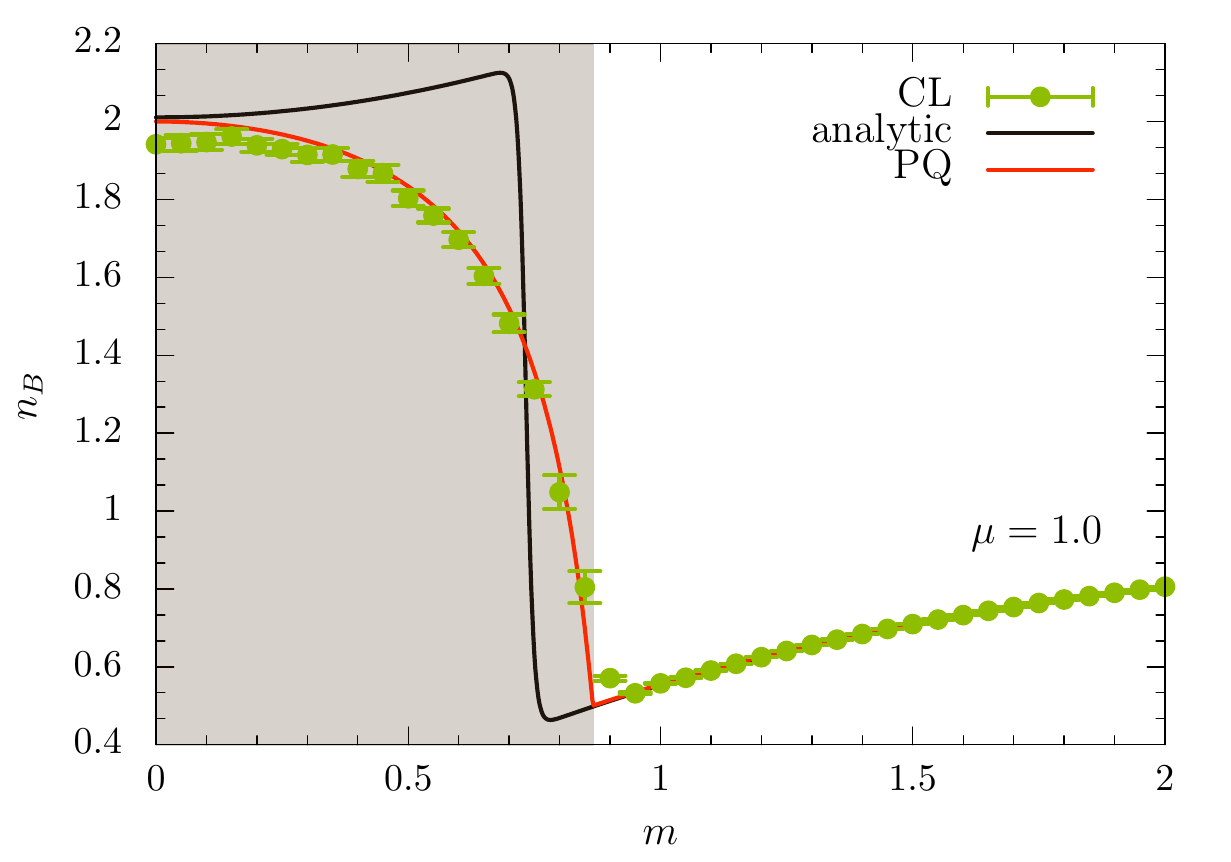} 
  \includegraphics[width=.48\linewidth]{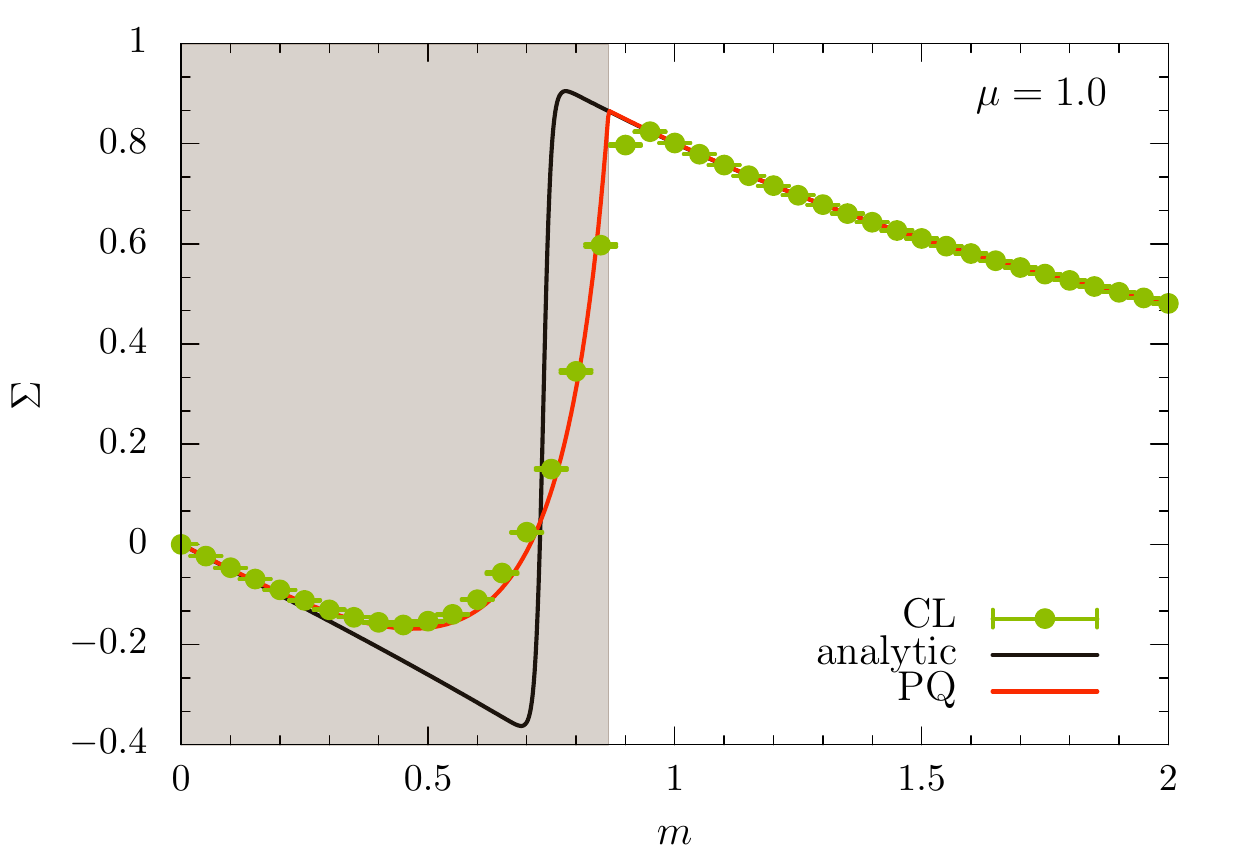} 
  \caption{Mass scan for $\mu=0.2$ (upper panels) and $\mu=1$ (lower panels) for
    the random matrix theory \eref{Dsteph} with matrix size  $N=48$. In both
    cases we plot the baryon number density on the left and the chiral
    condensate on the right.}
  \label{mscan}
\end{figure}

Having convinced ourselves that the CL algorithm has been implemented correctly
we now simulate the random matrix model at nonzero chemical potential and
compare the numerical data  to the analytical results, as well as the
corresponding large $N$ phase quenched ones. In Fig.~\ref{muscan} we show the
baryon density (left) and the chiral condensate (right) as a function of $\mu$
for $m=0$ (upper row), $m=0.2$ (middle row) and $m=1$ (bottom row). Quite
surprisingly, we find that our numerical CL results agree with the analytical
phase quenched results, and only see agreement with the dynamical one flavor
results when these coincide with the phase quenched results. This is the case
when the quark mass is outside the domain of the eigenvalues of the Dirac
operator, or equivalently, when the chemical potential is outside the domain of
the eigenvalues of $\gamma_0 D$. In Fig.~\ref{mscan} we show the baryon density
and the chiral condensate as a function of the quark mass for $\mu=0.2$ (top
row) and $\mu=1$ (bottom row), from which we draw similar conclusions.
The shaded regions in Figs.~\ref{muscan} and \ref{mscan} and further down in
the paper
denote the region
where the analytical one-flavor mean field results do not agree with the phase
quenched mean field results.

We thus conclude that the CL algorithm fails in the region when the baryon
number density and the chiral condensate are not holomorphic functions of the
matrix elements. We will discuss this in more detail in the next section where
we discuss the fermion determinant and the Dirac spectrum.

%%%%%%%%%%%%%%%%%%%%%%%%%%
%%%% pq-comparison

%%%%%%%%%%%%%%%%%%%%%%%%%%%%%%%%%%%
\section{The Dirac Spectrum and the Fermion Determinant}
%%%%%%%%%%%%%%%%%%%%%%%%%%%%%%%%%%%%%

\begin{figure}[t!]
  \centering
  \includegraphics[width=.95\linewidth]{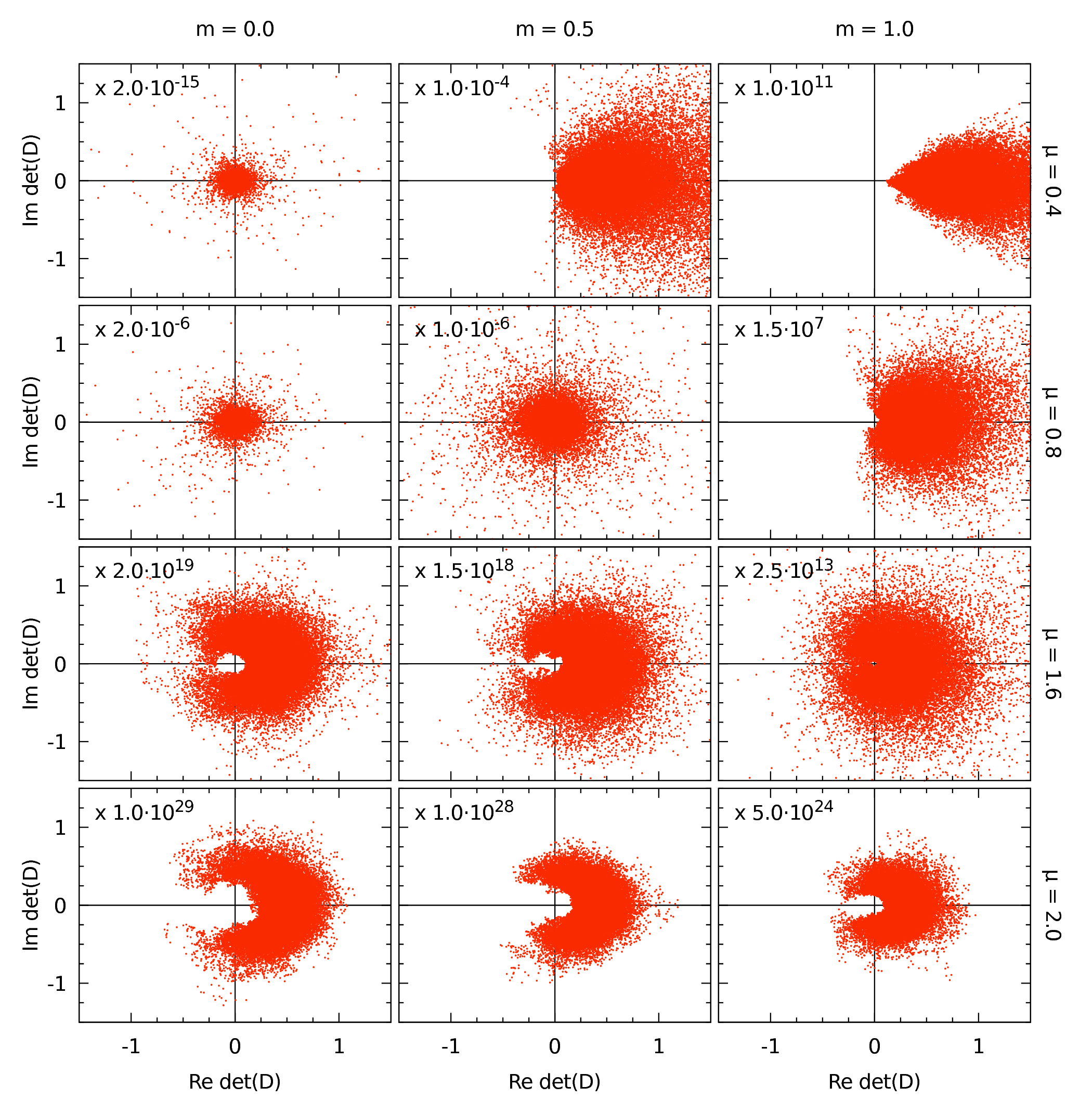}
  \caption{Scatter plots of the fermion determinant of the random matrix model
    \eref{Dsteph} with matrix size $N=48$ for  quark masses $m=\{0.0, 0.5,
    1.0\}$, and chemical potentials $\mu= \{0.4, 0.8, 1.6, 2.0\}$.}
  \label{detmnz} 
\end{figure}

One of the requirements for the correct convergence of CL is that the
``operator'', in our case
\begin{equation}
  {\rm Tr} \frac 1{D+m}\qquad {\rm and} \qquad {\rm Tr}\gamma_0 \frac 1{D+m},
\end{equation}
is a holomorphic function of the complexified variables. This is not the case
for the chiral condensate when the quark mass is inside the two-dimensional
locus of the eigenvalues of $D$, or equivalently,  for the baryon density if the
chemical potential is inside the spectral support of $\gamma_0D$, which is also
a two-dimensional domain.  So for the CL to be convergent we need to require
that $\det(D+\mu\gamma_0+ m ) =\det(\gamma_0(D+m) +\mu)>\epsilon $ with
$\epsilon$ a finite constant.  In Fig.~\ref{detmnz} we show scatter plots of the
determinant in the complex plane, obtained during the CL simulation for zero and
nonzero mass, respectively. Indeed, we will find that simulations converge well
when the flow of the determinant avoids the origin, as has also been observed,
for example, for the Osborn RMT model \cite{Mollgaard:2014mga} and for
two-dimensional QCD \cite{Bloch:2015coa}.

%%%%det-scatter plots
%%%%%%%%%%

\begin{figure}
  \centering
  \includegraphics[width=.90\linewidth]{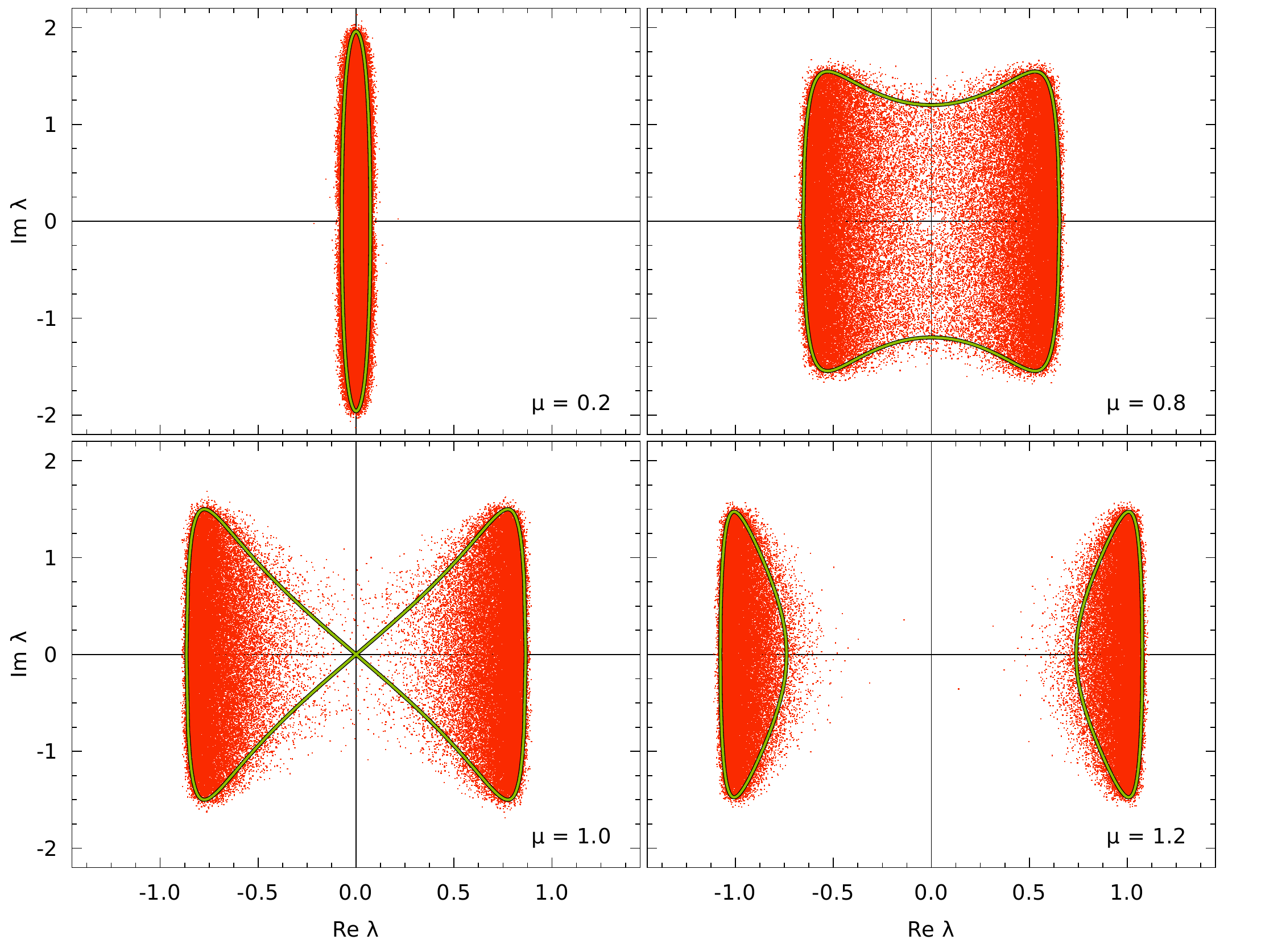} 
  \caption{The Dirac spectrum of the random matrix model \eref{Dsteph} with
    matrix size $N=48$, for twenty configurations, for $m=0$ and $\mu = \{0.2,
    0.8, 1.0, 1.2\}$.}
  \label{eigs}
\end{figure}

In random matrix theory the effect of the fermion determinant on the global
distribution of the eigenvalues is a $1/N$-correction, and will arrange only a
small number of eigenvalues near zero. Also for a nonzero imaginary chemical
potential the quenched and the dynamical eigenvalue distribution are the same for
large $N$. For real chemical potential the eigenvalue distribution is complex
because of the phase of the fermion determinant, but for large $N$ the spectral
support is still the same as for the quenched or phase quenched theory.  Since a
fermion determinant does not change the overall spectral density to leading order
in $1/N$, we expect that also for real chemical potential the distribution of
the eigenvalues will not be affected significantly by the CL evolution. Indeed,
as can be seen in Fig.~\ref{eigs}, where we plot the eigenvalues of the Dirac
operator, for various values of the chemical potential, the support of the Dirac
eigenvalues on the CL trajectory is still given by the quenched result (green
curve).  Since the CL algorithm is probabilistic we thus necessarily have that
the chiral condensate and the baryon number, which are determined by the
distribution of Dirac eigenvalues, will be given by the (phase-)quenched result,
as we have seen in Figs. \ref{muscan} and \ref{mscan}.  In the next section we
will show that the correct result can be obtained using a reweighting algorithm.

%%%%%%%%%%Es m=0

%%%%%%%%%%%%%%%%%%%%%%%%%%%%%%%%%%%%%

\section{Reweighted Complex Langevin}

\begin{figure}[t!] 
  \centering
  \includegraphics[width=.45\linewidth]{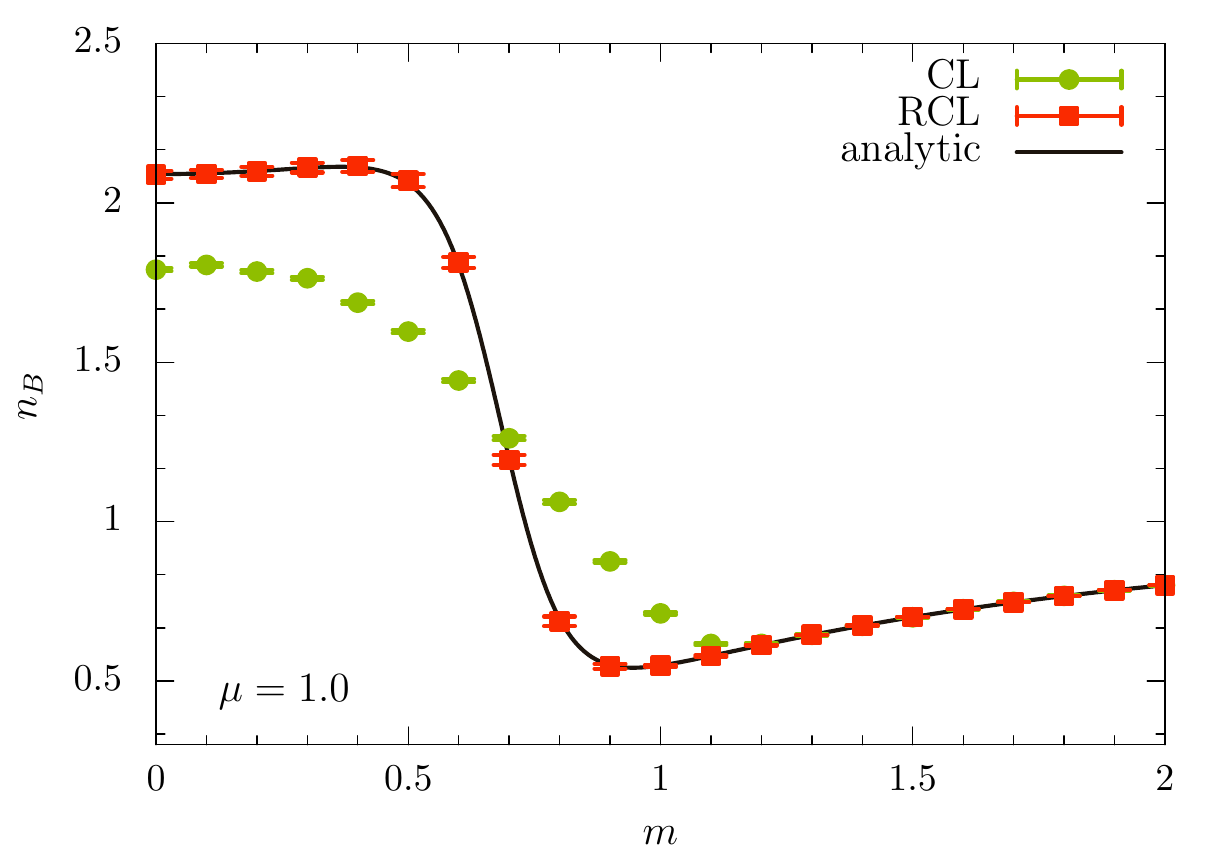} 
  \includegraphics[width=.45\linewidth]{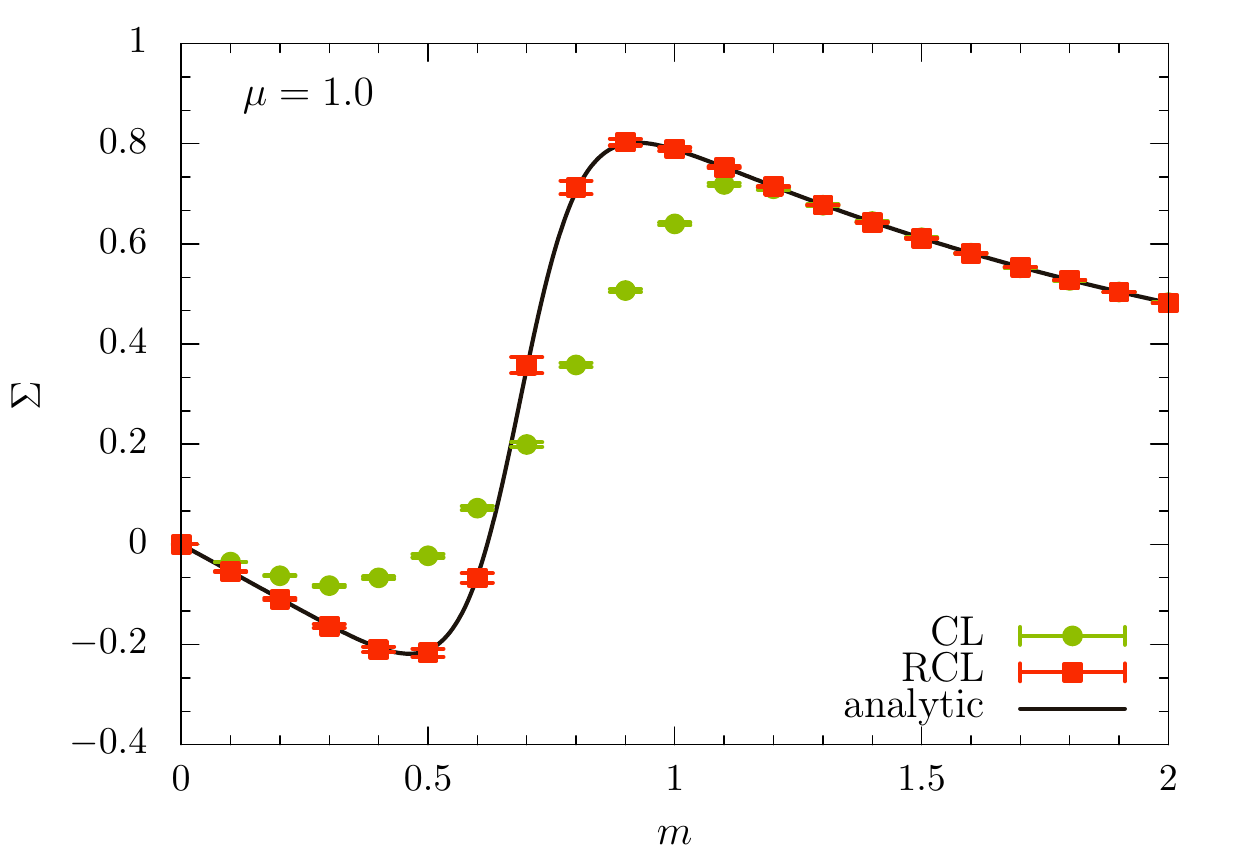} 
  \\[1ex]
  \includegraphics[width=.45\linewidth]{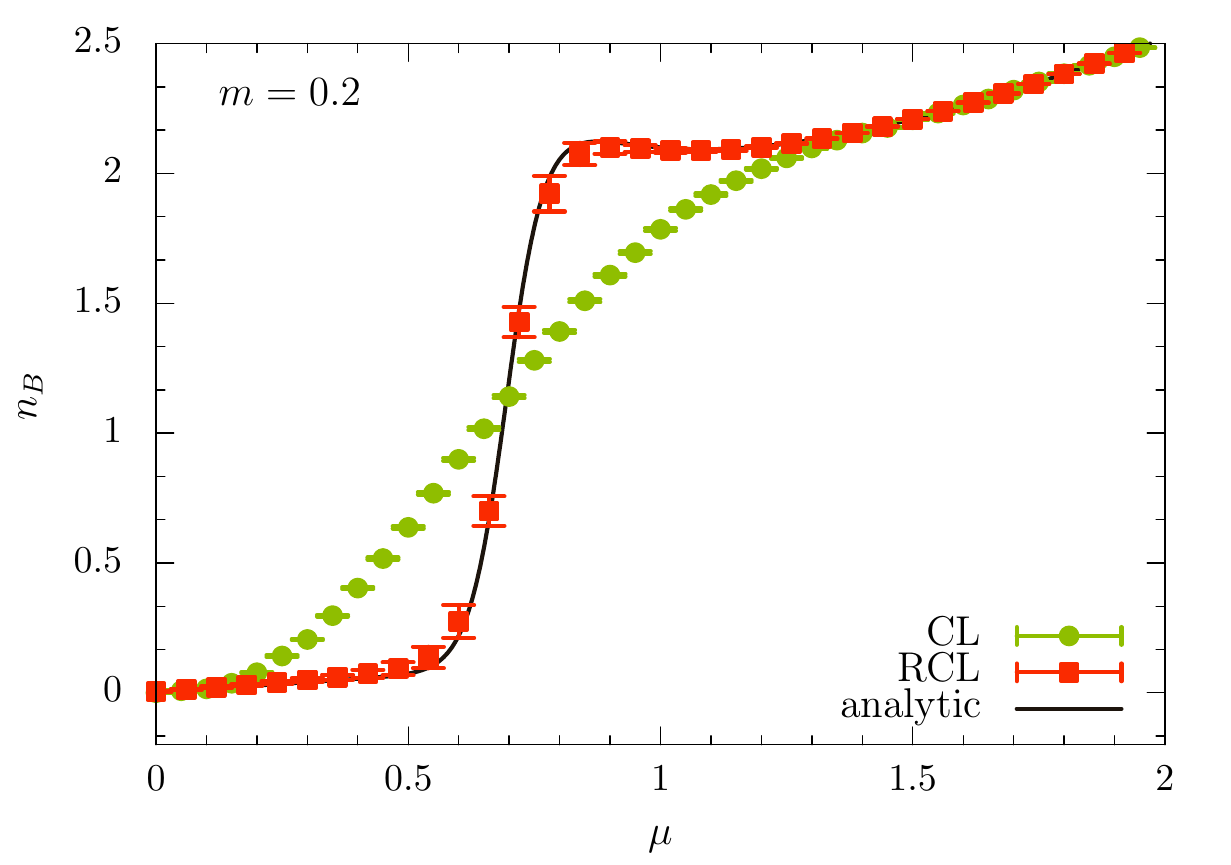} 
  \includegraphics[width=.45\linewidth]{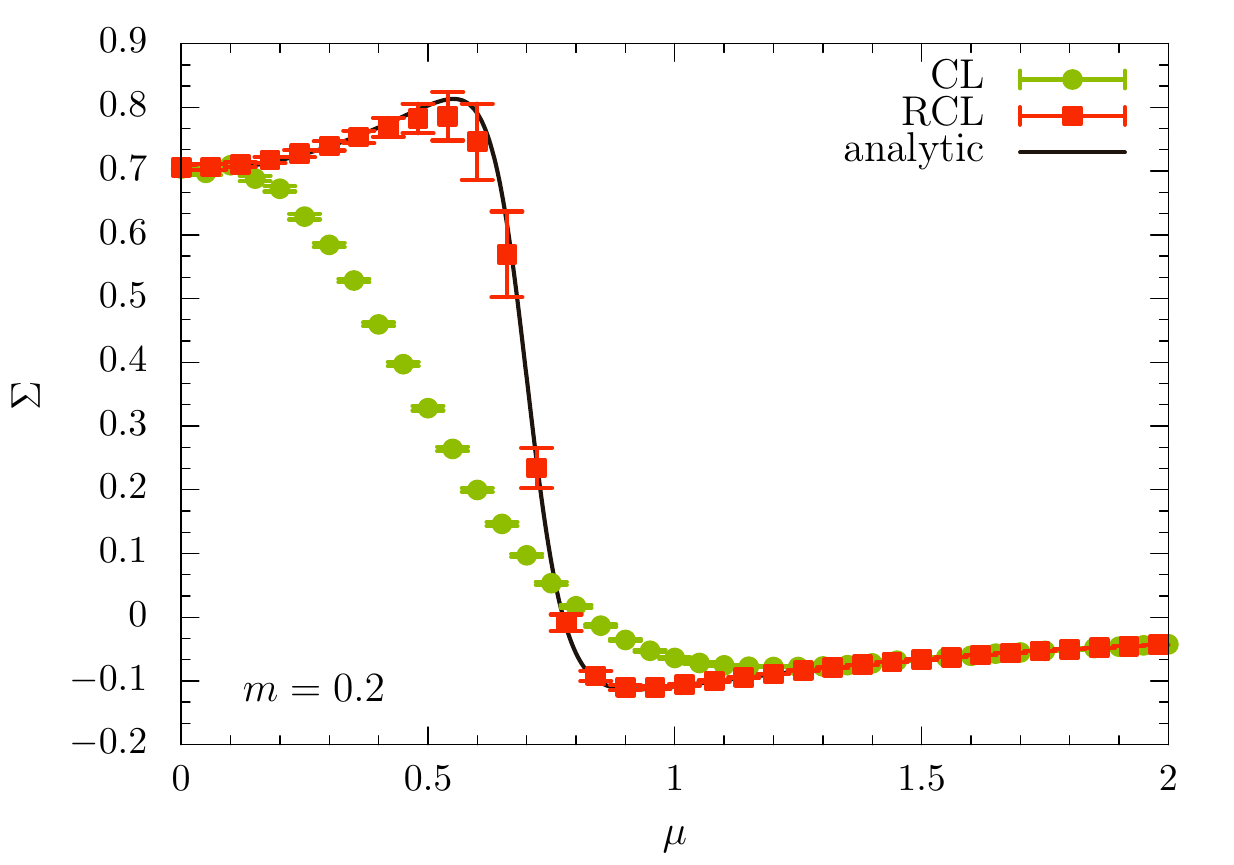} 
  \caption{Results for the RCL method for the random matrix model \eref{Dsteph}
    with matrix size $N=6$. The mass scan for $\mu=1$ (upper panels) is generated
    from an auxiliary ensemble with $m_0=4$ at the same $\mu$, while the
    $\mu$-scan for $m=0.2$ (lower panels) uses an auxiliary ensemble at $\mu_0=2$
    and same mass. We show the baryon number density on the left and the chiral
    condensate on the right.}
  \label{RCL-N=6}
\end{figure}

After having established that the CL algorithm fails to reproduce the known
analytical results of the random matrix model \eref{Dsteph}, the obvious
question to be asked is if something can be done to fix the pathologies of the
algorithm in regions of the parameter space where it fails. We apply the
reweighted complex Langevin (RCL) method \cite{Bloch:2017sfg,Bloch:2017ods}, and
we will show that one can significantly improve the convergence properties of
the algorithm. This will lead to correct results for values of the parameters
for which a naive implementation of the CL algorithm was giving wrong results.
Our efforts to test the algorithm will be focused on the region close to the
phase transition. 

The motivation of the method mainly comes from the expectation that reweighting
CL trajectories might work better than other traditional forms of reweighting,
mainly because the target ensemble with parameters ($\xi=m$, $\mu$) and the
auxiliary ensemble with parameters ($\xi_0=m_0$, $\mu_0$) are expected to have
larger overlap. In reweighting one computes the expectation value of an
observable $\mathcal{O}$ using
\begin{equation}
  \langle \mathcal{O}\rangle_{\xi}=\frac{\int dx \,w(x; \xi) \mathcal{O}(x ;\xi)}{\int dx \,w(x; \xi) }=
  \frac{\int dx\, w(x ; \xi_0)\left[ \frac{w(x; \xi)}{w(x ; \xi_0)}\mathcal{O}(x ; \xi)\right]}{\int dx\, w(x ; \xi_0)\left[ \frac{w(x; \xi)}{w(x ; \xi_0)}\right] }
  =\frac{\left\langle \frac{w(x; \xi)}{w(x ; \xi_0)}\mathcal{O}(x ; \xi) \right\rangle_{\xi_0}}{\left\langle \frac{w(x; \xi)}{w(x ; \xi_0)} \right\rangle_{\xi_0}}.
\end{equation}
However, contrary to the traditional forms of reweighting, the weight $w(x
;\xi_0)=e^{-S(x; \xi_0)}$ is complex, and thus, we need to employ the CL
algorithm to sample this auxiliary ensemble. Of course, this is performed
through a judicious selection of the reweighting parameters, chosen from the
region where the algorithm satisfies the CL convergence properties. In this case
the reweighting equation becomes, after complexification of the variables,
\begin{equation}
  \langle \mathcal{O}\rangle_{\xi} =
    \frac{%
      \int dx dy\, P(z;\xi_0)\left[ \frac{w(z;\xi)}{w(z; \xi_0)}\mathcal{O}(z;\xi)\right]%
    }{%
      \int dxdy\, P(z;\xi_0)\left[ \frac{w(z;\xi)}{w(z;\xi_0)}\right] },
\end{equation}
where $P(z;\xi_0)$ is the real probability in the complex variables $z=x+iy$,
generated by the CL trajectory. In practice, the configurations of the auxiliary
ensemble are sampled according to their probability $P(z,\xi_0)$ in the
complexified variables by evolving the CL equations for the auxiliary action. An
expectation value in the target ensemble is then computed as a ratio of the
average effective observable and the average reweighting factor, both measured
along the auxiliary CL trajectories.

We first test the method with small matrices ($N=6$) and we see in
Fig.~\ref{RCL-N=6} that reweighting the CL trajectories can fix all the
problematic issues of the algorithm. It is interesting to observe that for this
relatively small matrix size the analytical answer can be reproduced for the
whole range of parameters, as can be seen from the scans of the mass and of the
chemical potential. It is important to stress that this is already quite
intriguing since the naive implementation of the algorithm was failing to
reproduce the correct answer in the region where the operators are
non-holomorphic. Moreover, it is  interesting to observe that, even though the
auxiliary ensemble is chosen at one side of the phase transition, i.e., with
large $m_0$ for the mass scan or large $\mu_0$ for the $\mu$-scan, the RCL data
at the other side of the phase transition still agree very well with the
analytical results. Nevertheless, we already notice that the error bars start to
grow in the phase transition region, as is expected if the sign problem grows in
that region. 

\begin{figure}[t!] 
    \centering
    \includegraphics[width=.45\linewidth]{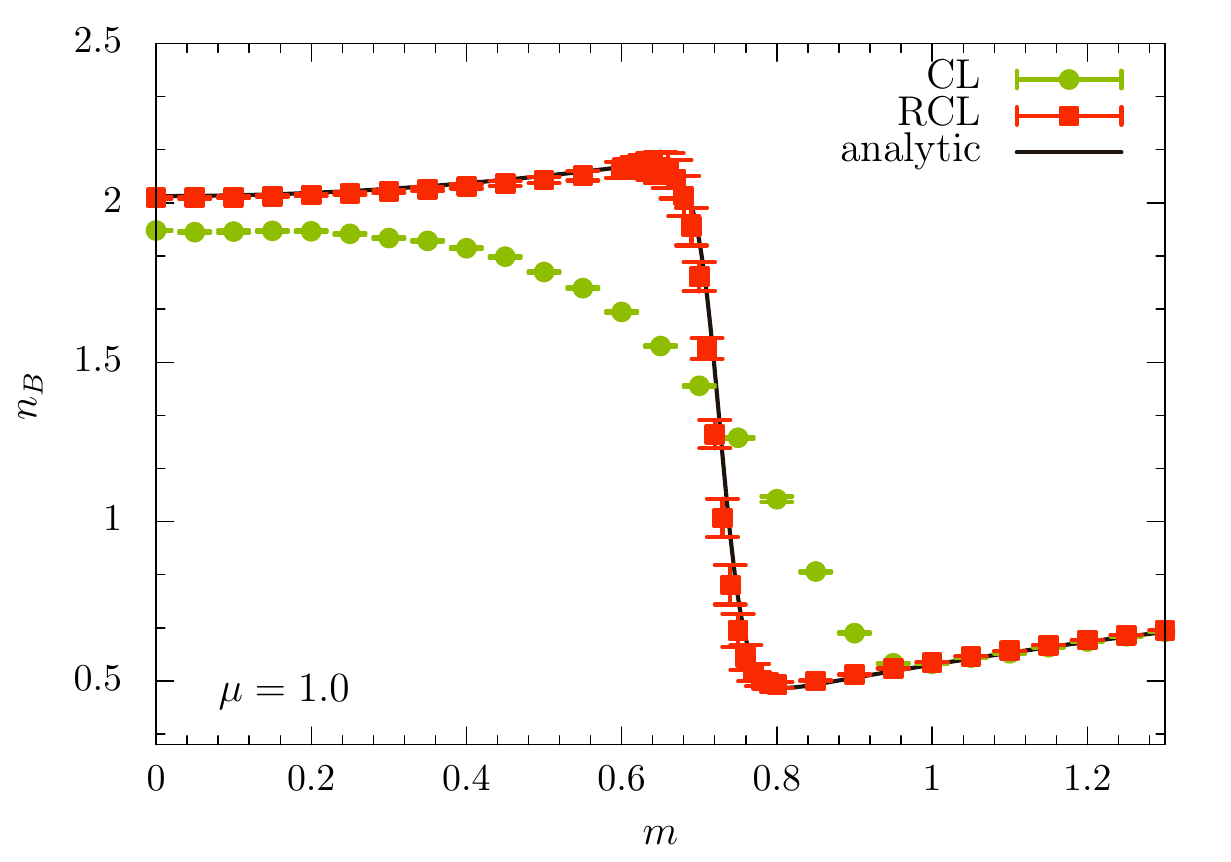} 
    \includegraphics[width=.45\linewidth]{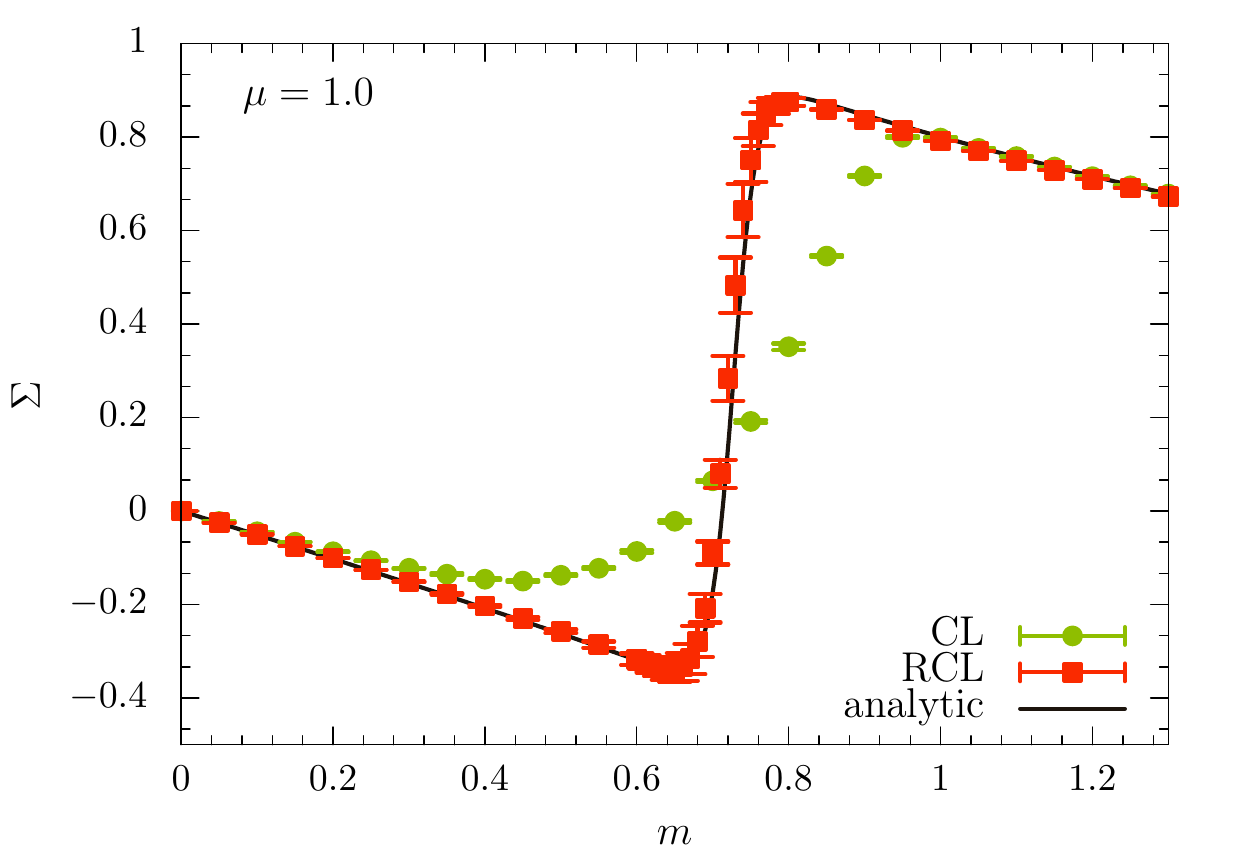} 
    \\[1ex]
    \includegraphics[width=.45\linewidth]{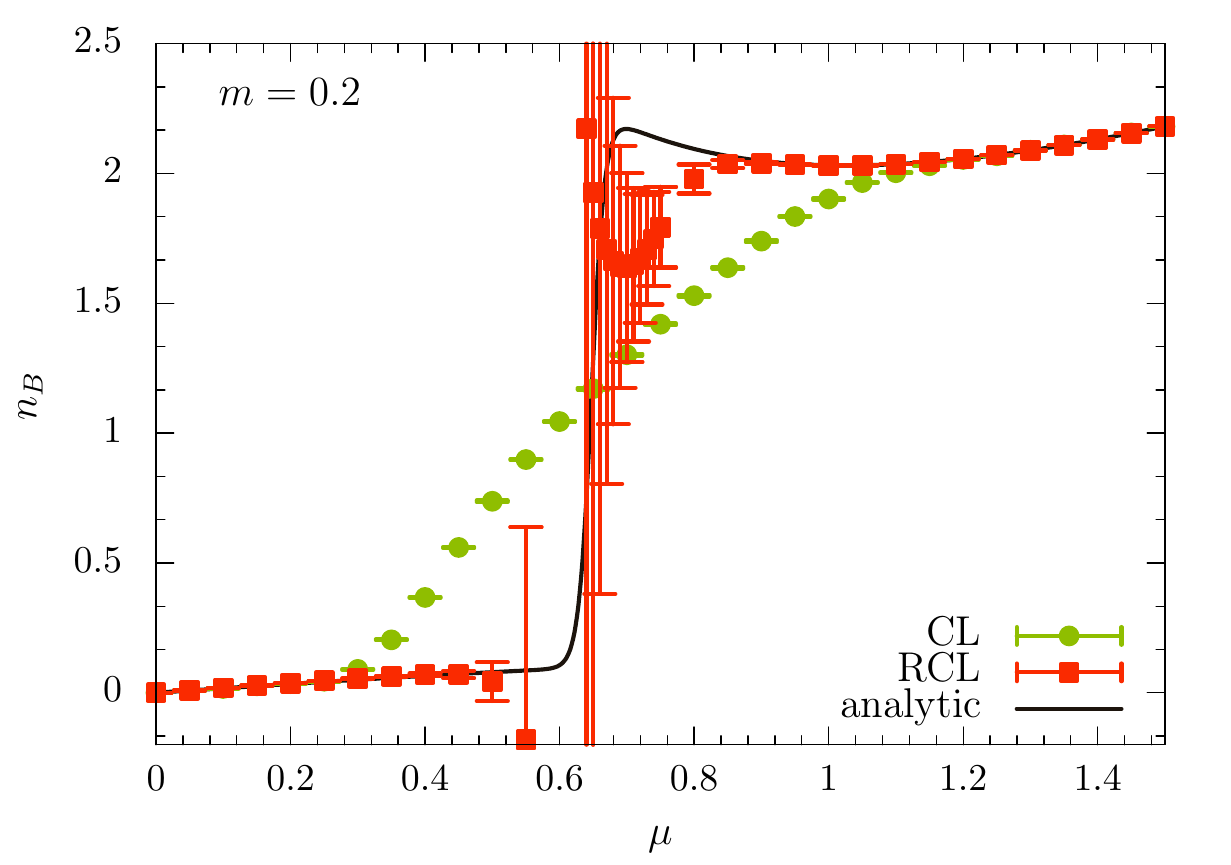} 
    \includegraphics[width=.45\linewidth]{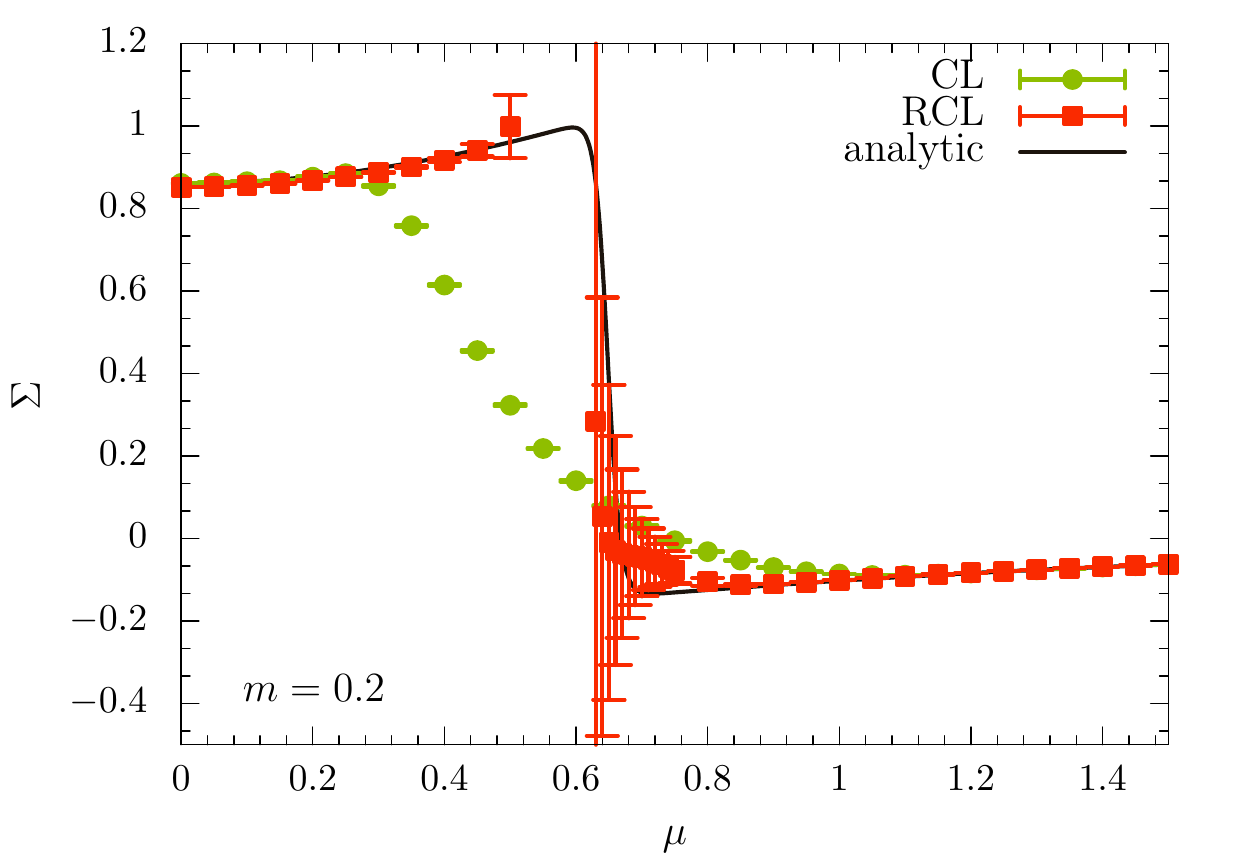} 
    \caption{Results for the RCL method for to the random matrix model
      \eref{Dsteph} with matrix size $N=24$. The mass scan for $\mu=1$ (upper
      panels) is generated from an auxiliary ensemble with $m_0=1.3$ and same
      $\mu$, while the $\mu$-scan for $m=0.2$ (lower panels) uses an auxiliary
      ensemble at $\mu_0=1.5$ and same mass. Again we show the baryon number
      density on the left and the chiral condensate on the right.}
    \label{RCL-N=24}
\end{figure}

\begin{figure}[th!] 
    \centering
    \includegraphics[width=.45\linewidth]{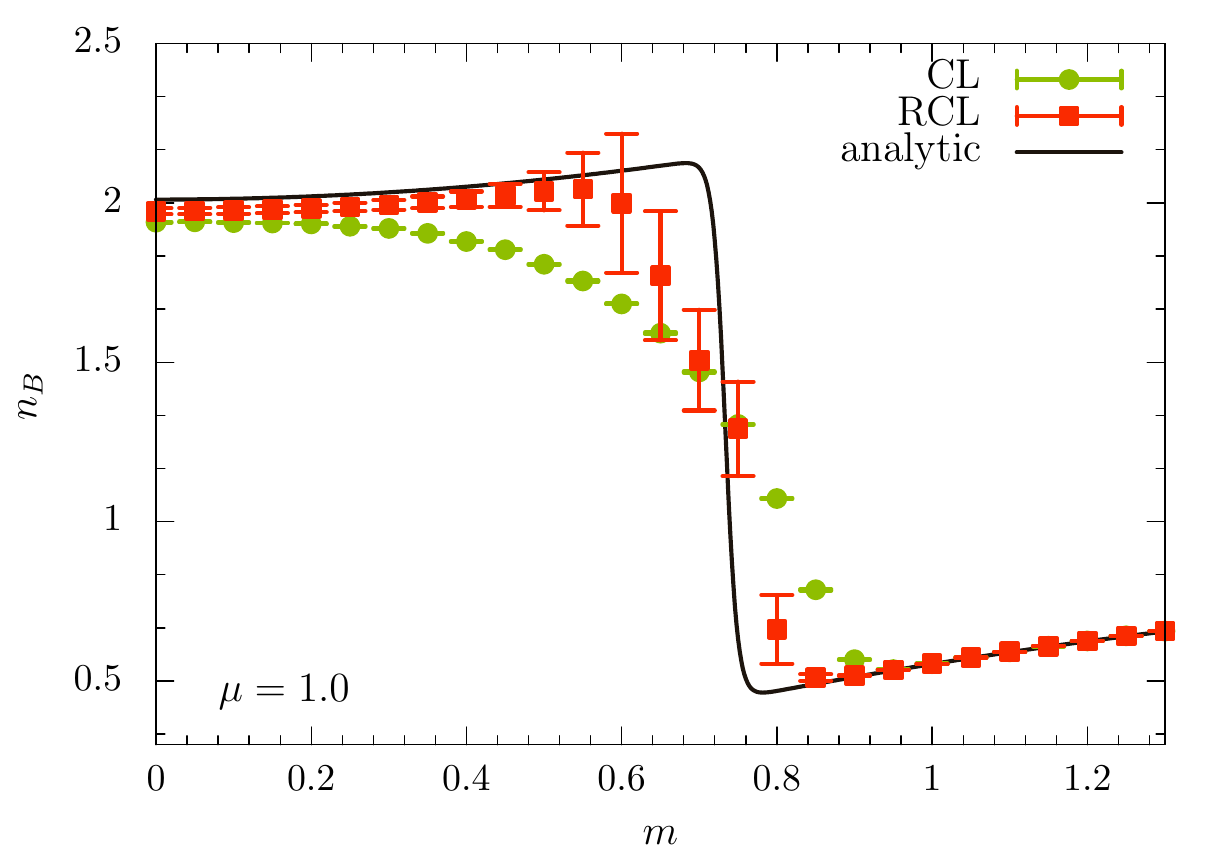} 
    \includegraphics[width=.45\linewidth]{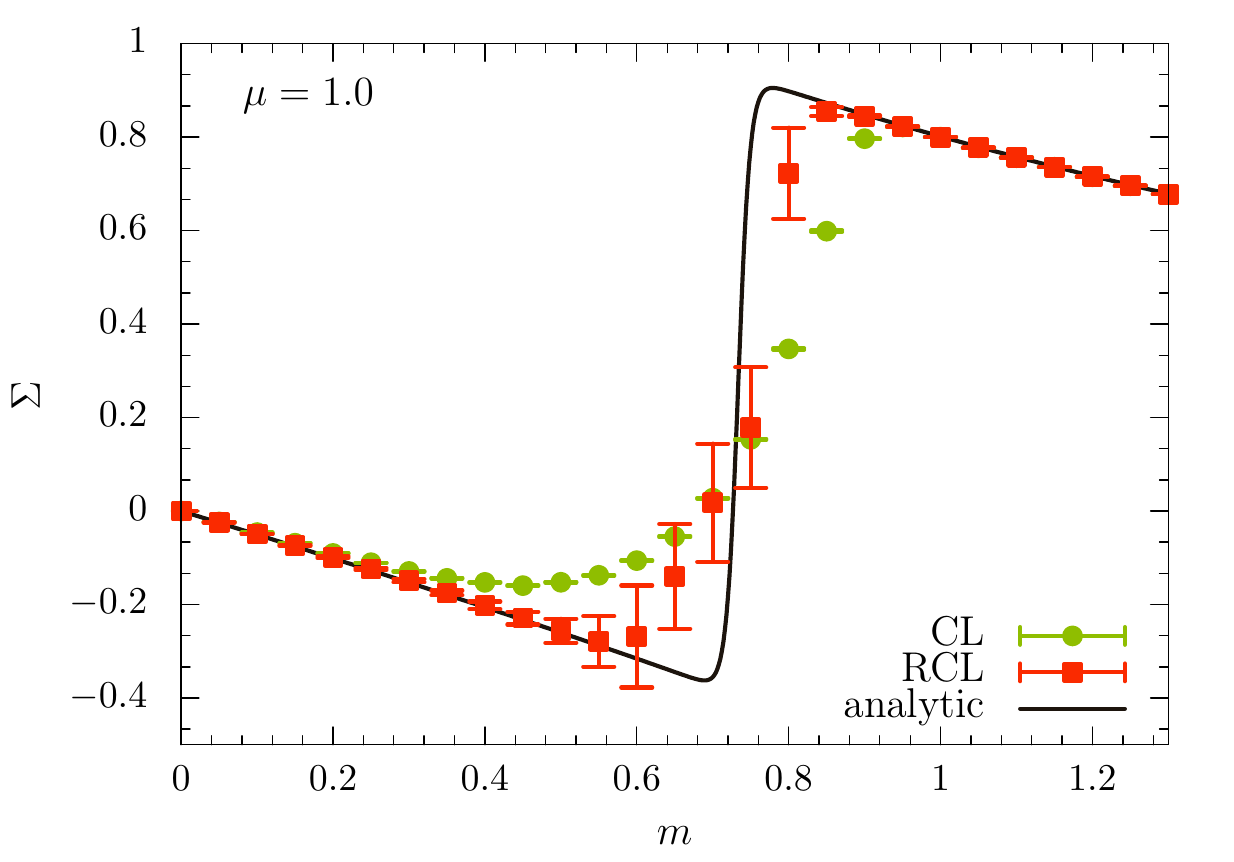} 
  \\[1ex]
    \includegraphics[width=.45\linewidth]{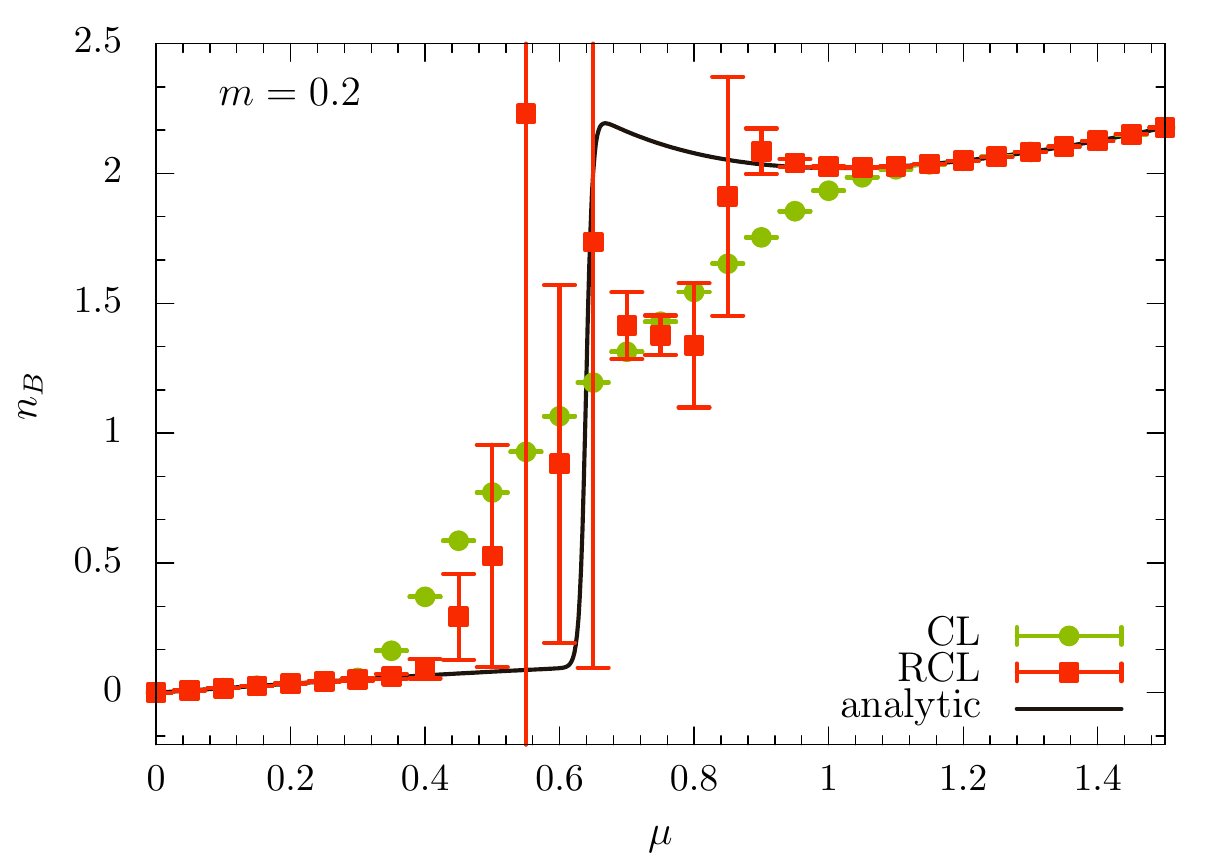} 
    \includegraphics[width=.45\linewidth]{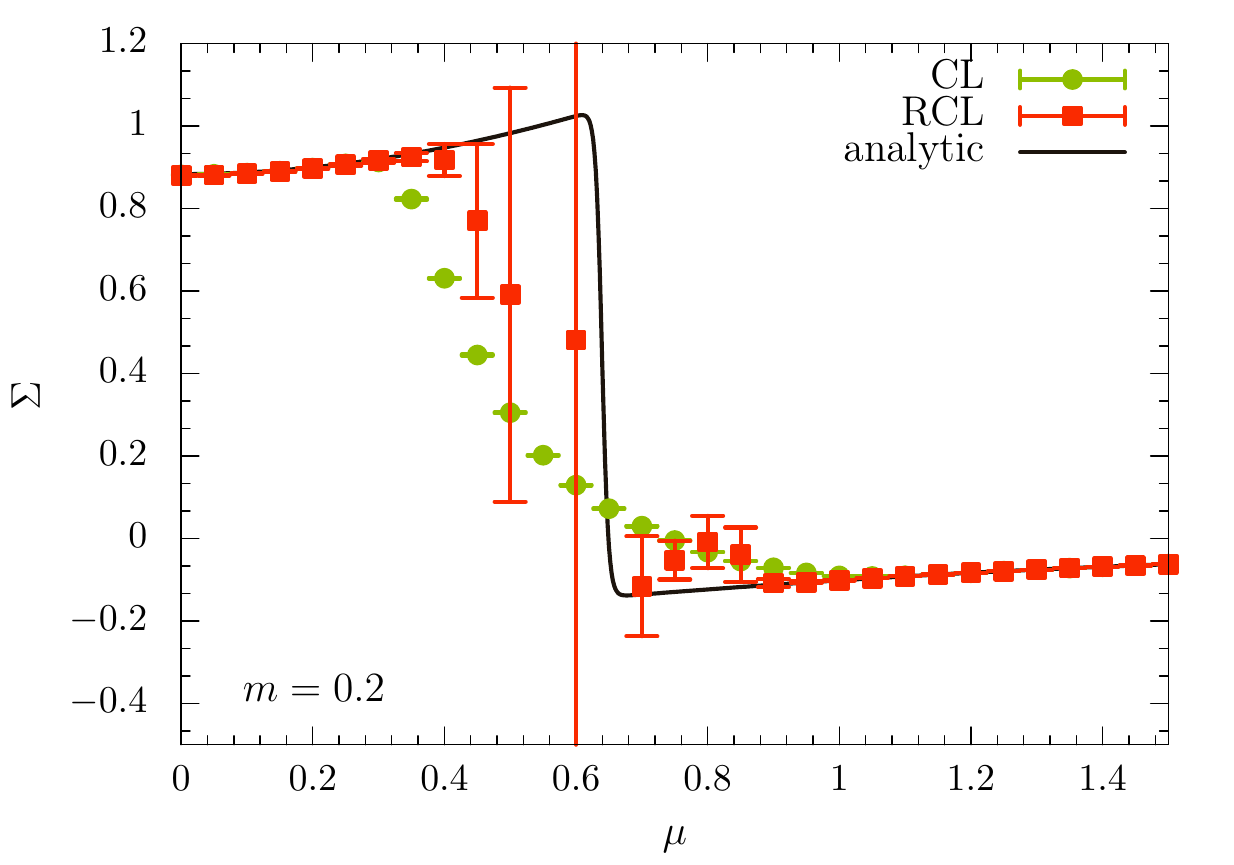} 
    \caption{Results for the RCL method for  the random matrix model\eref{Zst} with  $N=48$. The mass scan for $\mu=1$
      (upper panels) is generated from an auxiliary ensemble with $m_0=1.3$ and
      same $\mu$, while the $\mu$-scan for $m=0.2$ (lower panels) uses an
      auxiliary ensemble at $\mu_0=1.5$ and same mass. Again we show the baryon
      number density on the left and the chiral condensate on the right.}
    \label{RCL-N=48}
\end{figure}

Of course in order to claim to have solved the sign problem, which is
exponentially hard with respect to the volume of the system, we need to show
that the number of matrices needed to achieve the sought precision does not
scale exponentially with the matrix size $N$. For this reason we increased the
matrix size, in order to investigate if this method of reweighting actually
works and how it scales with respect to the matrix size $N$. In
Fig.~\ref{RCL-N=24} we show the RCL data for $N=24$. The upper graphs show a
mass scan for fixed $\mu$ and the lower graphs a $\mu$ scan for fixed mass. To
keep the error under control in the phase transition region the number of
configurations had to be increased by a factor of 100 compared to $N=6$. This
allows us to have small error bars in the mass scan for $\mu=1.0$, for all mass
values. In this scan the RCL always gives the correct value even though CL
obviously fails as soon as $m<1.0$.
In the $\mu$-scan we see that RCL is able to improve on the CL method outside
the phase transition, however, inside this region, i.e., for $0.6 <\mu< 0.8$,
the sign problem clearly reappears. 
The figure plainly shows that RCL performs qualitatively better than the CL
method in all cases, but with the caveat that the phase transition region is
still difficult to access.
Undoubtedly, the same is true for $N=48$ as can be seen in Fig.~\ref{RCL-N=48}.
The RCL works reasonably well in the mass scan, performed
for $\mu=1$, whereas the CL does poorly over most of the mass range.
Unfortunately, the $\mu$ scan distinctly shows that the reweighting only works
above and below the phase transition region. Again we observe the salient
feature that even though the auxiliary ensemble is taken at one side of the
phase transition, the reweighting procedure reproduces the data well also at the
other side of the transition. This seems to point to the absence of an overlap
problem in the RCL method, even though the sign problem is clearly present in
the phase transition region. One disturbing point of this investigation is that
it confirms how bad the original CL performs for a very large range of
parameters, even away from the phase transition. Indeed, the CL seems to fail in
regions where reweighting still works quite well and is not yet hampered by the
sign problem.

Using the data for $N=6,\;12,\;24,\;48$ we find a naive volume scaling for the
RCL method that is proportional to $\exp(0.3\times N)$ for the number of
configurations necessary to get the same accuracy for all values of $N$. This
shows that, even though this reweighting method rectifies the failing of the CL
method for a large range of parameter values, it still is exponential in the
volume in the phase transition region.

\section{Shifted Representation}

In an attempt to mend the problems due to the phase of the Dirac operator we now
shift the effect of the chemical potential away from the fermionic term. This
can be done with a simple shift of variables. Written out in the Cartesian
representation, the Dirac operator is
\begin{equation}
  D = %
  \begin{pmatrix}
    m & iA - B + \mu \\
    iA^T + B^T + \mu & m
  \end{pmatrix}.
\end{equation}
We can absorb $\mu$ into $A$ with a simple change of variables, $A' = A - i\mu$.
The action in terms of the matrices $A'$ and $B$ is
\begin{equation}
  S = N \tr \big( A'^T A' + 2i\mu A' - \mu^2 + B^T B\big) %
    - N_f \tr \log \big(m^2 + X' Y'\big),
\end{equation}
where $X' = A' + iB$ and $Y' = A'^T - iB^T$. In this representation the $\mu$ dependence has been shifted from the fermionic to the bosonic term. Computing the CL force term results in
\begin{subequations}
  \begin{align}
    \frac{\partial S}{\partial A'_{mn}} &= 2N(A'_{mn} + i\mu \delta_{mn})%
      - N_f \big(X' G' + (G' Y')^T \big)_{mn}, \\
    \frac{\partial S}{\partial B_{mn}} &= 2N B_{mn}%
      + i N_f \big(X' G' - (G' Y')^T \big)_{mn},
  \end{align}
\end{subequations}
where $G' = (m^2 + Y'X')^{-1}$ is defined in terms of the shifted fields. The
advantage of the shifted representation is that it starts in an anti-Hermitian
state, and due to the fact that CL is non-deterministic, the configurations
could potentially evolve to a different minimum.

\begin{figure}[t!]
  \centering
    \includegraphics[width=0.45\textwidth]{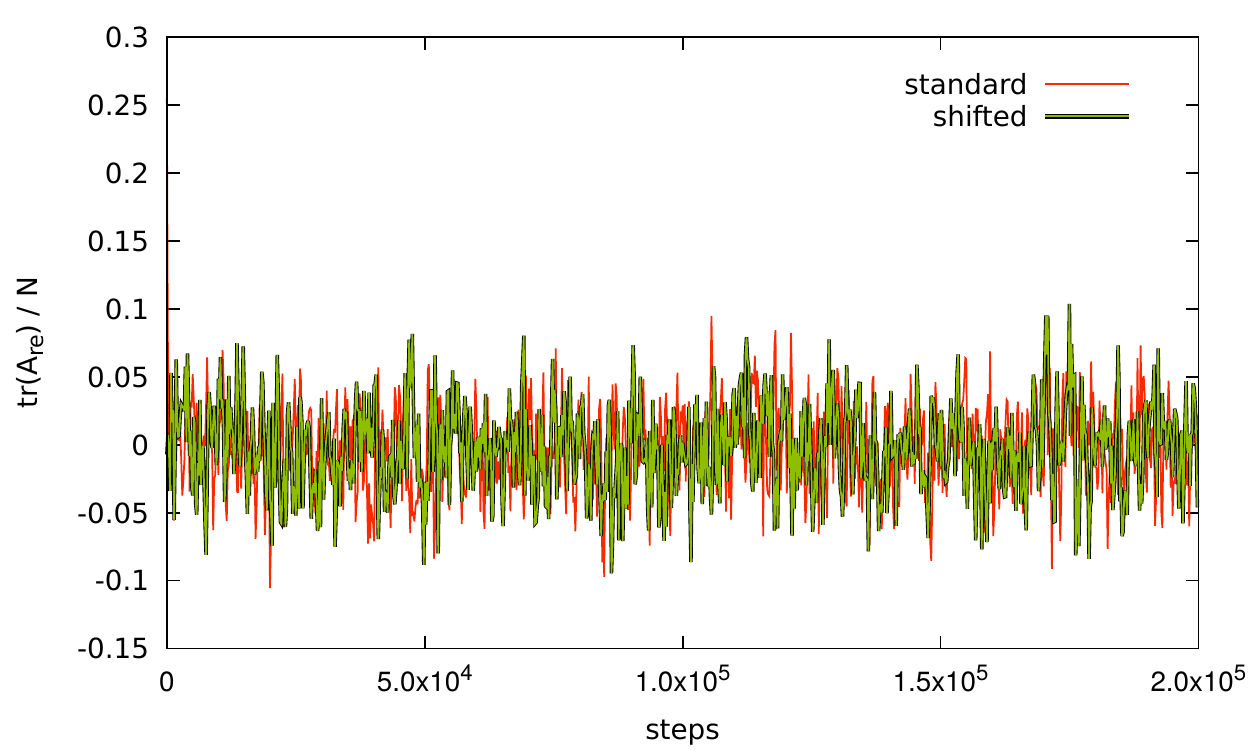}
    \includegraphics[width=0.45\textwidth]{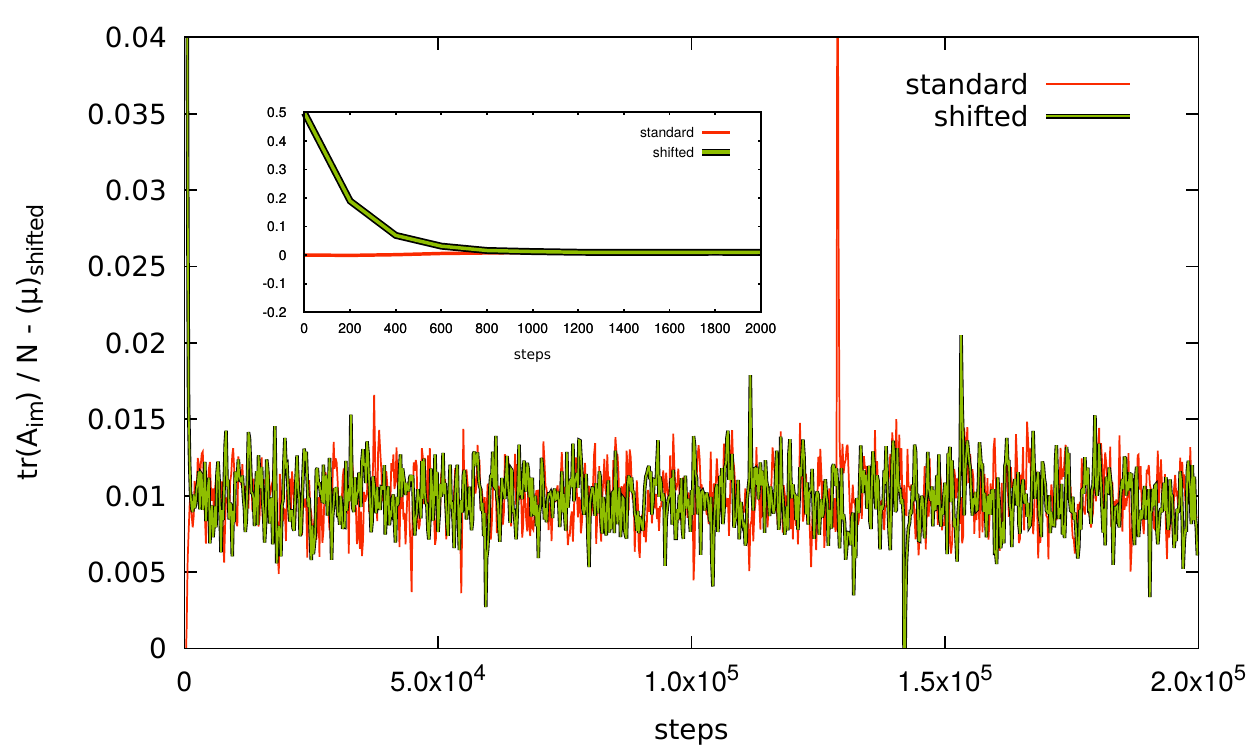}
    \caption{Complex Langevin evolution of the real (left) and imaginary (right)
      parts of the average diagonal entry of $A$ and $A'$ in the standard and
      shifted representation of the random matrix theory \eref{Dsteph},
      respectively.  The shifted $A'$ is subtracted by $i \mu$.}
  \label{fig:shifted_diag_average}
\end{figure}

%\subsection{CL results}

To analyze the dynamics of the shifted representation we analyze the elements of
the matrices $A$ and $A'$ during CL evolution; the real and imaginary part of
their average diagonal entry are shown in Fig.~\ref{fig:shifted_diag_average}.
Although the two matrices start out very differently, they are similar after
thermalization. This seems to indicate that 
\begin{equation}
  \big\langle A' \big\rangle_{\mathrm{CL},\text{shifted}} = \big\langle A
  \big\rangle_{\mathrm{CL},\text{standard}} - i \mu,
\end{equation}
and thus they converge to the same solution.  Since the Dirac operator in the
shifted representation starts the CL evolution at a chiral condensate and a
baryon number density for $\mu=0$, one might expect better convergence
properties at least below the critical value of the chemical potential. In the
next section we will use the shifted representation when analyzing the effect of
gauge cooling.

\section{Gauge Cooling}

The complexified action takes on redundant degrees of freedom which is evident
from the fact that the action is invariant under an enhanced symmetry group as
compared to the original RMT. One can utilize this enlarged symmetry in an
attempt to steer the Langevin flow towards more physical configurations due to
the fact that although the action is invariant under these transformations, the
flow itself is not.  This method is commonly referred to as \emph{gauge
  cooling}, and has been used to great effect in a plethora of models
\cite{Seiler:2012wz, Nagata:2016mmh}. Most relevant to our study is its
successful application to the Osborn RMT model \eref{Dosb}
\cite{Nagata:2016mmh}, which we will refer to for comparison.

The original RMT is invariant under the $\UN$ transformation
\begin{equation}
  W \to g W g^{\dagger}, \hspace{5mm} W^{\dagger} \to g W^{\dagger} g^{\dagger} %
    \hspace{5mm} \text{where} \hspace{5mm} g \in \UN.
\end{equation}
However the complexified action is invariant under the enlarged $\GLN$
transformation
\begin{equation}
  X \to h X h^{-1}, \hspace{5mm} Y \to h Y h^{-1} %
    \hspace{5mm} \text{where} \hspace{5mm} h \in \GLN.
  \label{eq:similarity_trafo}
\end{equation}
We stress that the cooling transformation does not change the eigenvalues of the
Dirac operators $D$ and $\gamma_0(D+m)$, and that the effect of cooling occurs in tandem
with the Langevin updates. Next we will look at how to choose $h$ in an
advantageous way.

\subsection{Cooling norms}

The transformation matrices $h$ are chosen such that a \emph{cooling norm} is
reduced. These norms are constructed to quantify an undesirable property of the
matrix configurations. The most basic of these is the Hermiticity norm
\cite{Nagata:2016mmh}
\begin{equation}
  \mathcal{N}_H = \frac{1}{N} \tr \Big[ \big(X - Y^{\dagger}\big)^{\dagger} %
    \big(X - Y^{\dagger}\big) \Big],
\end{equation}
which measures the deviation of the CL configuration from a valid RMT
configuration. It is zero when $X^{\dagger} = Y$, and grows when the matrices
$A$ and $B$ acquire imaginary parts.

We also introduce an eigenvalue norm \cite{Nagata:2016mmh}
\begin{equation}
  \mathcal{N}_{\mathrm{ev}} = \sum_{i = 1}^{n_{\mathrm{ev}}} e^{-\xi \gamma_i}
\end{equation}
where $\gamma_i$ are the $n_{\mathrm{ev}}$ lowest eigenvalues of the positive
definite matrix $D^{\dagger} D$, and $\xi$ is a real positive parameter. This
norm suppresses configurations with Dirac eigenvalues close to zero.

Finally, we will also use a generalization of the anti-Hermiticity norm,
\begin{equation}
  \mathcal{N}_{AH}^p %
    = \frac{1}{N} \tr \Big[\Big( \big(\phi + \psi^{\dagger}\big)^{\dagger} %
    \big(\phi + \psi^{\dagger}\big) \Big)^p\Big],
  \label{AH-Norm}
\end{equation}
which was introduced in \cite{Nagata:2016mmh} for $p=1$. The matrices $\psi$ and
$\phi$ are the off-diagonal elements of $D$. For the Stephanov model they are
given by $\psi = i X + \mu$ and $\phi = i Y + \mu$, so that the norm becomes
\begin{equation}
  \mathcal{N}_{AH}^p %
      = \frac{1}{N} \tr \Big[\Big( \big(iX -iY^\dagger +2\mu \big)
        \big(iY -iX^\dagger+2\mu \big) \Big)^p\Big].
\end{equation}
For $p=1$ the $\mu$-dependent terms do not depend on the similarity
transformation $h$, and the Dirac operator is generally not anti-Hermitian at
the minimum of the norm. Therefore will use the $p=2$ anti-Hermiticity norm
below.

As the different norms try to fix different problems one can also combine them
in aggregate norms. One useful choice is to combine the Hermiticity norm, which
quantifies how much the configurations drift into the imaginary plane, with
either the anti-Hermiticity or the eigenvalue norm, both of which handle
problematic configurations related to a singular behavior of the drift
\begin{equation}
  \mathcal{N}_{\mathrm{agg}} = (1-s) N_{AH/\mathrm{ev}} + s N_H,
  \hspace{.25cm}\text{where}\; s \in [0, 1].
\end{equation}

\subsection{Computing $h$}

We follow the procedure outlined in \cite{Nagata:2016mmh} to compute the
transformation matrix $h$. We can write $h$ in terms of the $\UN$ generators,
$\lambda_i \in \mathfrak{u}(N)$
\begin{equation}
  h = e^{a_i \lambda_i}, \hspace{.2cm} a_i \in \mathbb{C}.
\end{equation}
Because the RMT is invariant under $\UN$ transformations, we can choose
$a_i \in \mathbb{R}$ to only pick out the $\GLN/\UN$ transformations.
Assuming the norm is a function $\mathcal{N}(X,Y)$ we want to solve the following
equation
\begin{equation}
  \tilde{h} = \Big\{%
    e^{a_i \lambda_i} \:\Big|\: %
      a_i = \argmin_{a_i'} \,\mathcal{N}\big(%
        e^{a_i' \lambda_i} X e^{-a_i' \lambda_i}, %
        e^{a_i' \lambda_i} Y e^{-a_i' \lambda_i}\big) \Big\} .
\end{equation}
This can be reduced to a one dimensional minimization problem by first
computing the gradient descent vector of the transformation through
\begin{equation}
  \tilde{a}_i = -\frac{\partial}{\partial a_i} \mathcal{N} \,\Big|_{a_i = 0},
\end{equation}
and then solve the one parameter minimization problem
\begin{equation}
  \tilde{h} \approx %
    \Big\{ %
      e^{\beta \tilde{a}_i \lambda_i} \:\Big|\: %
      \beta = \argmin_{\beta'} \,\mathcal{N}\big(%
        e^{\beta' \tilde{a}_i \lambda_i} X e^{-\beta' \tilde{a}_i \lambda_i}, %
        e^{\beta' \tilde{a}_i \lambda_i} Y e^{-\beta' \tilde{a}_i \lambda_i}\big) \Big\} .
\end{equation}
where $\beta$ is a real positive quantity.  $\beta$ is computed by applying
Brent's method \cite{Brent1973}, to which we add an upper bound to avoid
a numerically unstable minimization. We take this upper bound to be $0.1$.  The
derivative of the norm with respect to $a_i$ can be computed either numerically
or analytically depending on the norm. After applying the similarity
transformation $X \to h X h^{-1}$ and $Y \to h Y h^{-1}$, the derivative of the
Hermiticity norm is 
\begin{equation}
  \frac{\partial}{\partial a_i} \mathcal{N}_H= \frac{2}{N} \tr\big( %
  Y^{\dagger} [\lambda_i, Y] + X^{\dagger} [\lambda_i, X] \big),
\end{equation}
where $[A,B]$ is the standard commutator. After applying the
similarity transformation to $\phi$ and $\psi$ in \eqref{AH-Norm}, the
derivative of the $p=2$ anti-Hermiticity norm is found to be
\begin{multline}
  \frac{\partial}{\partial a_i} \mathcal{N}_{AH}^{p=2} = 
  \frac{2}{N} \tr \Big( (\phi^{\dagger} + \psi)(\phi + \psi^{\dagger}) 
  \big( \phi^{\dagger} [\lambda_i, \phi] - \phi^{\dagger}[\lambda_i,\psi^\dagger]
  + \psi [\lambda_i,\phi] - \psi[\lambda_i, \psi^{\dagger}] \\
  - [\lambda_i, \phi^{\dagger}]\phi - [\lambda_i, \phi^{\dagger}] \psi^{\dagger}
  + [\lambda_i, \psi] \phi + [\lambda_i, \psi] \psi^{\dagger} \big) \Big).
\end{multline}
Finally, the derivative of the eigenvalue norm is computed numerically.

\subsection{Results}

Below we present the results obtained by applying the gauge cooling method to
the Stephanov model. We will also show results for the Osborn model using the
gauge cooling procedure outlined in \cite{Nagata:2016mmh}. For the runs we have
used a block size $N = 24$, a Langevin step size $\Delta t = 10^{-4}$, and a
total Langevin time $t_{\mathrm{end}} = 1$. Whenever cooling is involved, we
apply 10 cooling transformations between every Langevin update. For this
investigation, the Stephanov model is simulated with parameters $\{m = 0.2$,
$\mu = 0.5\}$, while the Osborn model is simulated with $\{m = 0.1$, $\mu =
0.25\}$. The two sets of parameters were chosen in a region where the full and
the phase quenched results deviate by an intermediate amount, and the two models
have a comparably severe sign problem.

\begin{figure}
  \centering
    \includegraphics[width=0.45\textwidth]{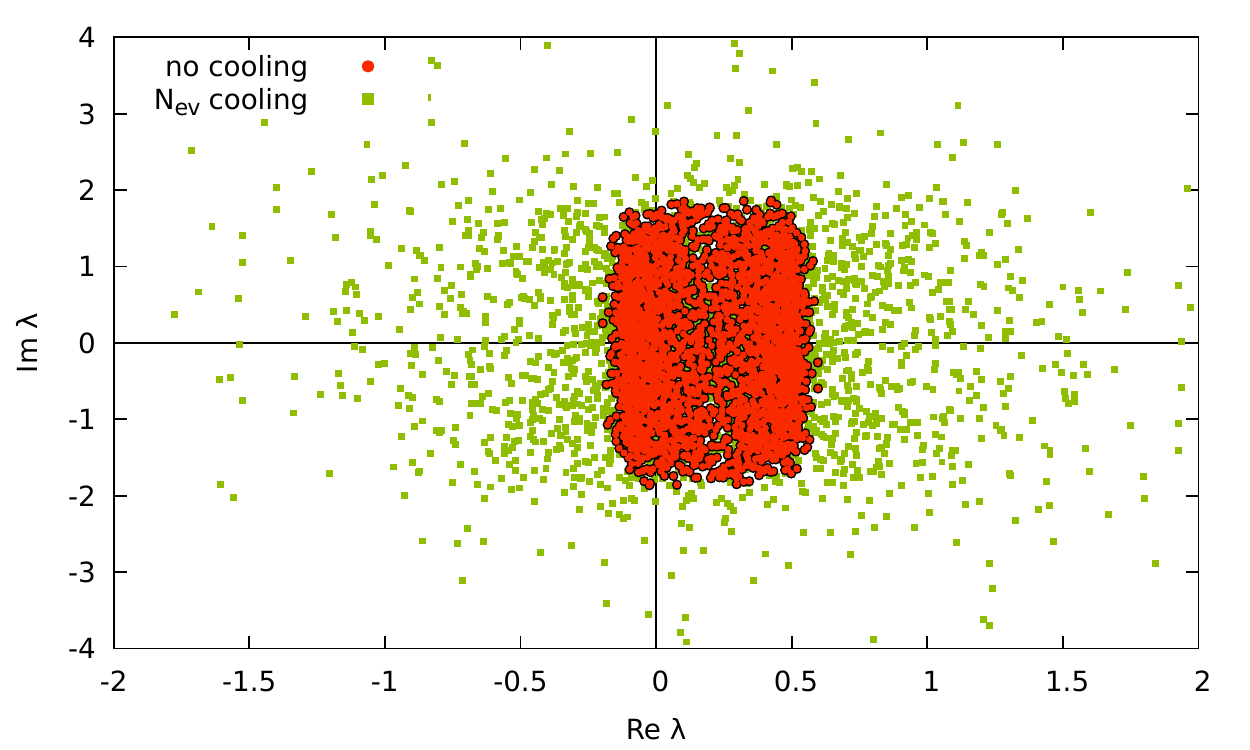}
    \includegraphics[width=0.45\textwidth]{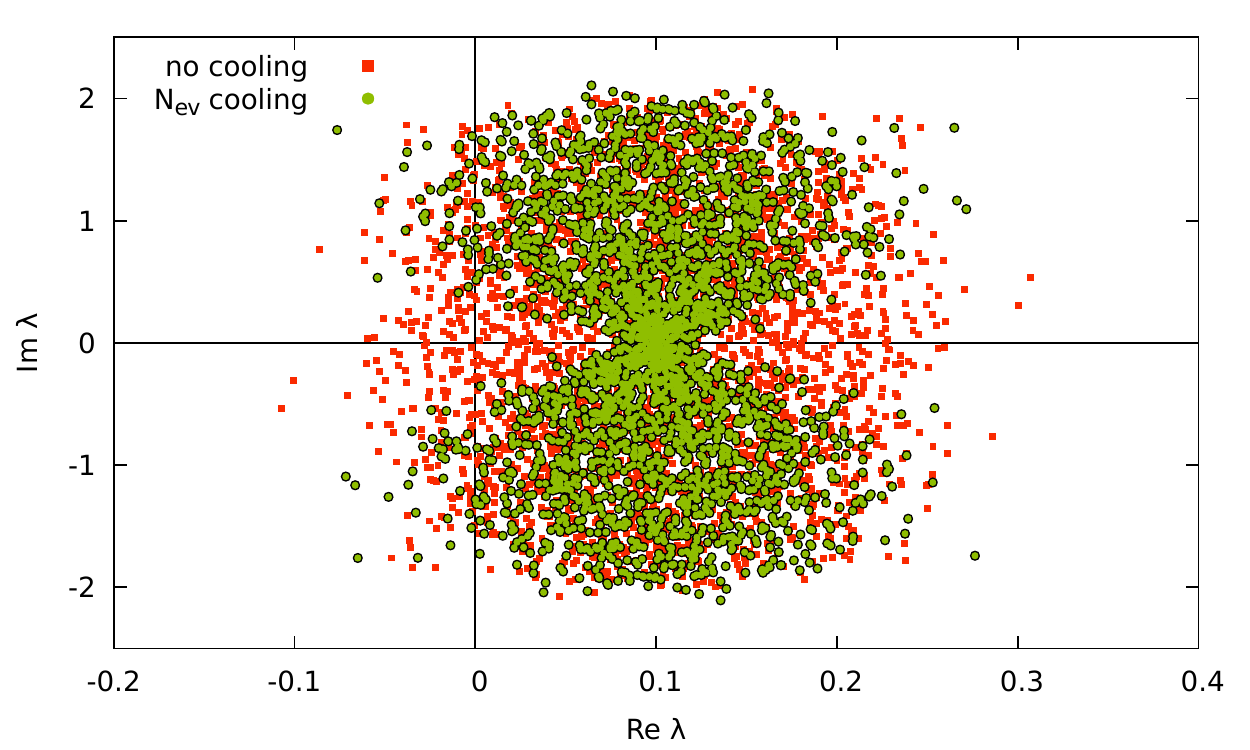}
  \caption{Scatter plots of the eigenvalues of the massive Dirac operator for a
    standard CL run together with the ones from a gauge cooled run. We chose the
    parameters $\{\xi = 100, n_\mathrm{ev} = 2\}$ for
    $\mathcal{N}_{\mathrm{ev}}$. The plots show the eigenvalues from the last 60
    trajectories, separated by 100 updates. The left hand plot shows the
    Stephanov model, while the Osborn model is shown to the right.}
  \label{fig:eval_cooling_eigenvalues}
\end{figure}

First, we discuss the effect of cooling on the distribution of the Dirac
eigenvalues. We start with results for the eigenvalue norm, see
Fig.~\ref{fig:eval_cooling_eigenvalues}. On the right hand side, we present
results for the Osborn model \cite{Nagata:2016mmh}, which show that applying
gauge cooling using the $\mathcal{N}_{\mathrm{ev}}$ norm results in the
eigenvalue distribution developing a "wedge" that excludes zero. In contrast,
this does not happen for the Stephanov model, as can be seen in the left figure.
In this case the distribution of the cooled CL evolution is even wider than that
of the uncooled CL evolution.

\begin{figure}[tb!]
  \centering
  \includegraphics[width=0.45\textwidth]{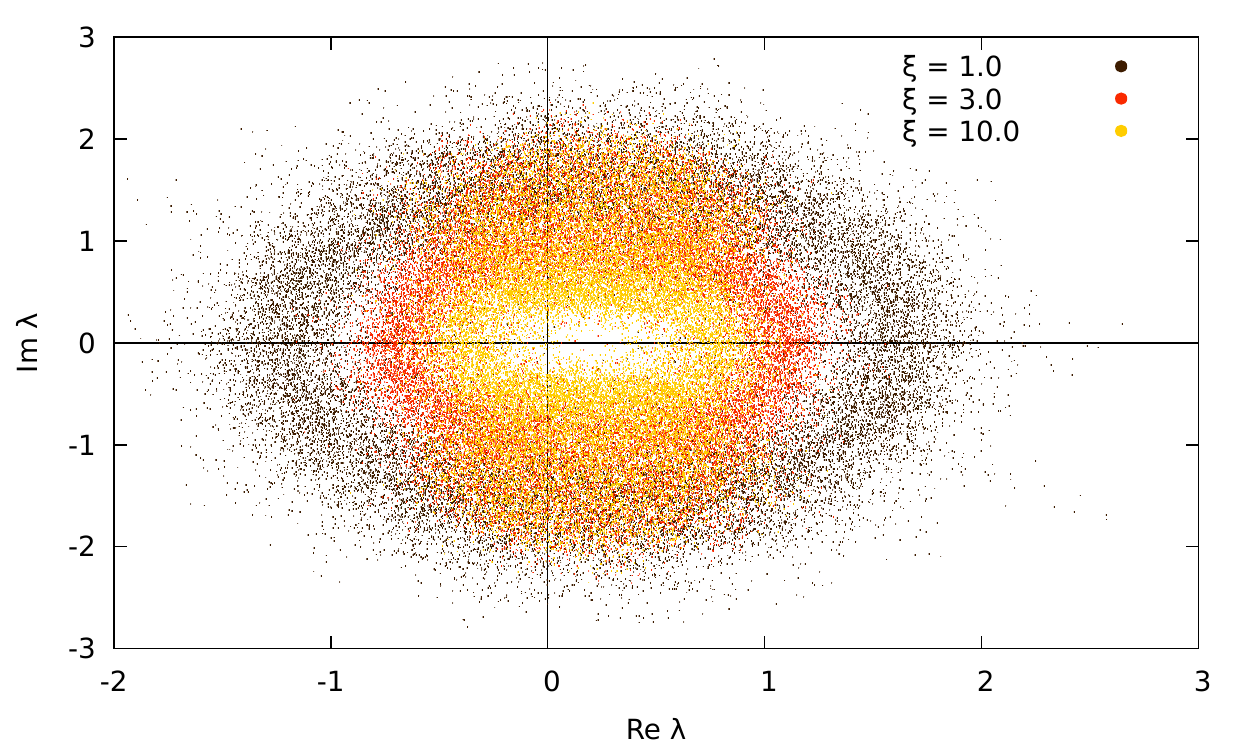}
  \includegraphics[width=0.45\textwidth]{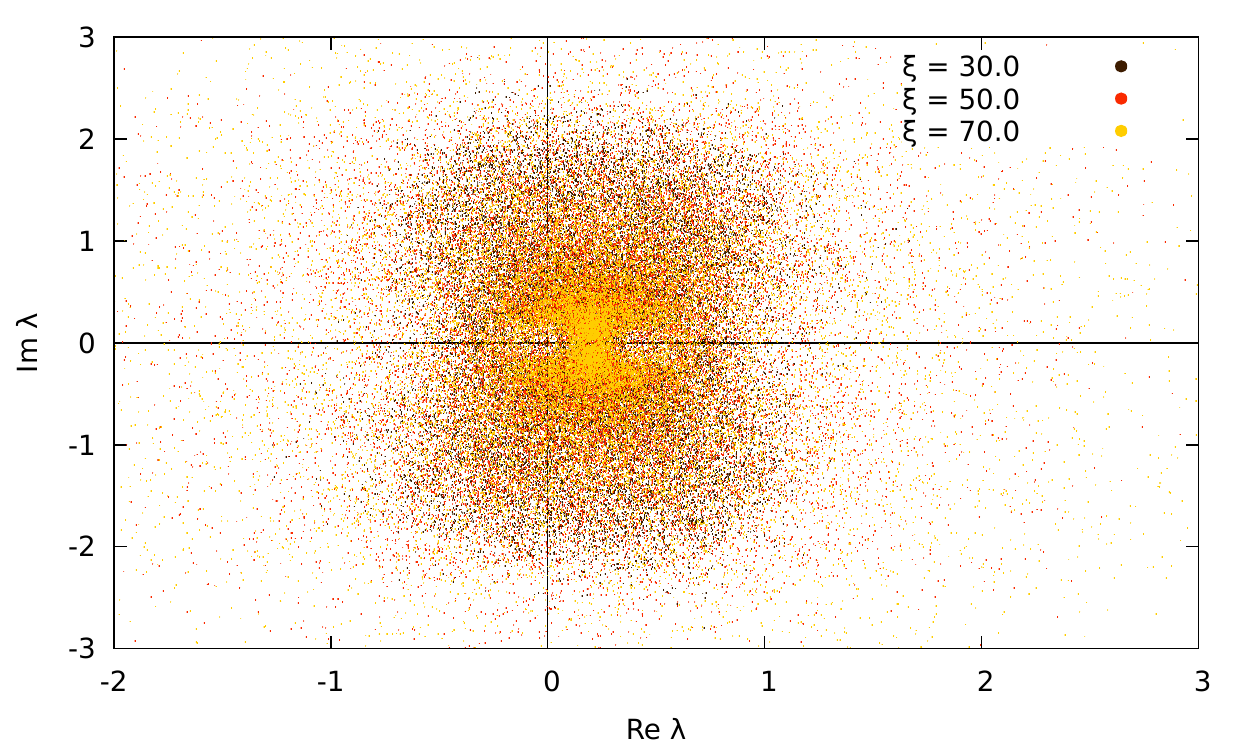}
  \caption{Scatter plots of the eigenvalues of the massive Dirac operator
    \eref{Dsteph} with $m=0.2$ and $\mu=0.5$ for various values of the cooling
    parameter using eigenvalue cooling. All eigenvalues were included in the
    cooling norm. In the left figure we show eigenvalues for $\xi=1$ (brown),
    $\xi=3$ (red) and $\xi=10$ (yellow) and in the right figure the
    eigenvalues are for $\xi=30$ (brown), $\xi=50$ (red) and $\xi=70$ (yellow).}
  \label{cool-ev}
\end{figure}

\begin{figure}[t!]
  \centering
  \includegraphics[width=0.45\textwidth]{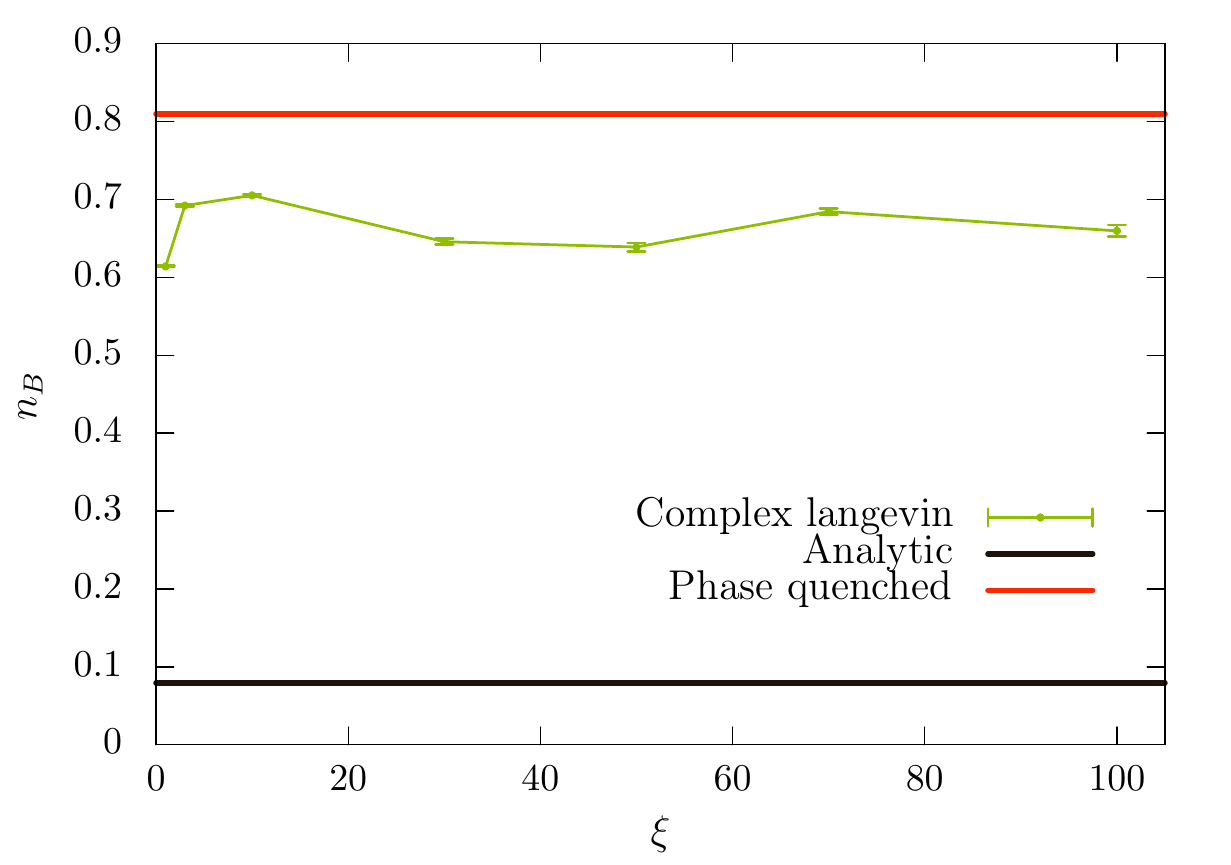}
  \includegraphics[width=0.45\textwidth]{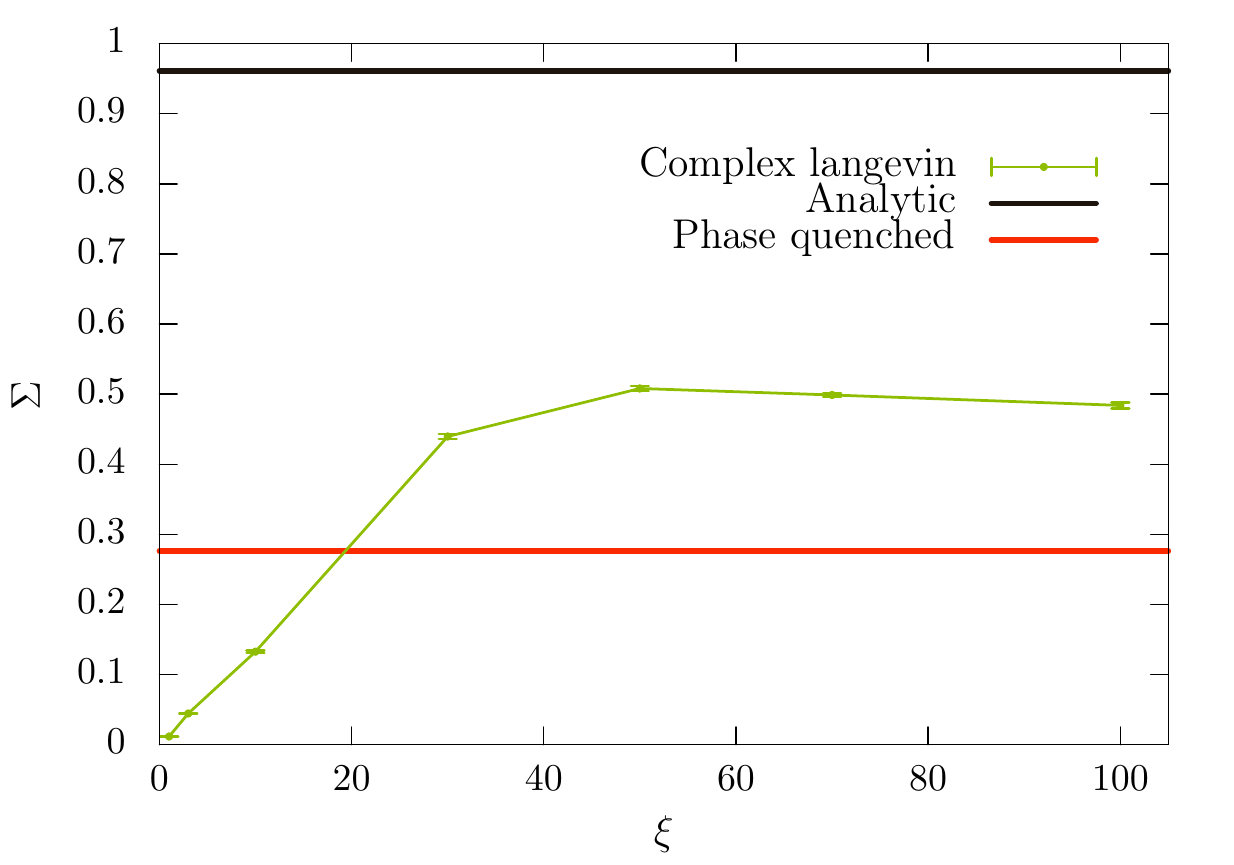}
  \caption{The baryon number (left) and the chiral condensate (right) for the
    random matrix model \eref{Dsteph} as a function of the eigenvalue cooling
    parameter $\xi$ for $m=0.2$ and $\mu=0.5$. The horizontal lines indicate
    the quenched value and the one-flavor value of the chiral condensate and the
    baryon number density.}
  \label{cond-ev}
\end{figure}

Since it is surprising that cooling with the eigenvalue norm results in a wider
eigenvalue distribution for the Stephanov model, we have studied the dependence
on the cooling parameters $\xi$ and the number of eigenvalues included in more
detail. In Fig.~\ref{cool-ev} we show scatter plots of the Dirac eigenvalues for
small $\xi$ (left) and large $\xi$ (right). For small values of $\xi$ the
eigenvalue distribution turns into a spherically symmetric ring, which gives
rise to a vanishing  chiral condensate; see Fig.~\ref{cond-ev} for the chiral
condensate and the baryon number density as a function of $\xi$. For increasing
$\xi$ the eigenvalue distribution becomes more elongated along the imaginary
axis, and at $\xi \approx 60$, the central hole in the eigenvalue distribution
disappears. For large $\xi$, see Fig.~\ref{cool-ev} right, the eigenvalue
distribution approaches the spectral domain of the quenched theory, albeit with
many more outlying eigenvalues. The chiral condensate and the baryon number
density in Fig.~\ref{cond-ev} show a continuous dependence on $\xi$ up to $\xi
\approx 60$, and take on approximately constant values beyond this point. The
chiral condensate  approaches its quenched value, while the baryon number
remains different from the quenched result. A possible interpretation of these
results is that for small $\xi$ the cooling process moves the CL trajectories to
a  Lefschetz thimble that does not give the correct dynamical result, while for
large $\xi$, many Lefschetz thimbles that contribute each with their own sign,
wove the result in the direction of the quenched result because CL does not take this phase factor into
account.

\begin{figure}
  \centering
  \includegraphics[width=0.45\textwidth]{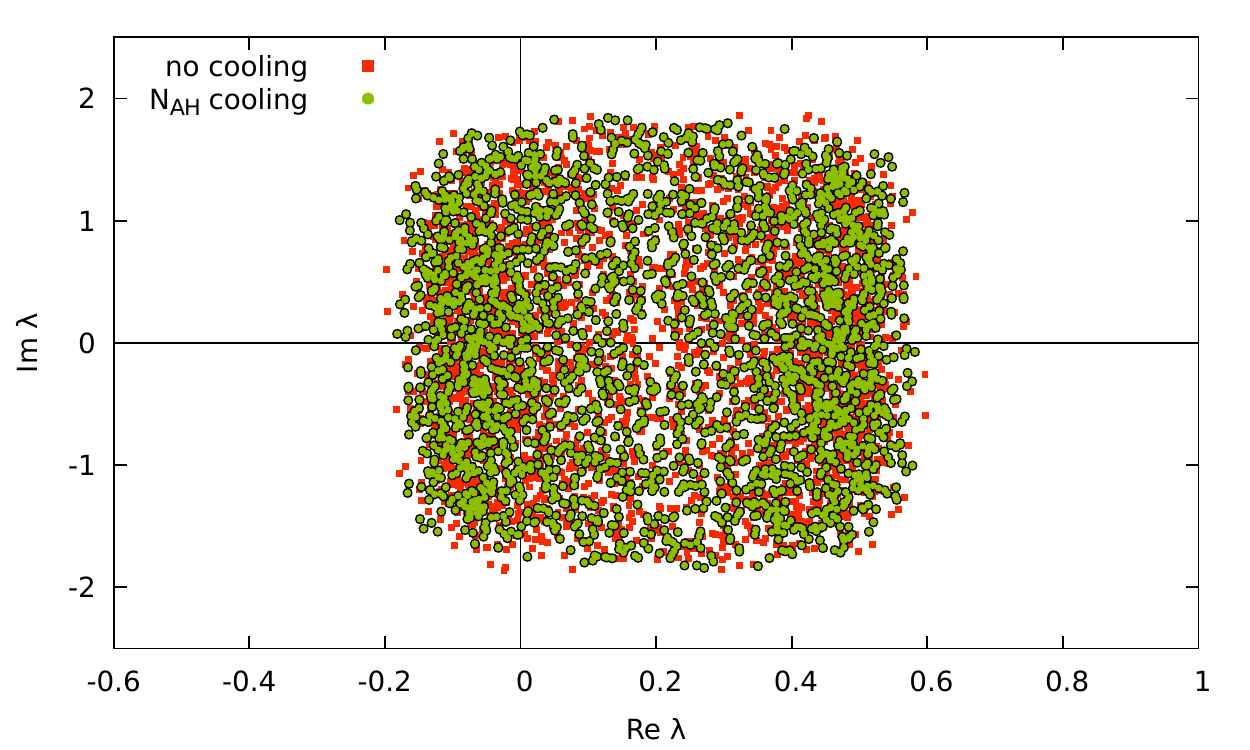}
  \includegraphics[width=0.45\textwidth]{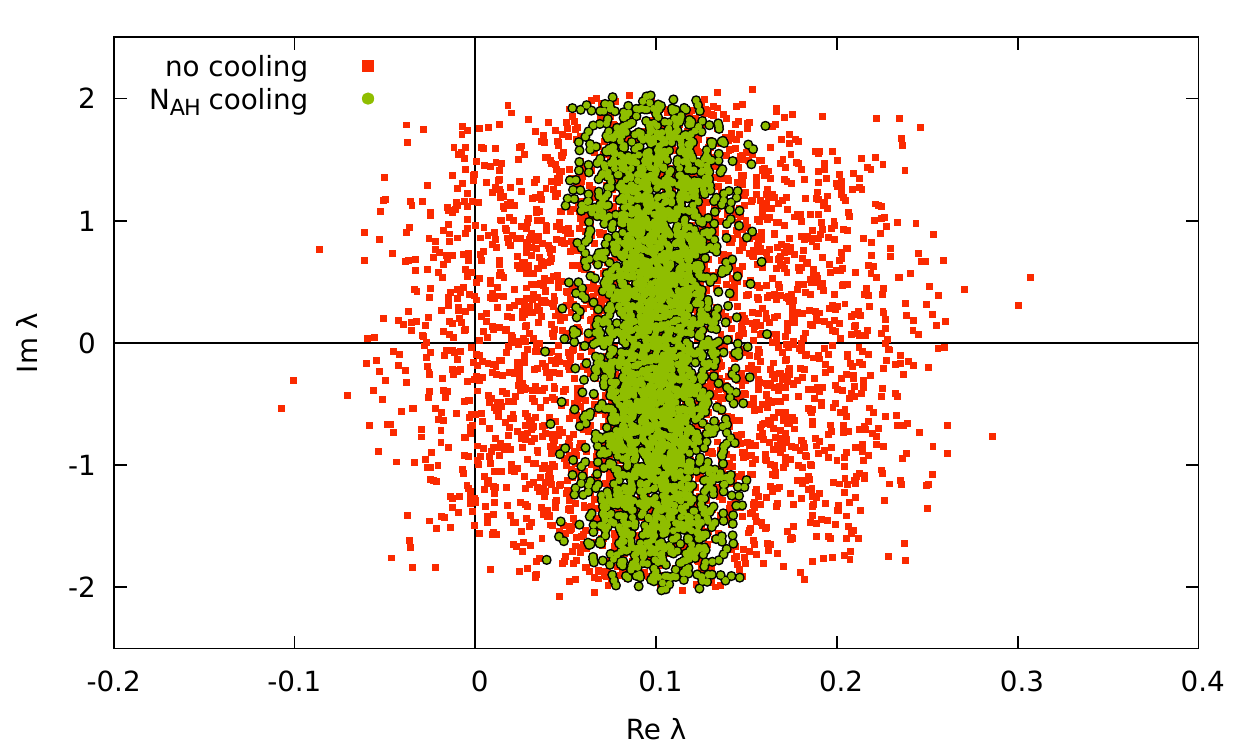}
  \caption{Scatter plots of the eigenvalues of the massive Dirac operator for a standard
    CL run together with the ones from a run cooled with the $\mathcal{N}_{AH}$
    cooling norm. The plots show the eigenvalues from the last 60 trajectories,
    separated by 100 updates. The left hand side plot shows the Stephanov model,
    while the Osborn model is shown to the right.}
  \label{fig:ah_cooling_eigenvalues}
\end{figure}

Results for the eigenvalue distributions obtained by cooling with the
anti-Hermiticity norm $\mathcal{N}_{AH}$ are shown in
Fig.~\ref{fig:ah_cooling_eigenvalues}. Once more we observe that the Osborn
model is susceptible to the effects of gauge cooling, while it has no effect on
the fermionic eigenvalues of the Stephanov model.

\begin{figure}
  \centering
  \includegraphics[width=0.45\textwidth]{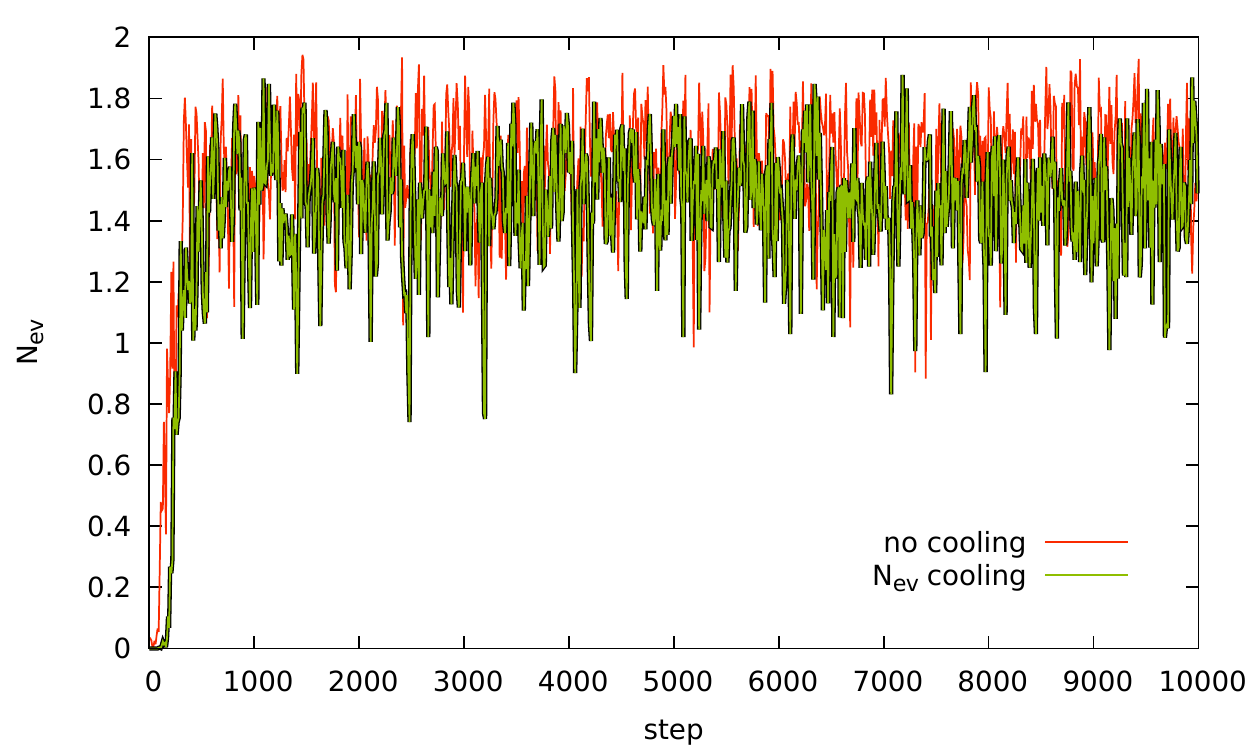}
  \includegraphics[width=0.45\textwidth]{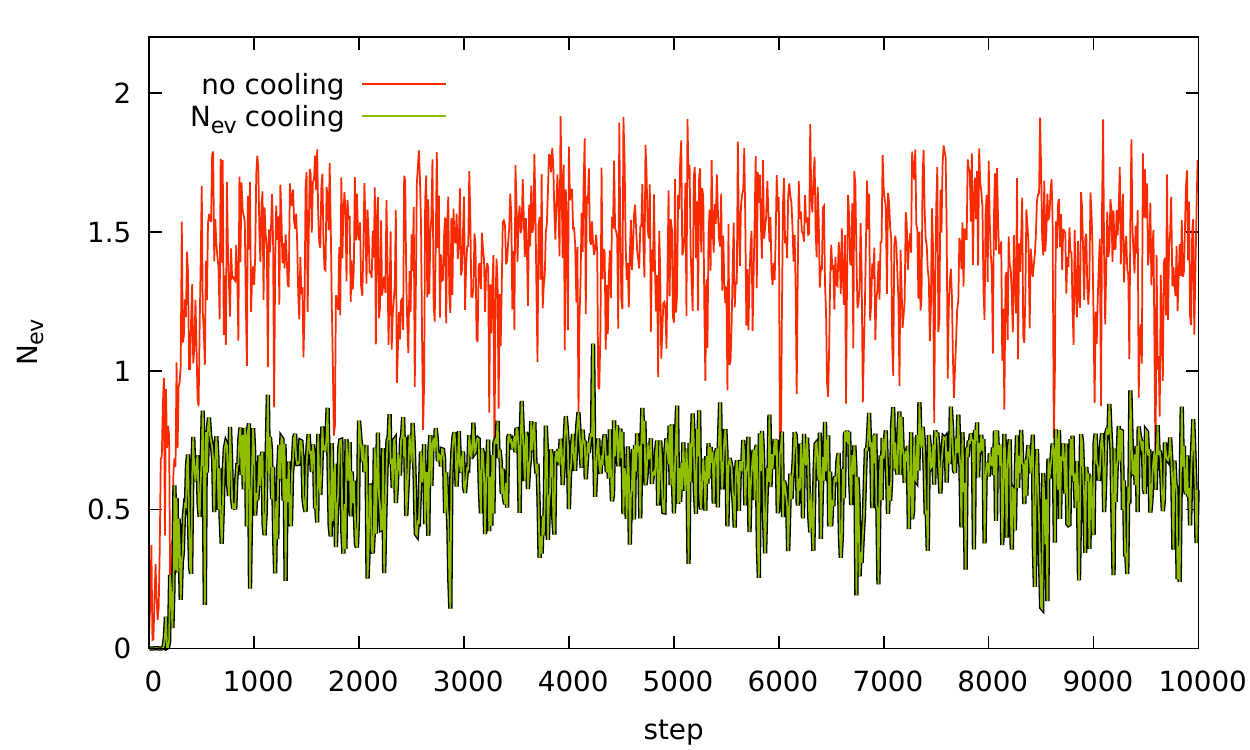}
  \caption{Value of $\mathcal{N}_{\mathrm{ev}}$ as a function of Langevin time.
    The Stephanov model is on the right and the  Osborn model is on the left.}
  \label{fig:eval_cooling_norm}
\end{figure}

\begin{figure}
  \centering
  \includegraphics[width=0.45\textwidth]{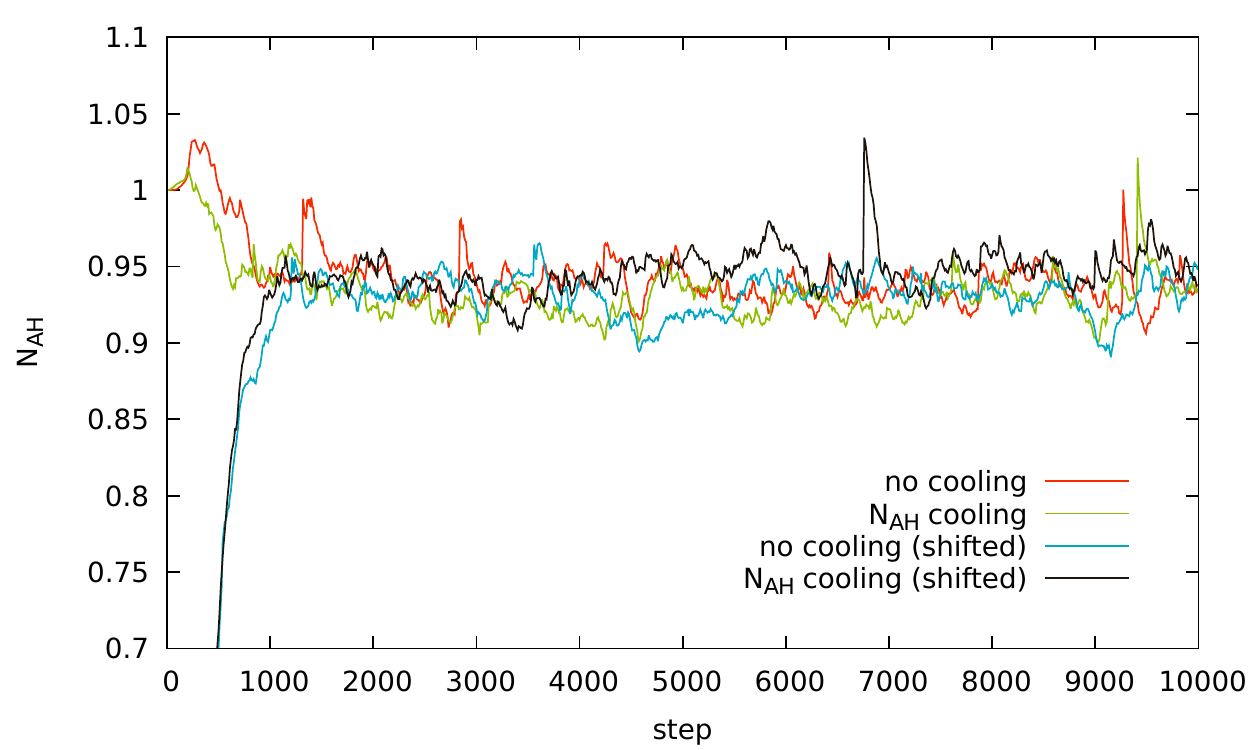}
  \includegraphics[width=0.45\textwidth]{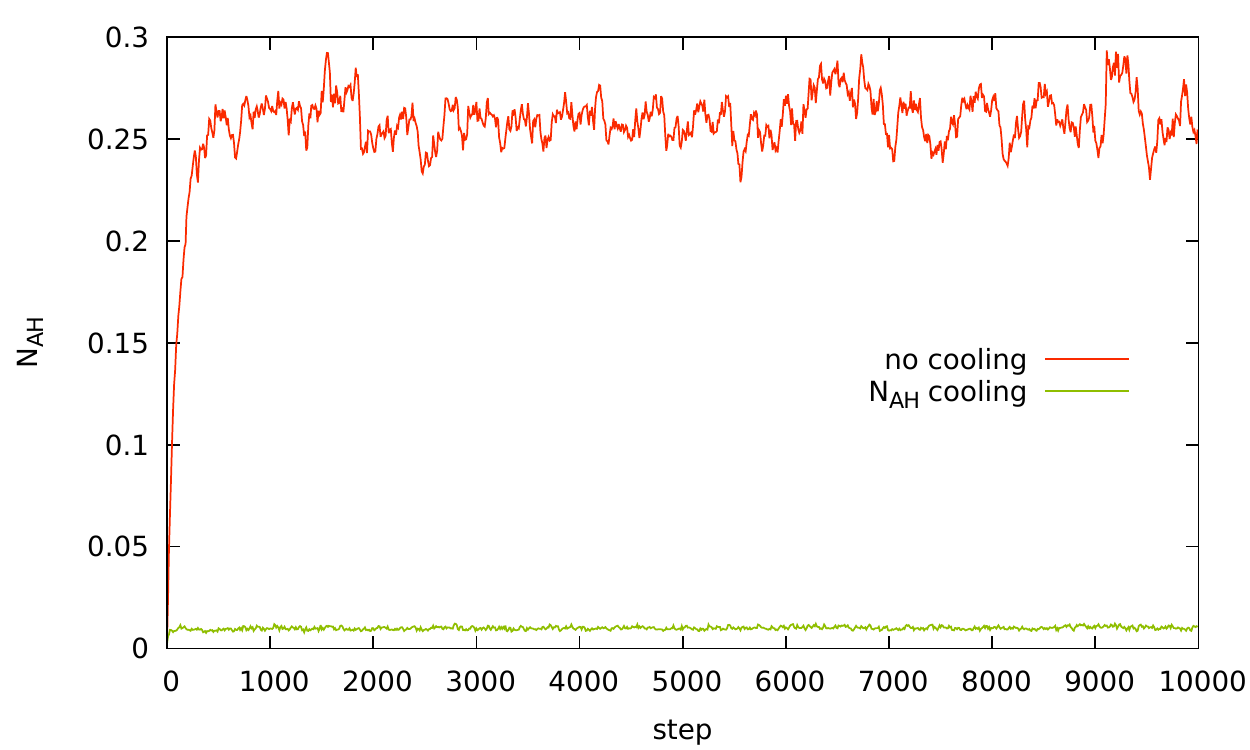}
  \caption{Value of $\mathcal{N}_{AH}$ as a function of Langevin time for the
    Stephanov model (left) and the  Osborn model (right).  The left figure
    also includes the history for the shifted representation. These start at 0
    for $t = 0$, but quickly shoot up to meet the unshifted curves.}
  \label{fig:ah_cooling_norm}
\end{figure}

We also looked at the evolution of the norms $\mathcal{N}_{AH}$ and
$\mathcal{N}_{\mathrm{ev}}$ as a function of the Langevin time, see
Figs.~\ref{fig:eval_cooling_norm} and \ref{fig:ah_cooling_norm}, respectively.
The plots show that the evolution of the norm reflects the eigenvalue situation.
Whereas for the Osborn model the norm is clearly reduced by the corresponding
cooling algorithm, no such improvement is seen for the Stephanov model. Even the
shifted representation, which could leverage its more advantageous initial
condition (the Dirac operator is anti-Hermitian) for cooling to work, simply
falls back to that of the uncooled, unshifted CL.  The difference between the
Osborn model and the Stephanov model is reminiscent of the difference in
convergence of the CL algorithm between U($N$) and SU($N$) one-dimensional
lattice QCD models \cite{Aarts:2010gr,Aarts:2012ft,Bloch:2015coa}.
 
\section{Deformation technique}

\begin{figure}
  \centering
  \includegraphics[width=0.45\textwidth]{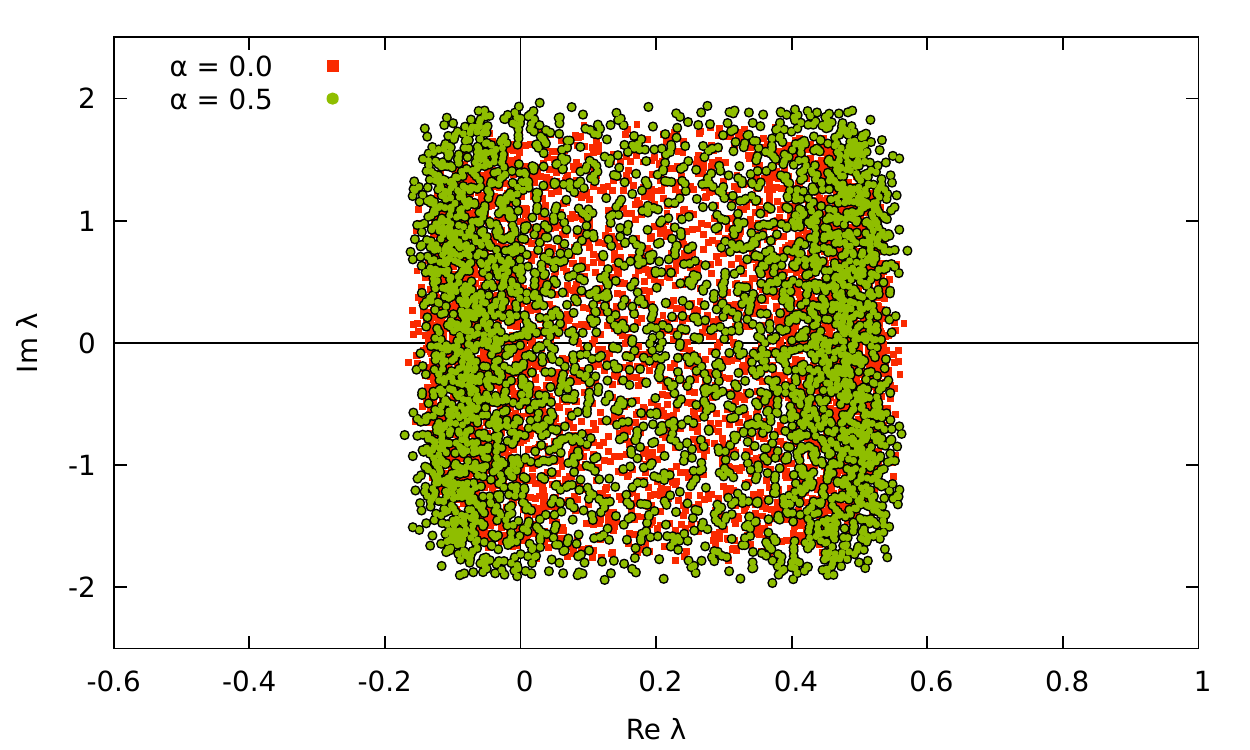}
  \includegraphics[width=0.45\textwidth]{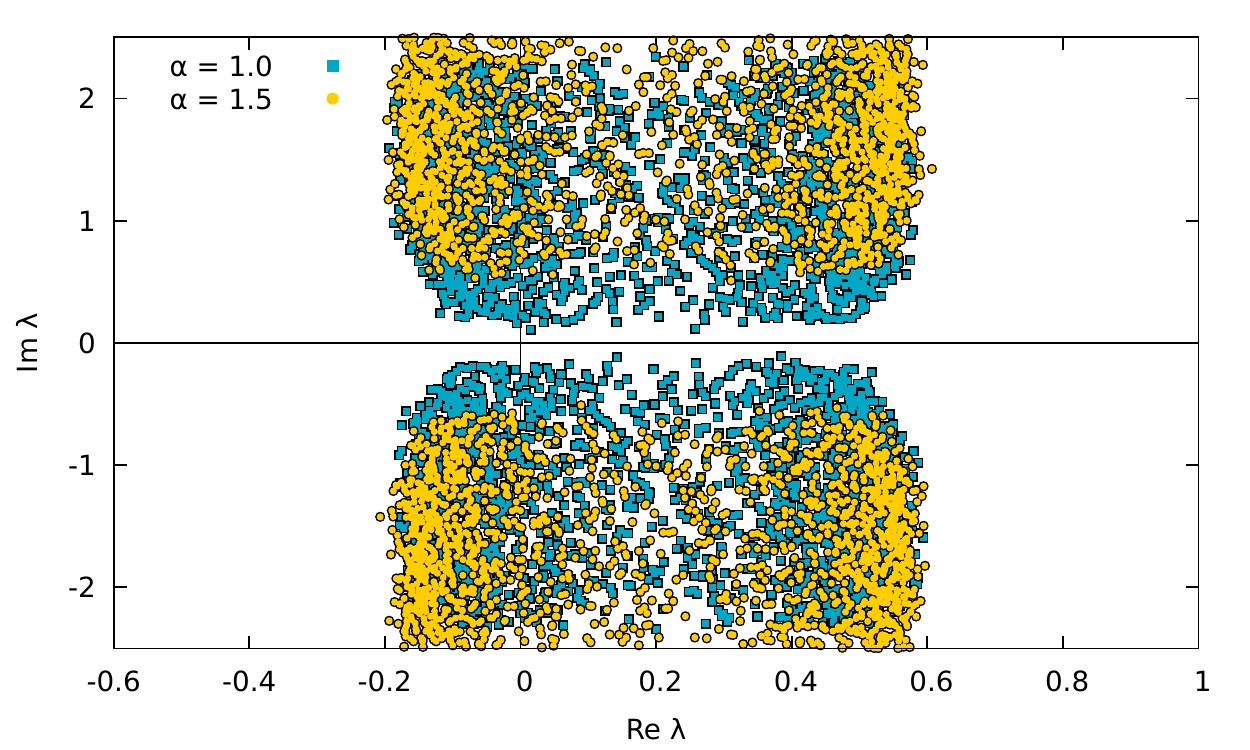}
  \caption{Scatter plots of eigenvalues from simulating
    \eqref{eq:imag-mu-action} with $N=96$, $\Delta t = 5\times 10^{-5}$,
    $t_{\text{end}} = 5.0$. Showing the last 20 configurations separated by
    $1000$ updates.  Both plots show $m = 0.2$ and $\mu = 0.5$ for varying
    values of the "temperature" $\alpha$.}
  \label{fig:imag_mu0.5_eval_movement}
\end{figure}

\begin{figure}
  \centering
  \includegraphics[width=0.45\textwidth]{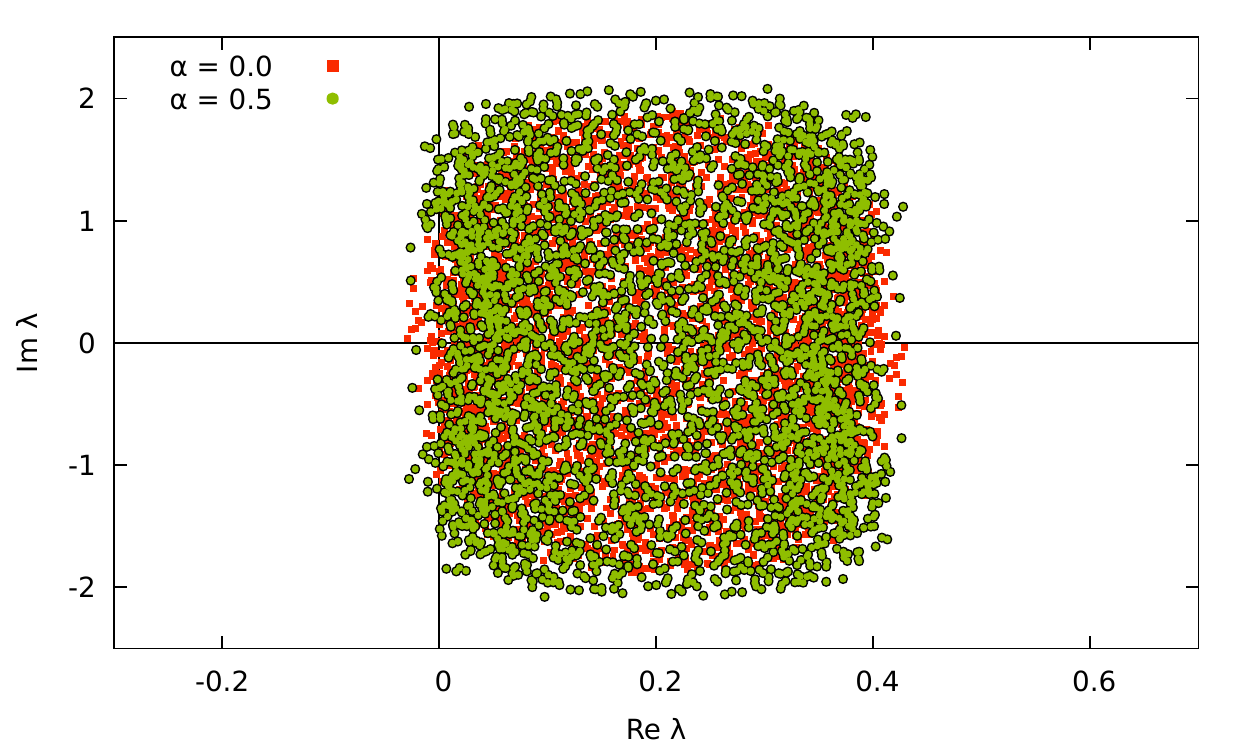}
  \includegraphics[width=0.45\textwidth]{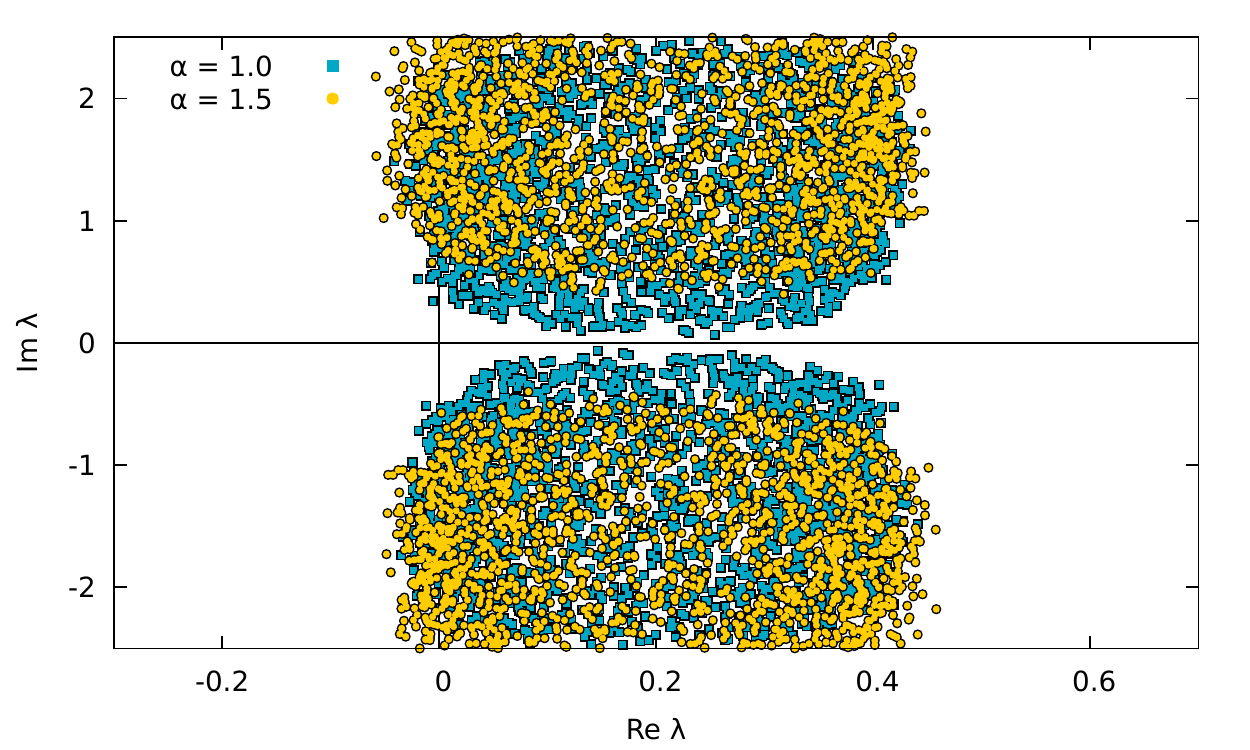}
  \caption{Same as Fig.~\ref{fig:imag_mu0.5_eval_movement} but with $m = 0.2$
    and $\mu = 0.35$.}
  \label{fig:imag_mu0.35_eval_movement}
\end{figure}

Another procedure which attempts to fix the issues CL has for simulating
systems at finite chemical potential was proposed in \cite{Ito:2017wun}. The
basic idea is to deform the Dirac operator such that its eigenvalues are removed
from the region around the origin, and then extrapolate the deformation
parameter to zero. We will deform the random matrix model by a finite
temperature term which in essence is given by the two lowest Matsubara
frequencies $\pm \pi T$ \cite{Jackson:1995nf,Halasz:1998qr}, 
\begin{equation} \label{eq:imag-mu-action}
  Z(m,\mu;\alpha) = \int \mathrm{d} X \mathrm{d} Y \; \det%
  \begin{pmatrix}
    m & X + \mu + i\Theta(\alpha) \\
    Y + \mu + i\Theta(\alpha) & m
  \end{pmatrix} %
  P(X,Y),
\end{equation}
where $\Theta(\alpha)$ is itself a block-matrix
\begin{equation}
  \Theta(\alpha) = %
  \begin{pmatrix}
    \alpha & 0 \\
    0 & -\alpha
  \end{pmatrix},
\end{equation}
and $\alpha$ can be thought of as the lowest Matsubara frequency. Following
\cite{Nagata:2017pgc}, we measure the physical quantities in question as a
function of $\alpha$, and then extrapolate $\alpha \to 0$. Beyond a critical
value of $\alpha$ the eigenvalue spectrum opens up in the imaginary direction at
which point chiral symmetry is restored. We can thus extrapolate from higher
values in $\alpha$ for which there are no eigenvalues at the origin. This
behavior is clearly demonstrated in Fig.~\ref{fig:imag_mu0.5_eval_movement} for
$(m, \mu) = (0.2,0.5)$, where we see a gap opening at $\alpha \approx 1.0$. This
is however a fairly large range to extrapolate over, and what is more, these
parameter values correspond to a different phase of the model. Since $N$ is
finite, the latter is not a fundamental problem though. The extrapolation
problem unfortunately does not really improve if we choose values of $(m,\mu)$
where the sign problem is milder. In Fig.~\ref{fig:imag_mu0.35_eval_movement} we
show a similar scatter plot for $(m,\mu) = (0.2, 0.35)$. As can be seen
from the location of the origin with respect to the eigenvalue cloud this is a
relatively \emph{mild} case, as the origin is close to the edge.

\begin{figure}
  \centering
  \includegraphics[width=0.98\textwidth]{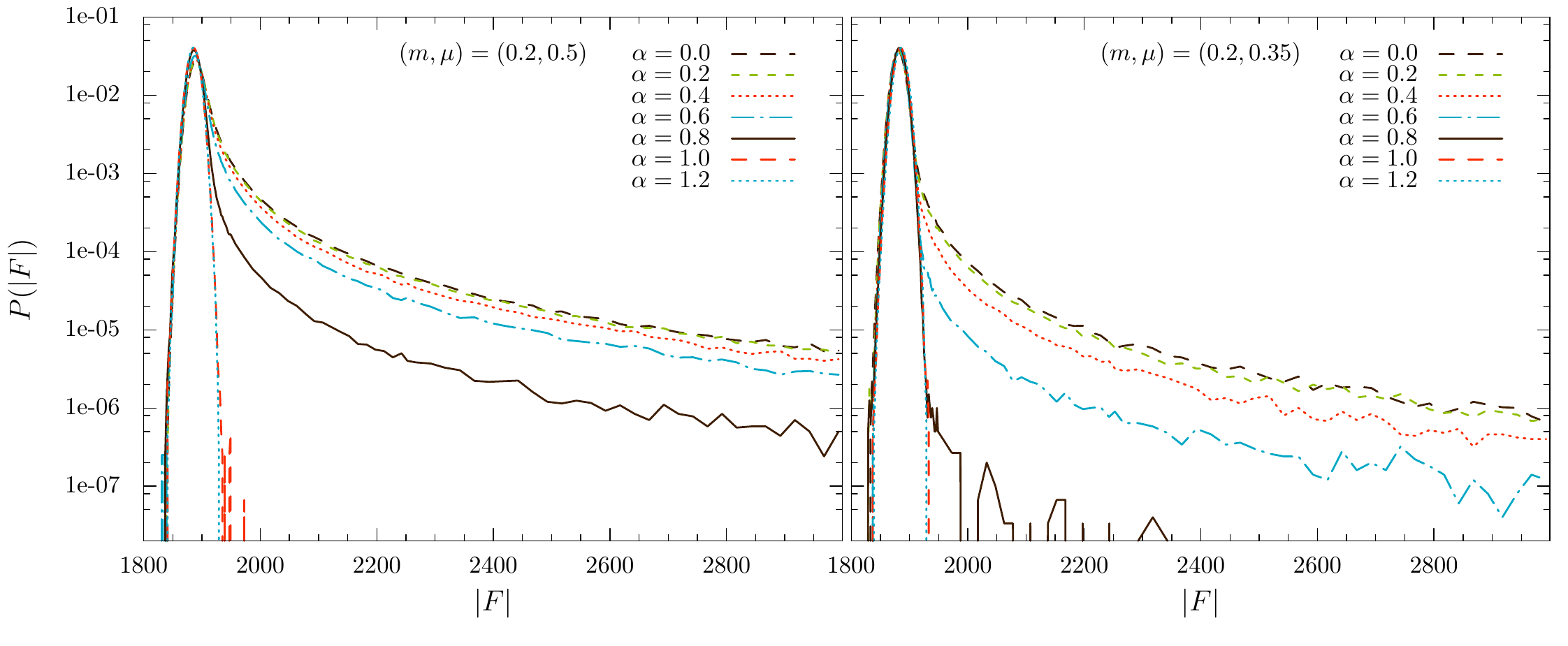}
  \caption{Histogram of the CL forces appearing in a simulation for $\mu = 0.5$
    (left) and $\mu = 0.35$ (right) for varying $\alpha$. Data gathered with a
    $t_{\mathrm{final}} = 100$ run using $\Delta t = 5\times 10^{-5}$.}
  \label{fig:imag_mu_force_hist}
\end{figure}

For a more quantitative approach we can also analyze the behavior of the force
norm as suggested by \cite{Nagata:2016vkn}.  It is postulated that if $P(|F|)$,
which is the density of the norm of the Langevin force, falls off at an
exponential rate (or faster), the Langevin algorithm will give the correct
result. However, if it falls off as a power law (or slower), we do not expect
Langevin to converge to the right answer \cite{Nagata:2016vkn}. Therefore we
define $\alpha_c$, from which one may extrapolate, as the first $\alpha$ for which
the Langevin force decays as a power law or slower. This is plotted in
Fig.~\ref{fig:imag_mu_force_hist} and demonstrates that the value of $\alpha_c$
does not change much as we move from a hard to a mild problem, depending on the
value of $\mu$; we also saw this in Figs.~\ref{fig:imag_mu0.5_eval_movement} and
\ref{fig:imag_mu0.35_eval_movement} which demonstrates that the gap does not
open until $\alpha \approx 1.0$.

\begin{figure}
  \centering
    \includegraphics[width=0.48\textwidth]{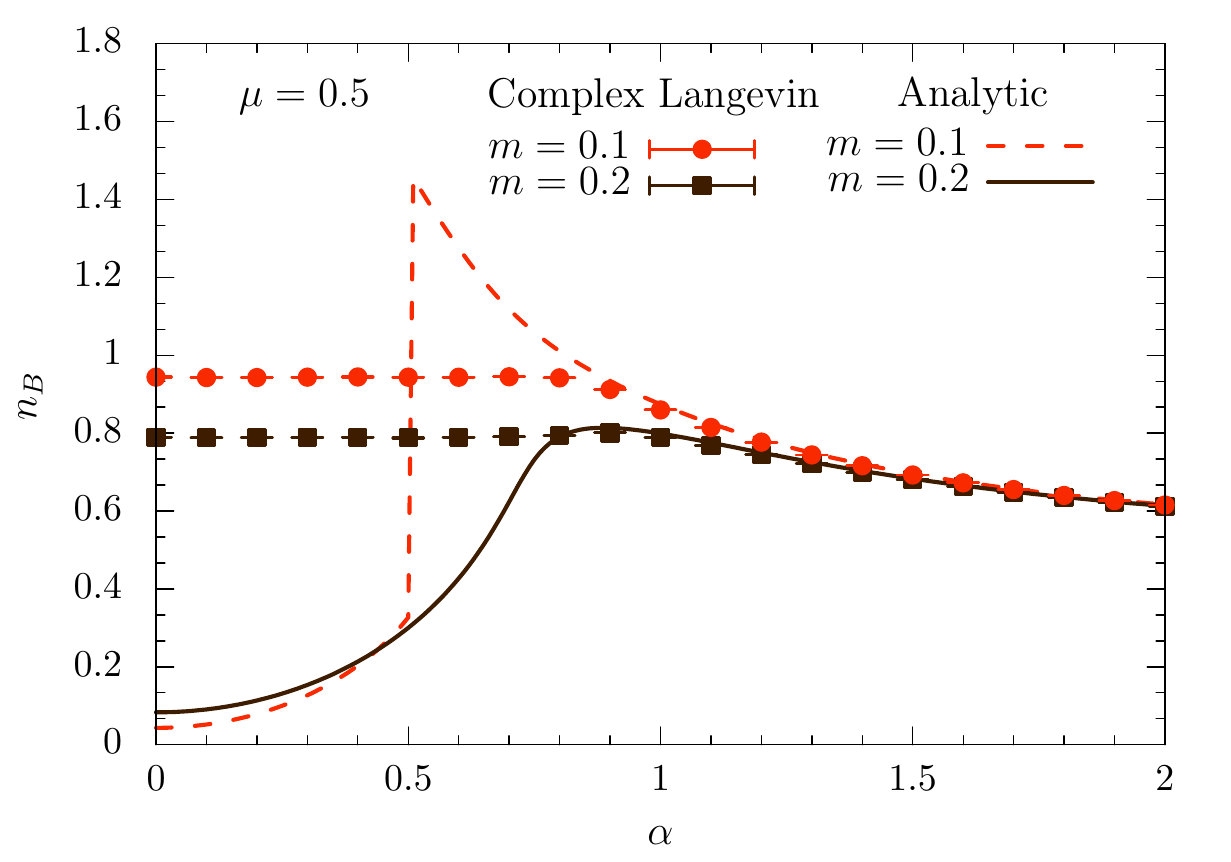} \;
    \includegraphics[width=0.48\textwidth]{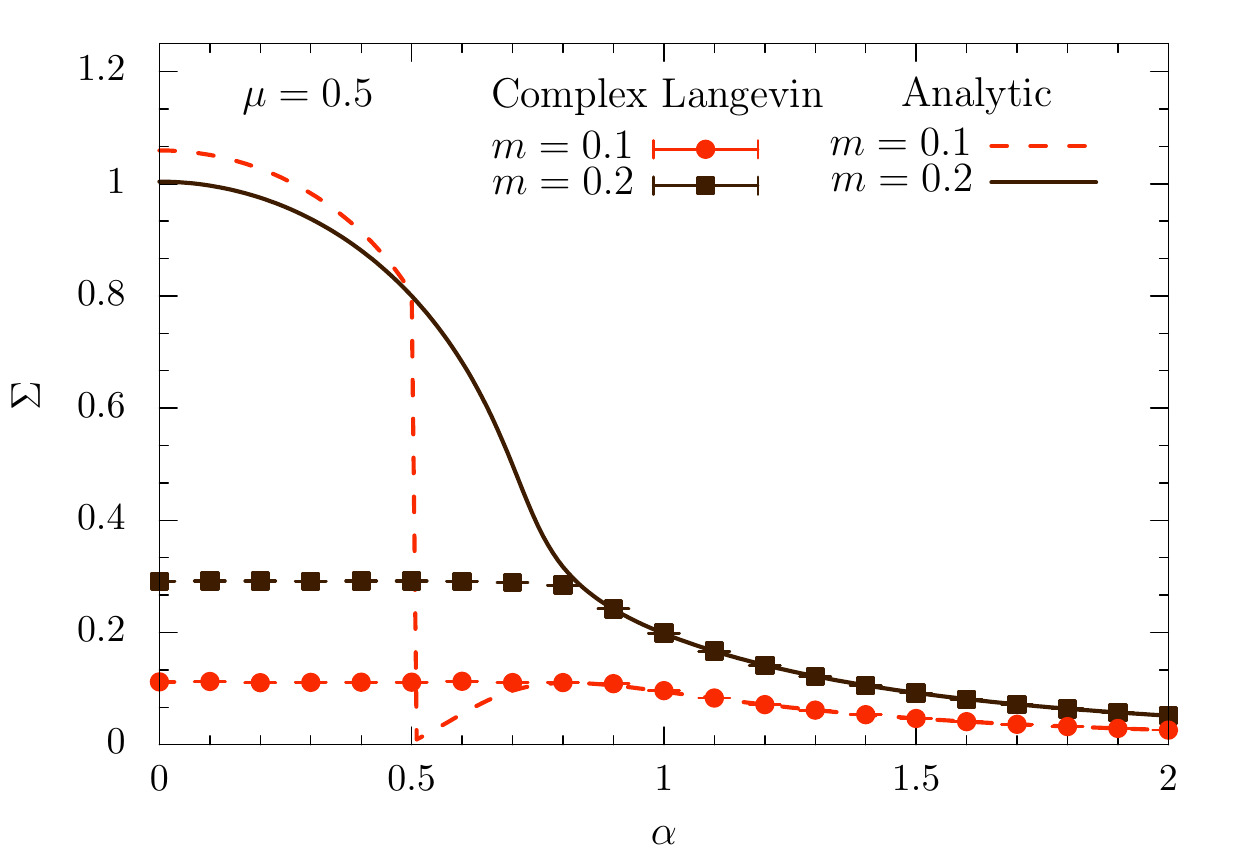}
    \caption{The baryon number (left) and the chiral condensate (right) as a
      function of the parameter $\alpha$ in \eref{eq:imag-mu-action} for $\mu = 0.5$ and $m = \{0.1, 0.2\}$.
      The curves are the analytic solutions in the thermodynamic limit while the
      points depict results obtained from simulations at $N=96$.}
  \label{fig:imag_mu_extrapolate}
\end{figure}

In Fig.~\ref{fig:imag_mu_extrapolate} we plot the analytic solution for the
random matrix theory \eqref{eq:imag-mu-action} in the thermodynamic limit, for
masses $m=\{0.1,0.2\}$ \cite{Halasz:1998qr}. We also plot the corresponding CL
results. As predicted by the histogram study of the previous paragraph we see
agreement with the analytic curve for $\alpha \gtrsim 1.0$.  There are however
two more crucial observations to be made.  First, looking at the $m=0.2$ data,
we can conclude that although theoretically possible, it is infeasible in
practice to extrapolate the values for the condensate and the baryon number to
$\alpha = 0$ due to the rapid change of these quantities in the region $\alpha
\in [0,1]$. Second, looking at the $m=0.1$ data, we observe that there is a
phase transition separating the $\alpha = 0$ and $\alpha \geq 1$ region, meaning
that the method has a limited range of convergence in mass. This means that even
if the issue of precision and statistics can be overcome to solve the first
issue, there is only a limited mass range this can work for.

\section{Conclusions}

In this work we have analyzed the complex Langevin algorithm for the Stephanov
model, which is a random matrix theory model for QCD at finite baryon density.
This model possesses a rich structure due to a phase transition that takes place
at a finite critical value of the baryon chemical potential, and separates two
distinct phases with zero and nonzero baryon density, respectively. The main
issue that was discussed in this paper is the convergence of the complex
Langevin algorithm which is particularly problematic in the cold and dense
regime as one is approaching the chiral limit.  We observed that a naive
implementation  of the complex Langevin algorithm yields  phase quenched results
for the chiral condensate and the baryon number density, whose analytical
expressions were also derived in this article.  The issues of the wrong
convergence was addressed by implementing several methods that have been
suggested in the literature  to rectify the pathologies of the complex Langevin
algorithm, a complex Langevin reweighting method, several cooling methods and an
extrapolation method.  In order to shed more light on the properties of some of
these methods, we also performed a direct comparison with a relatively similar
matrix model, the Osborn model, which however has a milder sign problem.

We were able to recover the correct solution with a novel reweighting technique
that uses complex Langevin trajectories chosen from the parameter regime where
the algorithm converges to the correct solution. A striking result of the
reweighting procedure was that by choosing an auxiliary ensemble at one side of
the transition we could reproduce the correct results on the other side of the
transition too. However, the cost of this method is exponential in the volume and
therefore it does not solve the sign problem.  Second, we tested the gauge
cooling method, where one utilizes the enhanced gauge symmetry of the
complexified action to modify the complex trajectories with the hope of
retrieving the correct solution.  While this method works remarkably well for
the Osborn model, as was already mentioned in the literature, it fails for the
Stephanov model.  We carefully studied the effect of cooling using different
norms on the Dirac spectrum, the chiral condensate and the baryon number
density.  Two of the norms yield the phase quenched results for these observables,
while the so called eigenvalue norm which depends on two parameters, gives results
that depend on these parameters and do not agree simultaneously with the correct
analytical result for any value of the parameters.  Another attempt to assist
cooling was done by shifting the entire $\mu$-dependence from the fermion determinant to
the ``gauge" part of the action and that was unsuccessful as well.  The hope was that a
different complexification, which actually starts off with the correct value of
the ``anti-Hermiticity norm", could potentially alleviate the convergence
problem, but to no avail.  Finally, we tested the so called deformation
technique, which is a novel idea that was introduced only very recently, and
which has produced some promising first results for QCD in small volumes.  The
idea is rather intriguing because the deformation parameter can be interpreted
as an imaginary chemical potential or a finite temperature in the matrix model
language, and it is well established by now that complex Langevin has far less
problems at high temperatures due to the much milder singular drift term
problem. However, it extrapolates from a parameter domain where the quark mass
is outside the spectral domain of the Dirac operator, to a parameter domain
where the quark mass is inside this domain, and it is not surprising that
the extrapolation to zero deformation parameter cannot be made in a controlled
and reliable way. 

In conclusion, we have shown that the complex Langevin algorithm including
cooling and deformation techniques cannot solve the sign problem of the random
matrix model originally proposed by Stephanov in the domain where the quark mass
is inside the spectrum of the Dirac operator.  The only method that gives the
correct solution is a complex Langevin reweighting method, but since the cost of
this method remains exponential in the volume we cannot claim that this method
solves the sign problem.  What distinguishes random matrix theory from QCD is
that it is a much stronger coupled theory, making it much harder for the drift
term, to evolve to the correct Langevin trajectory if it indeed exists. A
plausible explanation of our results is that there is no such trajectory, and
that correct results can only be obtained by taking into account multiple
thimbles, each with its own phase. These results do not necessarily generalize
to QCD which is a much weaker coupled theory, but do raise serious concerns that
the complex Langevin method does not work when the quark mass is inside the
spectral domain of the Dirac operator.

\section{Acknowledgments}
The authors acknowledge fruitful discussions with G. Aarts, Ph. de Forcrand, K.
Nagata, J. Nishimura, O. Philipsen, S. Shimasaki and I. O. Stamatescu.  SZ
acknowledges support by the National Science Foundation (USA) under grant
PHY-1516509, by the Jefferson Science Associates, LLC under  U.S. DOE Contract
\#DE-AC05-06OR23177 and by the DFG Collaborative Research Centre SFB 1225
(ISOQUANT). JB is supported by the DFG collaborative research center SFB/TRR-55.
JG has been supported by STFC grants ST/L000369/1 and ST/P00055X/1. We are
grateful for the computing resources made available by HPC Wales.

\appendix

\section{The Phase Quenched Partition Function}
\label{app1}

In this appendix we evaluate the mean field result for the free energy of the phase
quenched random matrix  partition function \eref{Zst} with Dirac operator \eref{Dsteph}.
Since in random matrix theory the number of flavors only enters as $O(N_f/N)$ we evaluate
the large $N$ limit of this partition function for the simplest case which
is $N_f =2$.
Some of the results on this appendix also appeared in
\cite{Stephanov:1996ki,Halasz:1999gc}

The phase quenched two-flavor partition function \eref{Zst} can be rewritten identically
as \cite{Stephanov:1996ki}
\be
Z_{\rm pq}(m,\mu) = \int d D \, e^{-N(|a-m|^2+|b-m^*|^2+|c|^2+|d|^2)} \, {\det}^n D
\label{pqint}
\ee
with
\be
D = 
\begin{pmatrix}
a & \mu &0& id\\
  \mu& a^* &  ic & 0\\
  0&id^*&b^*& \mu^*\\
  ic^*&0&\mu^*&b 
\end{pmatrix} .
\ee
The integration $d D$ is over the real and imaginary parts of $a$, $b$, $c$ and $d$. For large $n$, this partition function can be evaluated by a saddle point approximation. The determinant can be evaluated as
\be
\det D =(|a|^2 -z^2)(|b|^2-{z^*}^2) +a^*b^*c^*d+abcd^*+|z|^2(|c|^2+|d|^2)+|c|^2|d|^2.
\label{det0}
\ee
A variable and its complex conjugate are independent solutions, and the complex conjugate of the saddle point value of a solution may not be the solution of the complex conjugate variable.

The integral \eref{pqint} is an 8 dimensional integral with  saddle points determined by the equations
\be
F_1&\equiv&(a^*-m) \det D - a^*|b|^2 - cd^*b + a^* {\mu^*}^2=0,\\
F_2&\equiv&(a-m) \det D - a|b|^2 - c^*d b^* + a {\mu^*}^2=0,\\
F_3&\equiv&(b^*-m^*) \det D - b^*|a|^2 - cd^*a + b^* {\mu}^2=0,\\
F_4&\equiv&(b-m^*) \det D - b|a|^2 - c^*d a^* + b {\mu}^2=0,\\
F_5&\equiv&c^*\det D - abd^*-c^*|d|^2 -c^* |\mu|^2=0,\\
F_6&\equiv&c\det D - a^*b^*d-c|d|^2 -c |\mu|^2=0,\\
F_7&\equiv&d^*\det D -a^*b^*c^*-d^*|c|^2 -d^* |\mu|^2=0,\\
F_8&\equiv&d\det D -abc-d|c|^2 -d |\mu|^2=0.
\ee
The solution of these equations occur in two different phases, the normal phase with $c=d=0$ and the pion condensation phase with $c\ne 0$ and $d\ne 0$. In the first case, the equations decouple and are easier to solve.

Note that we did not use complex conjugation to prove the first relation. The equations $F_5=0, F_6=0, F_7=0, F_8=0$ are not linearly independent,
\be
F_5-F_6+F_7-F_8=0.
\ee
From $cF_5-c^*F_6=0$ or $dF_7-d^*F_8=0$ we obtain
\be
a^*b^*c^*d-abcd^*=0.
\label{abcd}
\ee
Combining this with $a^*F_1-aF_2=0$ and $b^*F_3 -bF_4=0$ we obtain
\be 
a = a^*,\quad {\rm and}\quad b = b^*.
\label{aast}
\ee
As mentioned above, this does not imply that $a$ and $b$ are real -- in fact, they are not as we will see below.

\subsection{Solution for the Condensed Phase}

In the condensed phase we have $c$ and $d$ nonzero. From $cF_5 -dF_7=0$  we obtain
\be
c c^* = d d^*.
\ee
which in combination with \eref{abcd} give
\be
c^2= d^2.
\ee
We can also use $F_5-F_7 =0$ to show that
\be
c=d,
\label{cd}
\ee
and from $F_8 = 0 $ we then find
\be
\det D = ab+|c|^2 +\mu \mu^*.
\label{ab}
\ee
Using the expression \eref{det0} for the determinant $\det D$ we find that it satisfies the equation,
\be
   {\det}^2 D -\det D -(a\mu^*+b\mu)^2 = 0.
\label{deteq}
\ee

From $F_2 =0$ and $F_4 =0$ and using (\ref{aast},\ref{ab},\ref{deteq})
 we  can then derive
\be
(a-b-m)\det D &=& -\mu^*(a \mu^* +b\mu),\\
(b-a-m^*)\det D&=& -\mu(a \mu +b\mu^*).
\label{abeq}
\ee
This results in
\be
   a-b =\frac{\mu m - \mu^* m^*}{\mu+\mu^*}.
\label{aminb}
\ee
From \eref{deteq} and the first equation of \eref{abeq} we obtain
\be
   \det D 
&=& \frac {(\mu+\mu^*)^2}{(\mu+\mu^*)^2-(m+m^*)^2}.
\ee
From \eref{abeq} we then find
\be
     a\mu^*+b\mu = \frac {(\mu+\mu^*)(m+m^*)}{(\mu+\mu^*)^2 -(m+m^*)^2}.
\label{aplusb}
\ee
From \eref{aminb} and \eref{aplusb} we obtain $a$ and $b$. The remaining unknowns then follow from (\ref{aast},\;\ref{cd}). Note that the phase of $c=d$ is not determined
by the saddle point equations.
  
\subsection{Normal Phase}
     
In the normal phase we have that $c=d=0$ and saddle point equations for $a, a^*$ and $b, b^*$ decouple. We still have $a=a^*$ and $b=b^*$ with $a$ and $b$ given by the solutions of
\be
(a-m)(a^2-\mu^2) = a,\nn \\
(b-m^*)(b^2-{\mu^*}^2) = b.
\label{cubic}
\ee
For $m >0$ and $\mu >0$ these equations have three real solutions and the correct solution is given by the one that minimizes the free energy at the saddle-point. Below the shaded area in Fig.~\ref{fig1} we use the solution that is continuously connected to the large $m$ solution of the saddle-point equation,
\be
a-m \to \frac 1m,
\ee
while above the shaded area we use the solution that is continuously connected to the large $ \mu$ solution.
\be
a-\mu \to -\frac m{\mu^2}.
\ee
\subsection{Free Energy}

We now take $m$ and $\mu$ real. Then the boundary of the region where $c\ne 0$ is given by
\be
c(m,\mu,\mu^*) =0.
\ee
On the real axis, $\mu = \mu_r$ this equation has four solutions, $\pm u_1$ and $\pm u_2$ with $0<u_1<u_2$. The free energy is given by
\be
F(m,\mu) = n(|a-m|^2+|b-m^*|^2+|c|^2+|d|^2) - n \log \det D
\ee
evaluated for the solutions of the saddle point equations.

The baryon number density and the chiral condensate are given by
\be
n_B &=& \frac 1{2n} \frac d{d\mu} F(m,\mu),\\
\Sigma &=& \frac 1{2n} \frac d{dm} F(m,\mu).
\ee

\bibliographystyle{jhep}
%\bibliography{biblio} 

\begin{thebibliography}{10}

\bibitem{Philipsen:2010gj}
O.~Philipsen, \emph{{Lattice QCD at non-zero temperature and baryon density}},
  in \emph{{Modern perspectives in lattice QCD: Quantum field theory and high
  performance computing. Proceedings, International School, 93rd Session, Les
  Houches, France, August 3-28, 2009}}, pp.~273--330, 2010,
  \href{https://arxiv.org/abs/1009.4089}{{\ttfamily 1009.4089}}.

\bibitem{Splittorff:2006fu}
K.~Splittorff and J.~J.~M. Verbaarschot, \emph{{Phase of the Fermion
  Determinant at Nonzero Chemical Potential}},
  \href{https://doi.org/10.1103/PhysRevLett.98.031601}{\emph{Phys. Rev. Lett.}
  {\bfseries 98} (2007) 031601},
  [\href{https://arxiv.org/abs/hep-lat/0609076}{{\ttfamily hep-lat/0609076}}].

\bibitem{Allton:2002zi}
C.~R. Allton et~al., \emph{{The QCD thermal phase transition in the presence of
  a small chemical potential}},
  \href{https://doi.org/10.1103/PhysRevD.66.074507}{\emph{Phys. Rev.}
  {\bfseries D66} (2002) 074507},
  [\href{https://arxiv.org/abs/hep-lat/0204010}{{\ttfamily hep-lat/0204010}}].

\bibitem{Barbour:1991vs}
I.~Barbour and A.~Bell, \emph{{Complex zeros of the partition function for
  lattice QCD}},
  \href{https://doi.org/10.1016/0550-3213(92)90324-5}{\emph{Nucl. Phys.}
  {\bfseries B372} (1992) 385--402}.

\bibitem{DElia:2002tig}
M.~D'Elia and M.-P. Lombardo, \emph{{Finite density QCD via imaginary chemical
  potential}}, \href{https://doi.org/10.1103/PhysRevD.67.014505}{\emph{Phys.
  Rev.} {\bfseries D67} (2003) 014505},
  [\href{https://arxiv.org/abs/hep-lat/0209146}{{\ttfamily hep-lat/0209146}}].

\bibitem{deForcrand:2002hgr}
P.~de~Forcrand and O.~Philipsen, \emph{{The QCD phase diagram for small
  densities from imaginary chemical potential}},
  \href{https://doi.org/10.1016/S0550-3213(02)00626-0}{\emph{Nucl. Phys.}
  {\bfseries B642} (2002) 290--306},
  [\href{https://arxiv.org/abs/hep-lat/0205016}{{\ttfamily hep-lat/0205016}}].

\bibitem{Kogut:2000ek}
J.~B. Kogut, M.~A. Stephanov, D.~Toublan, J.~J.~M. Verbaarschot and
  A.~Zhitnitsky, \emph{{QCD-like theories at finite baryon density}},
  \href{https://doi.org/10.1016/S0550-3213(00)00242-X}{\emph{Nucl. Phys.}
  {\bfseries B582} (2000) 477--513},
  [\href{https://arxiv.org/abs/hep-ph/0001171}{{\ttfamily hep-ph/0001171}}].

\bibitem{Kogut:2002zg}
J.~B. Kogut and D.~K. Sinclair, \emph{{Lattice QCD at finite isospin density at
  zero and finite temperature}},
  \href{https://doi.org/10.1103/PhysRevD.66.034505}{\emph{Phys. Rev.}
  {\bfseries D66} (2002) 034505},
  [\href{https://arxiv.org/abs/hep-lat/0202028}{{\ttfamily hep-lat/0202028}}].

\bibitem{Parisi:1984cs}
G.~Parisi, \emph{On complex probabilities},
  \href{https://doi.org/10.1016/0370-2693(83)90525-7}{\emph{Phys. Lett.}
  {\bfseries B131} (1983) 393--395}.

\bibitem{Klauder:1983nn}
J.~R. Klauder, \emph{{Stochastic Quantization}},
  \href{https://doi.org/10.1007/978-3-7091-7651-1_8}{\emph{Acta Phys. Austriaca
  Suppl.} {\bfseries 25} (1983) 251--281}.

\bibitem{Aarts:2009uq}
G.~Aarts, E.~Seiler and I.-O. Stamatescu, \emph{{The Complex Langevin method:
  When can it be trusted?}},
  \href{https://doi.org/10.1103/PhysRevD.81.054508}{\emph{Phys. Rev.}
  {\bfseries D81} (2010) 054508},
  [\href{https://arxiv.org/abs/0912.3360}{{\ttfamily 0912.3360}}].

\bibitem{Fodor:2015doa}
Z.~Fodor, S.~D. Katz, D.~Sexty and C.~Török, \emph{{Complex Langevin dynamics
  for dynamical QCD at nonzero chemical potential: A comparison with
  multiparameter reweighting}},
  \href{https://doi.org/10.1103/PhysRevD.92.094516}{\emph{Phys. Rev.}
  {\bfseries D92} (2015) 094516},
  [\href{https://arxiv.org/abs/1508.05260}{{\ttfamily 1508.05260}}].

\bibitem{Bloch:2017jzi}
J.~Bloch and O.~Schenk, \emph{{Selected inversion as key to a stable Langevin
  evolution across the QCD phase boundary}},  in \emph{{35th International
  Symposium on Lattice Field Theory (Lattice 2017) Granada, Spain, June 18-24,
  2017}}, 2017, \href{https://arxiv.org/abs/1707.08874}{{\ttfamily
  1707.08874}}.

\bibitem{Seiler:2012wz}
E.~Seiler, D.~Sexty and I.-O. Stamatescu, \emph{{Gauge cooling in complex
  Langevin for QCD with heavy quarks}},
  \href{https://doi.org/10.1016/j.physletb.2013.04.062}{\emph{Phys. Lett.}
  {\bfseries B723} (2013) 213--216},
  [\href{https://arxiv.org/abs/1211.3709}{{\ttfamily 1211.3709}}].

\bibitem{Sexty:2013ica}
D.~Sexty, \emph{{Simulating full QCD at nonzero density using the complex
  Langevin equation}},
  \href{https://doi.org/10.1016/j.physletb.2014.01.019}{\emph{Phys. Lett.}
  {\bfseries B729} (2014) 108--111},
  [\href{https://arxiv.org/abs/1307.7748}{{\ttfamily 1307.7748}}].

\bibitem{Aarts:2014bwa}
G.~Aarts, E.~Seiler, D.~Sexty and I.-O. Stamatescu, \emph{{Simulating QCD at
  nonzero baryon density to all orders in the hopping parameter expansion}},
  \href{https://doi.org/10.1103/PhysRevD.90.114505}{\emph{Phys. Rev.}
  {\bfseries D90} (2014) 114505},
  [\href{https://arxiv.org/abs/1408.3770}{{\ttfamily 1408.3770}}].

\bibitem{Aarts:2017vrv}
G.~Aarts, E.~Seiler, D.~Sexty and I.-O. Stamatescu, \emph{{Complex Langevin
  dynamics and zeroes of the fermion determinant}},
  \href{https://doi.org/10.1007/JHEP05(2017)044}{\emph{JHEP} {\bfseries 05}
  (2017) 044}, [\href{https://arxiv.org/abs/1701.02322}{{\ttfamily
  1701.02322}}].

\bibitem{Stephanov:1996ki}
M.~A. Stephanov, \emph{{Random matrix model of QCD at finite density and the
  nature of the quenched limit}},
  \href{https://doi.org/10.1103/PhysRevLett.76.4472}{\emph{Phys. Rev. Lett.}
  {\bfseries 76} (1996) 4472--4475},
  [\href{https://arxiv.org/abs/hep-lat/9604003}{{\ttfamily hep-lat/9604003}}].

\bibitem{Jackson:1995nf}
A.~D. Jackson and J.~J.~M. Verbaarschot, \emph{{A random matrix model for
  chiral symmetry breaking}},
  \href{https://doi.org/10.1103/PhysRevD.53.7223}{\emph{Phys. Rev.} {\bfseries
  D53} (1996) 7223--7230},
  [\href{https://arxiv.org/abs/hep-ph/9509324}{{\ttfamily hep-ph/9509324}}].

\bibitem{Osborn:2004rf}
J.~C. Osborn, \emph{Universal results from an alternate random matrix model for
  {QCD} with a baryon chemical potential},
  \href{https://doi.org/10.1103/PhysRevLett.93.222001}{\emph{Phys. Rev. Lett.}
  {\bfseries 93} (2004) 222001},
  [\href{https://arxiv.org/abs/hep-th/0403131}{{\ttfamily hep-th/0403131}}].

\bibitem{Bloch:2012bh}
J.~Bloch, F.~Bruckmann, M.~Kieburg, K.~Splittorff and J.~Verbaarschot,
  \emph{{Subsets of configurations and canonical partition functions}},
  \href{https://doi.org/10.1103/PhysRevD.87.034510}{\emph{Phys.Rev.} {\bfseries
  D87} (2013) 034510}, [\href{https://arxiv.org/abs/1211.3990}{{\ttfamily
  1211.3990}}].

\bibitem{Halasz:1997he}
A.~M. Halasz, A.~D. Jackson and J.~J.~M. Verbaarschot, \emph{{Fermion
  determinants in matrix models of QCD at nonzero chemical potential}},
  \href{https://doi.org/10.1103/PhysRevD.56.5140}{\emph{Phys. Rev.} {\bfseries
  D56} (1997) 5140--5152},
  [\href{https://arxiv.org/abs/hep-lat/9703006}{{\ttfamily hep-lat/9703006}}].

\bibitem{Damgaard:2010cz}
P.~H. Damgaard, K.~Splittorff and J.~J.~M. Verbaarschot, \emph{{Microscopic
  Spectrum of the Wilson Dirac Operator}},
  \href{https://doi.org/10.1103/PhysRevLett.105.162002}{\emph{Phys. Rev. Lett.}
  {\bfseries 105} (2010) 162002},
  [\href{https://arxiv.org/abs/1001.2937}{{\ttfamily 1001.2937}}].

\bibitem{Akemann:2010em}
G.~Akemann, P.~H. Damgaard, K.~Splittorff and J.~J.~M. Verbaarschot,
  \emph{{Spectrum of the Wilson Dirac Operator at Finite Lattice Spacings}},
  \href{https://doi.org/10.1103/PhysRevD.83.085014}{\emph{Phys. Rev.}
  {\bfseries D83} (2011) 085014},
  [\href{https://arxiv.org/abs/1012.0752}{{\ttfamily 1012.0752}}].

\bibitem{Kieburg:2011uf}
M.~Kieburg, J.~J.~M. Verbaarschot and S.~Zafeiropoulos, \emph{{Eigenvalue
  Density of the non-Hermitian Wilson Dirac Operator}},
  \href{https://doi.org/10.1103/PhysRevLett.108.022001}{\emph{Phys. Rev. Lett.}
  {\bfseries 108} (2012) 022001},
  [\href{https://arxiv.org/abs/1109.0656}{{\ttfamily 1109.0656}}].

\bibitem{Kieburg:2013xta}
M.~Kieburg, J.~J.~M. Verbaarschot and S.~Zafeiropoulos, \emph{{Spectral
  Properties of the Wilson Dirac Operator and random matrix theory}},
  \href{https://doi.org/10.1103/PhysRevD.88.094502}{\emph{Phys. Rev.}
  {\bfseries D88} (2013) 094502},
  [\href{https://arxiv.org/abs/1307.7251}{{\ttfamily 1307.7251}}].

\bibitem{Kieburg:2015vqa}
M.~Kieburg, J.~J.~M. Verbaarschot and S.~Zafeiropoulos, \emph{{Dirac Spectrum
  of the Wilson Dirac Operator for QCD with Two Colors}},
  \href{https://doi.org/10.1103/PhysRevD.92.045026}{\emph{Phys. Rev.}
  {\bfseries D92} (2015) 045026},
  [\href{https://arxiv.org/abs/1505.01784}{{\ttfamily 1505.01784}}].

\bibitem{Cichy:2016tyj}
K.~Cichy, K.~Splittorff and S.~Zafeiropoulos, \emph{{Twisted mass Dirac
  spectrum}},  \href{https://arxiv.org/abs/1612.01289}{{\ttfamily 1612.01289}}.

\bibitem{Verbaarschot:1994qf}
J.~J.~M. Verbaarschot, \emph{{The Spectrum of the QCD Dirac operator and chiral
  random matrix theory: The Threefold way}},
  \href{https://doi.org/10.1103/PhysRevLett.72.2531}{\emph{Phys. Rev. Lett.}
  {\bfseries 72} (1994) 2531--2533},
  [\href{https://arxiv.org/abs/hep-th/9401059}{{\ttfamily hep-th/9401059}}].

\bibitem{Bloch:2016jwt}
J.~Bloch, J.~Glesaaen, O.~Philipsen, J.~Verbaarschot and S.~Zafeiropoulos,
  \emph{{Complex Langevin simulations of a finite density matrix model for
  QCD}},  in \emph{{12th Conference on Quark Confinement and the Hadron
  Spectrum (Confinement XII) Thessaloniki, Greece, August 28-September 4,
  2016}}, 2016, \href{https://arxiv.org/abs/1612.04621}{{\ttfamily
  1612.04621}}.

\bibitem{Bloch:2017LAT}
J.~Bloch, J.~Glesaaen, J.~Verbaarschot and S.~Zafeiropoulos, \emph{{Progress on
  Complex Langevin simulations of a finite density matrix model for QCD}},  in
  \emph{{35th International Symposium on Lattice Field Theory (Lattice 2017)
  Granada, Spain, June 18-24, 2017}}, 2017.

\bibitem{Shuryak:1992pi}
E.~V. Shuryak and J.~J.~M. Verbaarschot, \emph{{Random matrix theory and
  spectral sum rules for the Dirac operator in QCD}},
  \href{https://doi.org/10.1016/0375-9474(93)90098-I}{\emph{Nucl. Phys.}
  {\bfseries A560} (1993) 306--320},
  [\href{https://arxiv.org/abs/hep-th/9212088}{{\ttfamily hep-th/9212088}}].

\bibitem{Bloch:2012ye}
J.~Bloch, \emph{{A subset solution to the sign problem in random matrix
  simulations}}, \href{https://doi.org/10.1103/PhysRevD.86.074505}{\emph{Phys.
  Rev.} {\bfseries D86} (2012) 074505},
  [\href{https://arxiv.org/abs/1205.5500}{{\ttfamily 1205.5500}}].

\bibitem{Bloch:2011jx}
J.~Bloch, \emph{{Evading the sign problem in random matrix simulations}},
  \href{https://doi.org/10.1103/PhysRevLett.107.132002}{\emph{Phys. Rev. Lett.}
  {\bfseries 107} (2011) 132002},
  [\href{https://arxiv.org/abs/1103.3467}{{\ttfamily 1103.3467}}].

\bibitem{Mollgaard:2013qra}
A.~Mollgaard and K.~Splittorff, \emph{{Complex Langevin Dynamics for chiral
  Random Matrix Theory}},
  \href{https://doi.org/10.1103/PhysRevD.88.116007}{\emph{Phys. Rev.}
  {\bfseries D88} (2013) 116007},
  [\href{https://arxiv.org/abs/1309.4335}{{\ttfamily 1309.4335}}].

\bibitem{Mollgaard:2014mga}
A.~Mollgaard and K.~Splittorff, \emph{{Full simulation of chiral random matrix
  theory at nonzero chemical potential by complex Langevin}},
  \href{https://doi.org/10.1103/PhysRevD.91.036007}{\emph{Phys. Rev.}
  {\bfseries D91} (2015) 036007},
  [\href{https://arxiv.org/abs/1412.2729}{{\ttfamily 1412.2729}}].

\bibitem{Nagata:2016alq}
K.~Nagata, J.~Nishimura and S.~Shimasaki, \emph{{Gauge cooling for the
  singular-drift problem in the complex Langevin method - a test in Random
  Matrix Theory for finite density QCD}},
  \href{https://doi.org/10.1007/JHEP07(2016)073}{\emph{JHEP} {\bfseries 07}
  (2016) 073}, [\href{https://arxiv.org/abs/1604.07717}{{\ttfamily
  1604.07717}}].

\bibitem{Toublan:1999hx}
D.~Toublan and J.~Verbaarschot, \emph{{Effective low-energy theories and QCD
  Dirac spectra}},
  \href{https://doi.org/10.1142/S0217979201005908}{\emph{Int.J.Mod.Phys.}
  {\bfseries B15} (2001) 1404--1415},
  [\href{https://arxiv.org/abs/hep-th/0001110}{{\ttfamily hep-th/0001110}}].

\bibitem{Alford:1998sd}
M.~G. Alford, A.~Kapustin and F.~Wilczek, \emph{{Imaginary chemical potential
  and finite fermion density on the lattice}},
  \href{https://doi.org/10.1103/PhysRevD.59.054502}{\emph{Phys. Rev.}
  {\bfseries D59} (1999) 054502},
  [\href{https://arxiv.org/abs/hep-lat/9807039}{{\ttfamily hep-lat/9807039}}].

\bibitem{Son:2000xc}
D.~Son and M.~A. Stephanov, \emph{{QCD at finite isospin density}},
  \href{https://doi.org/10.1103/PhysRevLett.86.592}{\emph{Phys.Rev.Lett.}
  {\bfseries 86} (2001) 592--595},
  [\href{https://arxiv.org/abs/hep-ph/0005225}{{\ttfamily hep-ph/0005225}}].

\bibitem{Bloch:2015coa}
J.~Bloch, J.~Mahr and S.~Schmalzbauer, \emph{{Complex Langevin in
  low-dimensional QCD: the good and the not-so-good}}, {\emph{PoS} {\bfseries
  LATTICE2015} (2016) 158}, [\href{https://arxiv.org/abs/1508.05252}{{\ttfamily
  1508.05252}}].

\bibitem{Bloch:2017sfg}
J.~Bloch, J.~Meisinger and S.~Schmalzbauer, \emph{{Reweighted complex Langevin
  and its application to two-dimensional QCD}}, {\emph{PoS} {\bfseries
  LATTICE2016} (2017) 046}, [\href{https://arxiv.org/abs/1701.01298}{{\ttfamily
  1701.01298}}].

\bibitem{Bloch:2017ods}
J.~Bloch, \emph{{Reweighting complex Langevin trajectories}},
  \href{https://doi.org/10.1103/PhysRevD.95.054509}{\emph{Phys. Rev.}
  {\bfseries D95} (2017) 054509},
  [\href{https://arxiv.org/abs/1701.00986}{{\ttfamily 1701.00986}}].

\bibitem{Nagata:2016mmh}
K.~Nagata, H.~Matsufuru, J.~Nishimura and S.~Shimasaki, \emph{{Gauge cooling
  for the singular-drift problem in the complex Langevin method --- an
  application to finite density QCD}}, {\emph{PoS} {\bfseries LATTICE2016}
  (2016) 067}, [\href{https://arxiv.org/abs/1611.08077}{{\ttfamily
  1611.08077}}].

\bibitem{Brent1973}
R.~Brent, \emph{Algorithms for Minimization without Derivatives}, ch.~4: : An
  Algorithm with Guaranteed Convergence for Finding a Zero of a Function.
\newblock Prentice-Hall, Englewood Cliffs, NJ, 1973.

\bibitem{Aarts:2010gr}
G.~Aarts and K.~Splittorff, \emph{{Degenerate distributions in complex Langevin
  dynamics: one-dimensional QCD at finite chemical potential}},
  \href{https://doi.org/10.1007/JHEP08(2010)017}{\emph{JHEP} {\bfseries 1008}
  (2010) 017}, [\href{https://arxiv.org/abs/1006.0332}{{\ttfamily 1006.0332}}].

\bibitem{Aarts:2012ft}
G.~Aarts, F.~A. James, J.~M. Pawlowski, E.~Seiler, D.~Sexty et~al.,
  \emph{{Stability of complex Langevin dynamics in effective models}},
  \href{https://doi.org/10.1007/JHEP03(2013)073}{\emph{JHEP} {\bfseries 1303}
  (2013) 073}, [\href{https://arxiv.org/abs/1212.5231}{{\ttfamily 1212.5231}}].

\bibitem{Ito:2017wun}
Y.~Ito and J.~Nishimura, \emph{{Comparative studies of the deformation
  techniques for the singular-drift problem in the complex Langevin method}},
  in \emph{{35th International Symposium on Lattice Field Theory (Lattice 2017)
  Granada, Spain, June 18-24, 2017}}, 2017,
  \href{https://arxiv.org/abs/1710.07929}{{\ttfamily 1710.07929}}.

\bibitem{Halasz:1998qr}
A.~M. Halasz, A.~D. Jackson, R.~E. Shrock, M.~A. Stephanov and J.~J.~M.
  Verbaarschot, \emph{{On the phase diagram of {QCD}}},
  \href{https://doi.org/10.1103/PhysRevD.58.096007}{\emph{Phys. Rev.}
  {\bfseries D58} (1998) 096007},
  [\href{https://arxiv.org/abs/hep-ph/9804290}{{\ttfamily hep-ph/9804290}}].

\bibitem{Nagata:2017pgc}
K.~Nagata, J.~Nishimura and S.~Shimasaki, \emph{{Complex Langevin simulation of
  QCD at finite density and low temperature using the deformation technique}},
  in \emph{{35th International Symposium on Lattice Field Theory (Lattice 2017)
  Granada, Spain, June 18-24, 2017}}, 2017,
  \href{https://arxiv.org/abs/1710.07416}{{\ttfamily 1710.07416}}.

\bibitem{Nagata:2016vkn}
K.~Nagata, J.~Nishimura and S.~Shimasaki, \emph{{The argument for justification
  of the complex Langevin method and the condition for correct convergence}},
  \href{https://doi.org/10.1103/PhysRevD.94.114515}{\emph{Phys. Rev.}
  {\bfseries D94} (2016) 114515},
  [\href{https://arxiv.org/abs/1606.07627}{{\ttfamily 1606.07627}}].

\bibitem{Halasz:1999gc}
A.~M. Halasz, J.~Osborn, M.~A. Stephanov and J.~Verbaarschot, \emph{{Random
  matrices and the convergence of partition function zeros in finite density
  QCD}}, \href{https://doi.org/10.1103/PhysRevD.61.076005}{\emph{Phys.Rev.}
  {\bfseries D61} (2000) 076005},
  [\href{https://arxiv.org/abs/hep-lat/9908018}{{\ttfamily hep-lat/9908018}}].

\end{thebibliography}

\providecommand{\href}[2]{#2}\begingroup\raggedright\endgroup

\end{document}